\documentclass[12pt]{article}
\usepackage{times}

\usepackage[a4paper,width=150mm,top=25mm,bottom=25mm]{geometry}
\usepackage[T1]{fontenc}
\usepackage[utf8]{inputenc}
\usepackage{lmodern}
\usepackage{hyperref}
\usepackage{graphicx}
\usepackage{newfloat}
\usepackage{amsmath}
\usepackage{amssymb}
\usepackage[english]{algorithm2e}
\usepackage[framemethod=TikZ]{mdframed}
\usepackage{imakeidx}
\usepackage{wrapfig}
\usepackage{titling}

\usepackage{times}

\usepackage[framemethod=TikZ]{mdframed}
\mdfdefinestyle{MyFrame}{%
    linecolor=black!15,
    outerlinewidth=0.5pt,
    roundcorner=10pt,
    innertopmargin=0.5\baselineskip,
    innerbottommargin=0.5\baselineskip,
    innerrightmargin=2pt,
    innerleftmargin=2pt,
    outermargin=20pt,
    backgroundcolor=white}

\usepackage{graphbox} % for aligning images, especially in tables
\usepackage[labelformat = empty,position=top]{subcaption} 
\usepackage[export]{adjustbox}
\usepackage{array, booktabs, cellspace, hhline, slashbox, diagbox} % for tables

\graphicspath{ {figures/} } % folder for pictures

\usepackage{hyperref} % for clickable table of contents, figs. and equations
\hypersetup{
	colorlinks,
	linkcolor={red!50!black},
	citecolor={blue!50!black},
	urlcolor={blue!80!black}
}

\numberwithin{equation}{section} % for numbering of equations according to sections

\usepackage{bbm} % for \mathbbm{1}
\newcommand{\bm}[1]{\boldsymbol{#1}}
\usepackage{titlesec} % for starting each section from a new page and changing fonts of (sub)sections

\titleformat*{\section}{\LARGE\bfseries}
\titleformat*{\subsection}{\Large\bfseries}
\titleformat*{\subsubsection}{\large\bfseries}

\usepackage{listings}
\DeclareFixedFont{\ttb}{T1}{txtt}{bx}{n}{12} % for bold
\DeclareFixedFont{\ttm}{T1}{txtt}{m}{n}{12}  % for normal
\usepackage{color}
\definecolor{deepblue}{rgb}{0,0,0.5}
\definecolor{deepred}{rgb}{0.6,0,0}
\definecolor{deepgreen}{rgb}{0,0.5,0}
\newcommand\pythonstyle{\lstset{
		language=Python,
		basicstyle=\ttm,
		otherkeywords={self},             % Add keywords here
		keywordstyle=\ttb\color{deepblue},
		emph={MyClass,__init__},          % Custom highlighting
		emphstyle=\ttb\color{deepred},    % Custom highlighting style
		stringstyle=\color{deepgreen},
		frame=tb,                         % Any extra options here
		showstringspaces=false            % 
}}
\newcommand\pythoninline[1]{{\pythonstyle\lstinline!#1!}}

\newcounter{mybox}%[section] 
\renewcommand{\themybox}{\arabic{mybox}}
\newenvironment{mybox}[2][]{%
\refstepcounter{mybox}%
\ifstrempty{#1}%
{\mdfsetup{%
frametitle={%
\tikz[baseline=(current bounding box.east),outer sep=0pt]
\node[anchor=east,rectangle,fill=blue!20]
{\strut Box~\themybox};}}
}%
{\mdfsetup{%
frametitle={%
\tikz[baseline=(current bounding box.east),outer sep=0pt]
\node[anchor=east,rectangle,fill=black!15]
{\strut Box~\themybox:~#1};}}%
}%
\mdfsetup{innertopmargin=5pt,linecolor=black!15,%
linewidth=2pt,topline=true,%
frametitleaboveskip=\dimexpr-\ht\strutbox\relax
}
\begin{mdframed}[]\relax%
\label{#2}}{\end{mdframed}}

\newcounter{exbox}%[section] 
\renewcommand{\theexbox}{\arabic{exbox}}
\newenvironment{exbox}[2][]{%
\refstepcounter{exbox}%
\ifstrempty{#1}%
{\mdfsetup{%
frametitle={%
\tikz[baseline=(current bounding box.east),outer sep=0pt]
\node[anchor=east,rectangle,fill=blue!20]
{\strut Example~\theexbox};}}
}%
{\mdfsetup{%
frametitle={%
\tikz[baseline=(current bounding box.east),outer sep=0pt]
\node[anchor=east,rectangle,fill=black!15]
{\strut Example~\theexbox:~#1};}}%
}%
\mdfsetup{innertopmargin=5pt,linecolor=black!15,%
linewidth=2pt,topline=true,%
frametitleaboveskip=\dimexpr-\ht\strutbox\relax
}
\begin{mdframed}[]\relax%
\label{#2}}{\end{mdframed}}

\newcounter{algbox}%[section] 
\renewcommand{\thealgbox}{\arabic{algbox}}
\newenvironment{algbox}[2][]{%
\refstepcounter{algbox}%
\ifstrempty{#1}%
{\mdfsetup{%
frametitle={%
\tikz[baseline=(current bounding box.east),outer sep=0pt]
\node[anchor=east,rectangle,fill=blue!20]
{\strut Algorithm~\thealgbox};}}
}%
{\mdfsetup{%
frametitle={%
\tikz[baseline=(current bounding box.east),outer sep=0pt]
\node[anchor=east,rectangle,fill=black!15]
{\strut Algorithm~\thealgbox:~#1};}}%
}%
\mdfsetup{innertopmargin=5pt,linecolor=black!15,%
linewidth=2pt,topline=true,%
frametitleaboveskip=\dimexpr-\ht\strutbox\relax
}
\begin{mdframed}[]\relax%
\label{#2}}{\end{mdframed}}
\makeatletter
\DeclareMathOperator{\sign}{sign}
\DeclareFontFamily{OMX}{MnSymbolE}{}
\DeclareSymbolFont{MnLargeSymbols}{OMX}{MnSymbolE}{m}{n}
\SetSymbolFont{MnLargeSymbols}{bold}{OMX}{MnSymbolE}{b}{n}
\DeclareFontShape{OMX}{MnSymbolE}{m}{n}{
	<-6>  MnSymbolE5
	<6-7>  MnSymbolE6
	<7-8>  MnSymbolE7
	<8-9>  MnSymbolE8
	<9-10> MnSymbolE9
	<10-12> MnSymbolE10
	<12->   MnSymbolE12
}{}
\DeclareFontShape{OMX}{MnSymbolE}{b}{n}{
	<-6>  MnSymbolE-Bold5
	<6-7>  MnSymbolE-Bold6
	<7-8>  MnSymbolE-Bold7
	<8-9>  MnSymbolE-Bold8
	<9-10> MnSymbolE-Bold9
	<10-12> MnSymbolE-Bold10
	<12->   MnSymbolE-Bold12
}{}
\let\llangle\@undefined
\let\rrangle\@undefined
\DeclareMathDelimiter{\llangle}{\mathopen}%
{MnLargeSymbols}{'164}{MnLargeSymbols}{'164}
\DeclareMathDelimiter{\rrangle}{\mathclose}%
{MnLargeSymbols}{'171}{MnLargeSymbols}{'171}
\makeatother

\date{\normalsize \today}

\usepackage{fancyhdr} % for fancy headers
\begin{document}
\pagenumbering{roman}	
	\begin{titlepage}
		\begin{center}
			\vspace*{1cm}
			
			\LARGE
			\textbf{Lecture Notes: \\ Machine Learning for the Sciences}
			
			\vspace{1.5cm}
			
			\large
			Titus Neupert,$^{1}$ Mark H Fischer,$^{1}$ Eliska Greplova,$^{2,3}$ Kenny Choo,$^{1}$ and Michael Denner$^{1}$\vspace{15pt}\\
			
			\small
			${}^{1}$Department of Physics, University of Zurich, 8057 Zurich, Switzerland\\
			${}^{2}$Kavli Institute of Nanoscience, Delft University of Technology, 2600 GA Delft, The Netherlands\\
			${}^{3}$Institute for Theoretical Physics, ETH Zurich, CH-8093, Switzerland 
			
			\vspace{1.5cm}
			\small
			\thedate
			
			\vfill
						
			\vspace{0.5cm}
			
			\footnotesize
			This lecture notes including exercises are online available at \href{https://www.ml-lectures.org}{ml-lectures.org}.\\
			If you notice mistakes or typos, please report them to \href{mailto:comments@ml-lectures.org}{comments@ml-lectures.org}.
		\end{center}
	\end{titlepage}

%-----Introduction----------------------------------------------%
	\section*{Preface to v2}

This is an introductory machine-learning course specifically developed with STEM students in mind. Our goal is to provide the interested reader with the basics to employ machine learning  in their own projects and to familiarize themself with the terminology as a foundation for further reading of the relevant literature.
\vspace{5pt}

 In these lecture notes, we discuss supervised, unsupervised, and reinforcement learning. The notes start with an exposition of machine learning methods without neural networks, such as principle component analysis, t-SNE, clustering, as well as linear regression and linear classifiers. We continue with an introduction to both basic and advanced neural-network structures such as dense feed-forward and conventional neural networks,  recurrent neural networks, restricted Boltzmann machines, (variational) autoencoders, generative adversarial networks. Questions of interpretability are discussed for latent-space representations and using the examples of dreaming and adversarial attacks. The final section is dedicated to reinforcement learning, where we introduce basic notions of value functions and policy learning. 
\vspace{5pt}

These lecture notes are based on a course taught at ETH Zurich and the University of Zurich for the first time in the fall of 2021 by Titus Neupert and Mark H Fischer. The lecture notes are further based on a shorter German version of the lecture notes published in the Springer essential series, ISBN 978-3-658-32268-7, doi:\href{https://doi.org/10.1007/978-3-658-32268-7}{https://doi.org/10.1007/978-3-658-32268-7}. The content of these lecture notes together with exercises is available under \href{https://www.ml-lectures.org}{ml-lectures.org}.
\vspace{5pt}

This second version of the lecture notes is an updated version for the course taught at the University of Zurich in spring 2022 by Mark H Fischer. In particular, the notation should be more consistent, a short discussion of the maximal-likelihood principle was introduced in Sec.~\ref{sec:maxlikelihood}, and the interpretation of latent-space variables was extended in Sec.~\ref{sec:latent-space}.

\newpage

\tableofcontents
\newpage

\pagenumbering{arabic}
%-----Introduction----------------------------------------------%
	\section[Introduction]{Introduction}
\subsection{Why machine learning for the sciences?}
Machine learning (ML) and artificial neural networks are everywhere and change our daily life more profoundly than we might be aware of.
However, these concepts are not a particularly recent invention. Their foundational principles emerged already in the 1940s. The \emph{perceptron}, the predecessor of the artificial neuron, the basic unit of many neural networks to date, was invented by Frank Rosenblatt in 1958, and even cast into a hardware realization by IBM. 

It then took half a century for these ideas to become technologically relevant. Now, artificial intelligence based on neural-network algorithms has become an integral part of data processing with widespread applications. Famous milestones include image based tasks such as recognizing a cat and distinguishing it from a dog (2012), the generation of realistic human faces (2018), but also sequence-to-sequence (seq2seq) tasks, such as language translation, or developing strategies to play and beat the best human players in games from chess to Go.

The reason for ML's tremendous success is twofold. First, the availability of big and structured data caters to machine-learning applications. Second, while deep (feed-forward) networks (made from many ``layers'' of artificial neurons) with many variational parameters are tremendously more powerful than few-layer ones, it only recently, in the last decade or so, became feasible to train such networks. This big leap is known as the ``deep learning revolution''.

Machine learning refers to algorithms that infer information from data in an implicit way. If the algorithms are inspired by the functionality of neural activity in the brain, the term \emph{cognitive} or \emph{neural} computing is used. \emph{Artificial neural networks} refer to a specific, albeit most broadly used, ansatz for machine learning.
Another field that concerns iteself with inferring information from data is statistics. In that sense, both machine learning and statistics have the same goal. However, the way this goal is achieved is markedly different: while statistics uses insights from mathematics to extract information in a well defined and deterministic manner, machine learning aims at optimizing a variational function using available data through learning.

The mathematical foundations of machine learning with neural networks are poorly understood: we do not know why deep learning works. Nevertheless, there are some exact results for special cases. For instance, certain classes of neural networks are a complete basis of smooth functions, that is, when equipped with enough variational parameters, they can approximate any smooth high-dimensional function with arbitrarily precision. Other variational functions with this property we commonly use are Taylor or Fourier series (with the coefficients as ``variational'' parameters). We can think of neural networks as a class or variational functions, for which the parameters can be efficiently optimized---learned---with respect to a desired objective.

As an example, this objective can be the classification of handwritten digits from `0' to `9'. The input to the neural network would be an image of the number, encoded in a vector of grayscale values. The output is a probability distribution saying how likely it is that the image shows a `0', `1', `2', and so on. The variational parameters of the network are adjusted until it accomplishes that task well. This is a classical example of \emph{supervised learning}. To perform the network optimization, we need data consisting of input data (the pixel images) and labels (the integer number shown on the respective image).

Our hope is that the optimized network also recognizes handwritten digits it has not seen during the learning. This property of a network is called \emph{generalization}. It stands in opposition to a tendency called \emph{overfitting}, which means that the network has learned specificities of the data set it was presented with, rather than the abstract features necessary to identify the respective digit. 
An illustrative example of overfitting is fitting a polynomial of degree $9$ to $10$ data points, which will always be a perfect fit. Does this mean that this polynomial best characterizes the behavior of the measured system? Of course not.
Fighting overfitting and creating algorithms that generalize well are key challenges in machine learning. We will study several approaches to achieve this goal.

\begin{figure}
\centering
\includegraphics[scale=0.4]{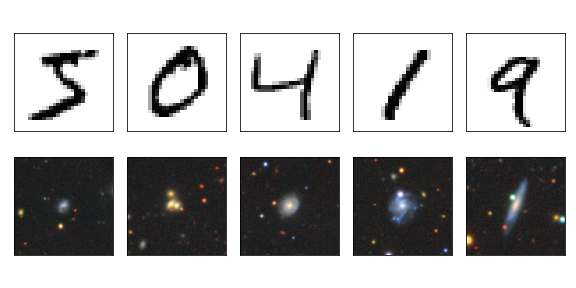}
\caption{{\bf Datasets for image classifications:} Examples of the digits from the handwritten MNIST dataset (top row) and images from the Galaxy Zoo project (bottom row).}
\label{fig:MNIST}
\end{figure}

Handwritten digit recognition has become one of the standard benchmark problems in the field. Why so? The reason is simple: there exists a very good and freely available data set for it, the MNIST database~\footnote{\href{http://yann.lecun.com/exdb/mnist}{http://yann.lecun.com/exdb/mnist}}, see top row of Fig.~\ref{fig:MNIST}. This curious fact highlights an important aspect of machine learning: it is all about data. The most efficient way to improve machine learning results is to provide more and better data. Thus, one should keep in mind that despite the widespread applications, machine learning is not the hammer for every nail. Machine learning is most beneficial if large and \emph{balanced} data sets in a machine-readable way are available, such that the algorithm can learn all aspects of the problem equally.

\begin{figure}
  \centering
  \includegraphics{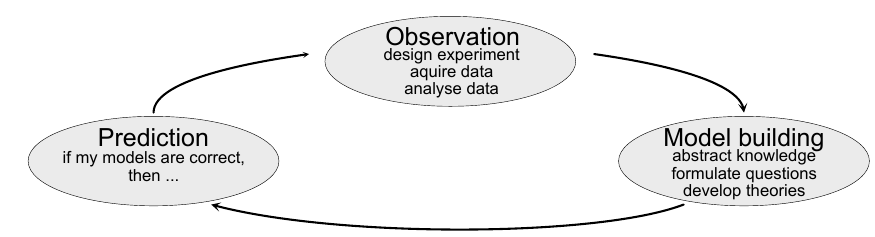}
  \caption{{\bf Simplified scientific workflow.} From observations, via abstraction to building and testing hypothesis or laws, to finally making predictions}
  \label{fig:scientific_workflow}
\end{figure}

This lecture is an introduction specifically targeting the use of machine learning in different domains of science. In scientific research, we see a vastly increasing number of applications of machine learning, mirroring the developments in industrial technology.
With that, machine learning presents itself as a universal new tool for the exact sciences, standing side-by-side with methods such as calculus, traditional statistics, and numerical simulations. This poses the question, where in the scientific workflow, summerized in Fig.~\ref{fig:scientific_workflow}, these novel methods are best employed.

While machine learning has been used in various fields that have been dealing with 'big data' problems---famous examples include the search of the Higgs boson in high-energy collision data or finding and classifying galaxies in astronomy---the last years have witnessed a host of new applications in the sciences: from high-accuracy precipitation prediction~\footnote{\href{https://www.nature.com/articles/s41586-021-03854-z}{Ravuri \emph{et al.}, Nature \textbf{597}, 672 (2021)}} to guiding human intuition in mathematics~\footnote{\href{https://www.nature.com/articles/s41586-021-04086-x}{Davis \emph{et al.}, Nature \textbf{600}, 70 (2021)}} and protein-folding simulations~\footnote{\href{https://www.nature.com/articles/s41586-021-03819-2}{Jumper \emph{et al.}, \textbf{596}, 7873 (2021)}}. The high-accuracy protein-folding prediction achieved with machine learning was named last years Science's breakthrough of the year and one of Nature's `Seven technologies to watch in 2022'.

Once a specific task has been identified, applying machine learning to the sciences does hold its very specific challenges: (i) scientific data has often very particular structure, such as the nearly perfect periodicity in an image of a crystal; (ii) typically, we have specific knowledge about correlations in the data which should be reflected in a machine learning analysis; (iii) we want to understand why a particular algorithm works, seeking a fundamental insight into mechanisms and laws of nature, in other words, developing models; (iv) in the sciences we are used to algorithms and laws that provide deterministic answers while machine learning is intrinsically probabilistic - there is no absolute certainty. Nevertheless, quantitative precision is paramount in many areas of science and thus a critical benchmark for machine learning methods.
\vspace{10pt}

\noindent{\bf A note on the concept of a model}\\
In both machine learning and the sciences, models play a crucial role. However, it is important to recognize the difference in meaning: In the (natural) sciences, a model is a conceptual representation of a phenomenon. A scientific model does not try to represent the whole world, but only a small part of it. A model is thus a simplification of the phenomenon and can be both a theoretical construct, for example the ideal gas model or the Bohr model of the atom, or an experimental simplification, such as a small version of an airplane in a wind channel.

In machine  learning, on the other hand, we most often use a complicated variational function, for example a neural network, to try to approximate a statistical model. But what is a model in statistics? Colloquially speaking, a statistical model comprises a set of statistical assumptions which allow us to calculate the probability $P(x)$ of \emph{any} event $x$. The statistical model does not correspond to the true distribution of all possible events, it simply approximates the distribution.
Scientific and statistical models thus share an important property: neither claims to be a representation of reality.

\subsection{Overview and learning goals}
This lecture is an introduction to basic machine learning algorithms for scientists and students of the sciences.
After this lecture, you will
\begin{itemize}
\setlength\itemsep{-0.2em}
    \item understand the basic terminology of the field,
    \item be able to apply the most fundamental machine learning algorithms,
    \item understand the principles of supervised and unsupervised learning and why it is so successful,
    \item know various architectures of artificial neural networks and be able to select the ones suitable for your problems,
    \item know how we find out what the machine learning algorithm uses to solve a problem.
\end{itemize}

The field of machine learning is full of lingo which to the uninitiated obscures what is at the core of the methods. Being a field in constant transformation, new terminology is being introduced at a fast pace. Our aim is to cut through slang with mathematically precise and concise formulations in order to demystify machine learning concepts for someone with an understanding of calculus and linear algebra.

As mentioned above, data is at the core of most machine learning approaches discussed in this lecture. With raw data in many cases very complex and extremely high dimensional, it is often crucial to first understand the data better and reduce their dimensionality. Simple algorithms that can be used before turning to the often heavy machinery of neural networks will be discussed in the next section, Sec.~\ref{sec:structuring_data}

\begin{figure}
  \centering
  \includegraphics{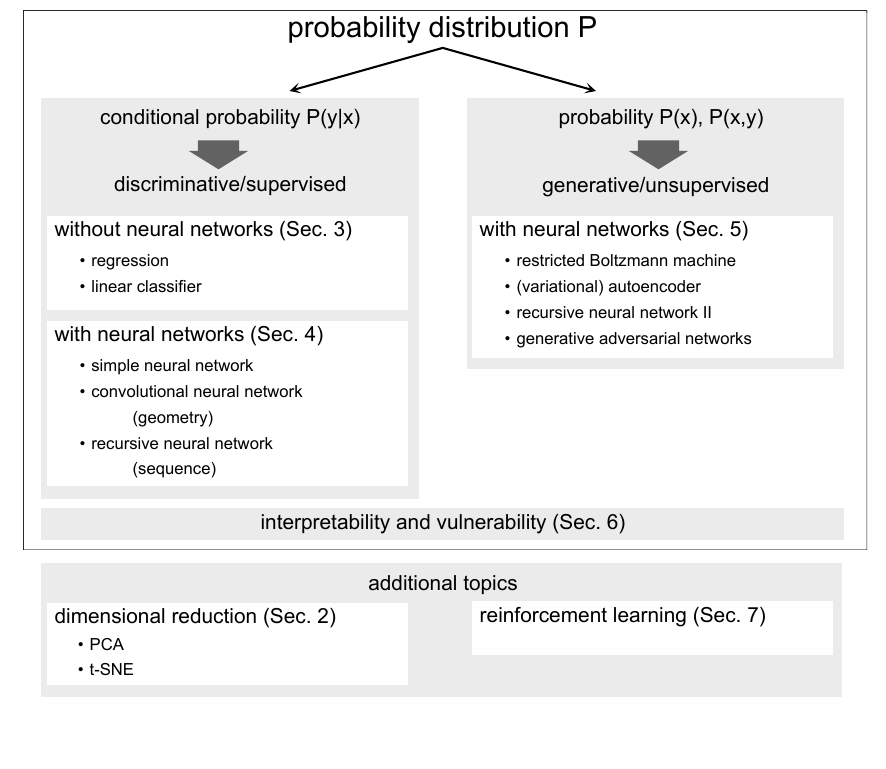}
  \caption{Overview over the plan of the lecture from the perspective of learning probability distributions.}
  \label{fig:overview}
\end{figure}

The machine learning algorithms we will focus on most can generally be divided into two classes of algorithms, namely \emph{discriminative} and \emph{generative} algorithms as illustrated in Fig.~\ref{fig:overview}. Examples of discriminative tasks include classification problems, such as the aforementioned digit classification or the classification into solid, liquid and gas phases given some experimental observables. Similarly, regression, in other words estimating relationships between variables, is a discriminative problem. Put differently, we try to approximate the conditional probability distribution $P(y|x)$ of some variable $y$ (the label) given some input data $x$. As data is provided in the form of input and target data for most of these tasks, these algorithms usually employ supervised learning. Discriminative algorithms are most straight-forwardly applicable in the sciences and we will discuss them in Secs.~\ref{sec:supervized_noNN} and \ref{sec:supervized}.

Generative algorithms, on the other hand, model a probability distribution $P(x)$. These approaches are---once trained---in principle more powerful, since we can also learn the joint probability distribution $P(x,y)$ of both the data $x$ and the labels $y$ and infer the conditional probability of $y$. Still, the more targeted approach of discriminative learning is better suited for many problems. However, generative algorithms are useful in the natural sciences, as we can sample from a known probability distribution, for example for image denoising, or when trying to find new compounds/molecules resembling known ones with given properties. These algorithms are discussed in Sec.~\ref{sec:unsupervized}.

The promise of artificial \emph{intelligence} may trigger unreasonable expectations in the sciences. After all, scientific knowledge generation is one of the most complex intellectual processes.
Computer algorithms are certainly far from achieving anything on that level of complexity and will in the near future not formulate new laws of nature independently. Nevertheless, researchers study how machine learning can help with individual segments of the scientific workflow (Fig.~\ref{fig:scientific_workflow}).
While the type of abstraction needed to formulate Newton's laws of classical mechanics seems incredibly complex, neural networks are very good at \emph{
implicit knowledge representation}. To understand precisely how they achieve certain tasks, however, is not an easy undertaking. We will discuss this question of \emph{interpretability} in Sec.~\ref{sec:interpretability}.

A third class of algorithms, which does not neatly fit the framework of approximating a statistical model and thus the distinction into discriminative and generative algorithms is known as reinforcement learning. Instead of approximating a statistical model, reinforcement learning tries to optimize strategies (actions) for achieving a given task. Reinforcement learning has gained a lot of attention with Google's AlphaGo Zero, a computer program that beat the best Go players in the world. As an example for an application in the sciences, reinforcement learning can be used to decide on what experimental configuration to perform next or to control a complex experiment. While the whole topic is beyond the scope of this lecture, we will give an introduction to the basic concepts of reinforcement learning in Sec.~\ref{sec:RL}.

A final note on the practice of learning. While the machine learning machinery is extremely powerful, using an appropriate architecture and the right training details, captured in what are called \emph{hyperparameters}, is crucial for its successful application. Though there are attempts to learn a suitable model and all hyperparameters as part of the overall learning process, this is not a simple task and requires immense computational resources. A large part of the machine learning success is thus connected to the experience of the scientist using the appropriate algorithms. 
We strongly encourage solving the accompanying exercises carefully and taking advantage of the exercise classes.

\subsection{Resources}
While it may seem that implementing ML tasks is computationally challenging, actually almost any ML task one might be interested in can be done with relatively few lines of code simply by relying on external libraries or mathematical computing systems such as Mathematica or Matlab. At the moment, most of the external libraries are written for the Python programming language.
Here are some useful Python libraries:
\begin{enumerate}
    \item \textbf{TensorFlow.} Developed by Google, Tensorflow is one of the most popular and flexible library for machine learning with complex models, with full GPU support. \textbf{Keras}, a high-level API for TensorFlow, further greatly simplifies employing even complicated models.
    \item \textbf{PyTorch.} Developed by Facebook, Pytorch is the biggest rival library to Tensorflow, with pretty much the same functionalities.
    \item \textbf{Scikit-Learn.} Whereas TensorFlow and PyTorch are catered for deep learning practitioners, Scikit-Learn provides much of the traditional machine learning tools, including linear regression and PCA.
    \item \textbf{Pandas.} Modern machine learning is largely reliant on big datasets. This library provides many helpful tools to handle these large datasets.
\end{enumerate}

\subsection{Prerequisites}
This course is aimed at students of the (natural) sciences with a basic mathematics education and some experience in programming. In particular, we assume the following prerequisites:
\begin{itemize}
  \item Basic knowledge of calculus and linear algebra.
  \item Rudimentary knowledge of statistics and probability theory (advantageous).
  \item Basic knowledge of a programming language. For the teaching assignments, you are free to choose your preferable one. The solutions will typically be distributed in Python in the form of Jupyter notebooks.
\end{itemize}
Please, don't hesitate to ask questions if any notions are unclear.

\subsection{References}
For further reading, we recommend the following books:
\begin{itemize}
	\item {\bf ML without neural networks}: \emph{The Elements of Statistical Learning}, T. Hastie, R. Tisbshirani, and J. Friedman (Springer)
	\item {\bf ML with neural networks}: \emph{Neural Networks and Deep Learning}, M. Nielson (\href{http://neuralnetworksanddeeplearning.com}{http://neuralnetworksanddeeplearning.com})
	\item {\bf Deep Learning Theory}: \emph{Deep Learning}, I. Goodfellow, Y. Bengio and A. Courville (\href{http://www.deeplearningbook.org}{http://www.deeplearningbook.org})
	\item {\bf Reinforcement Learning}: \emph{Reinforcement Learning}, R. S. Sutton and A. G. Barto (MIT Press)
\end{itemize}

%-----ML without neural networks-------------------------%
	%%%%%%%%%%%%%%%%%%%%%%%%%%%%%%%%%%%%%%%%%%%%%%%%%%%%%%%%%%%%%%%%%%%%%%%%%%%
% Machine Learning Basics %
%%%%%%%%%%%%%%%%%%%%%%%%%%%%%%%%%%%%%%%%%%%%%%%%%%%%%%%%%%%%%%%%%%%%%%%%%%%
\section{Structuring Data without Neural Networks}
\sectionmark{Structuring Data}
\label{sec:structuring_data}
Deep learning with neural networks is very much at the forefront of the recent renaissance in machine learning. However,
machine learning is not synonymous with neural networks.
There is a wealth of machine learning approaches without neural networks, and the boundary between them and conventional statistical analysis is not always sharp.

It is a common misconception that neural-network techniques would always outperform these approaches. In fact, in some cases, a simple linear method could achieve faster and better results. Even when we might eventually want to use a deep network, simpler approaches may help to understand the problem we are facing and the specificity of the data so as to better formulate our machine learning strategy. In this chapter, we shall explore machine-learning approaches without the use of neural networks. This will further allow us to introduce basic concepts and the general form of a machine-learning workflow.

\subsection{Principle component analysis}

At the heart of any machine learning task is data. In order to choose the most appropriate machine learning strategy, it is essential that we understand the data we are working with. However, very often, we are presented with a dataset containing many types of information, called \emph{features}\index{features} of the data. Such a dataset is also described as being high-dimensional. Techniques that extract information from such a dataset are broadly summarised as \emph{high-dimensional inference}.
For instance, we could be interested in predicting the progress of diabetes in patients given features such as age, sex, body mass index, or average blood pressure. Extremely high-dimensional data can occur in biology, where we might want to compare gene expression pattern in cells.
Given a multitude of features, it is neither easy to visualise the data nor pick out the most relevant information. This is where \textit{principle component analysis} (PCA) can be helpful.

Very briefly, PCA is a systematic way to find out which feature or combination of features vary the most across the data samples. We can think of PCA as approximating the data with a high-dimensional ellipsoid, where the principal axes of this ellipsoid correspond to the principal components. A feature, which is almost constant across the samples, in other words has a very short principal axis, might not be very useful. PCA then has two main applications: (1) It helps to visualise the data in a low dimensional space and (2) it can reduce the dimensionality of the input data to an amount that a more complex algorithm can handle.
\vspace{12pt}

\noindent
\textbf{PCA algorithm}\vspace{3pt}\\
Given a dataset of $m$ samples with $n$ data features, we can arrange our data in the form of a $m$ by $n$ matrix $X$ where the element $x_{ij}$ corresponds to the value of the $j$th data feature of the $i$th sample. We will also use the \emph{feature vector} $\bm{x}_i$ for all the $n$ features of one sample $i=1,\ldots,m$. The vector $\bm{x}_i$ can take values in the \emph{feature space}, for example $\bm{x}_i \in \mathbb{R}^n$. Going back to our diabetes example, we might have $10$ data features. Furthermore if we are given information regarding $100$ patients, our data matrix $X$ would have $100$ rows and $10$ columns.

The procedure to perform PCA can then be described as follows:

\begin{algbox}[Principle Component Analysis]{alg:pca}
\begin{enumerate}
    \item 
    Center the data by subtracting from each column the mean of that column,
    \begin{equation}
        \bm{x}_i \mapsto \bm{x}_{i} - \frac{1}{m} \sum_{i=1}^{m} \bm{x}_{i}.
    \end{equation}
    This ensures that the mean of each data feature is zero.
    \item 
    Form the $n$ by $n$ (unnormalized) covariance matrix
    \begin{equation}
        C = {X}^{T}{X} = \sum_{i=1}^{m} \bm{x}_{i}\bm{x}_{i}^{T}.
        \label{eqn: PCA Covariance Matrix}
    \end{equation} 
    \item 
    Diagonalize the matrix to the form $C = {X}^{T}{X} = W\Lambda W^{T}$, where the columns of $W$ are the normalized eigenvectors, or principal components, and $\Lambda$ is a diagonal matrix containing the eigenvalues. It will be helpful to arrange the eigenvalues from largest to smallest.
    \item 
    Pick the $l$ largest eigenvalues $\lambda_1, \dots \lambda_l$, $l\leq n$ and their corresponding eigenvectors $\bm{v}_1 \dots \bm{v}_l$. Construct the $n$ by $l$ matrix $\widetilde{W} = [\bm{v}_1 \dots \bm{v}_l]$.
    \item 
    Dimensional reduction: Transform the data matrix as
    \begin{equation} \label{eqn: PCA Dimensional Reduction}
        \widetilde{X} = X\widetilde{W}.
    \end{equation} 
    The transformed data matrix $\widetilde{X}$ now has dimensions $m$ by $l$.
\end{enumerate}
\end{algbox}

We have thus reduced the dimensionality of the data from $n$ to $l$. Notice that there are actually two things happening: First, of course, we now only have $l$ data features. But second, the $l$ data features are new features and not simply a selection of the original data. Rather, they are a linear combination of them. Using our diabetes example again, one of the ``new'' data features could be the sum of the average blood pressure and the body mass index. These new features are automatically extracted by the algorithm.

But why did we have to go through such an elaborate procedure to do this instead of simply removing a couple of features? The reason is that we want to maximize the \textit{variance} in our data. We will give a precise definition of the variance later in the chapter, but briefly the variance just means the spread of the data. Using PCA, we have essentially obtained $l$ ``new'' features which maximise the spread of the data when plotted as a function of this feature. We illustrate this with an example.
\vspace{12pt}

\begin{figure*}[t]
  \centering
    \includegraphics{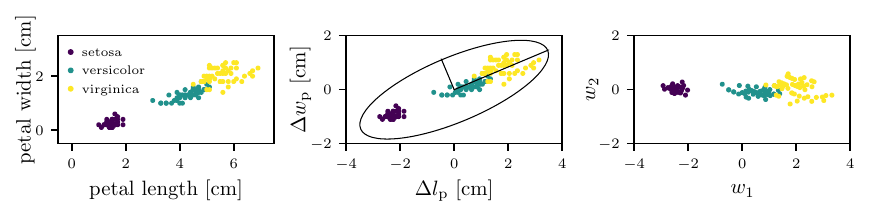}
    \caption{\textbf{PCA on Iris Dataset.} The original data (left) is first centered (center), and then, PCA 'fits an ellipsoid around the data' to extract the new feature axes (right).}
\label{fig: Iris PCA}
\end{figure*}

\noindent
\textbf{Example: The Iris dataset}\vspace{3pt}\\
Let us consider a very simple dataset with just $2$ data features. We have data, from the Iris dataset~\footnote{\href{https://archive.ics.uci.edu/ml/datasets/iris}{https://archive.ics.uci.edu/ml/datasets/iris}}, a well known dataset on 3 different species of flowers. We are given information about the petal length and petal width. Since there are just $2$ features, it is easy to visualise the data. In Fig.~\ref{fig: Iris PCA}, we show how the data is transformed under the PCA algorithm.

Notice that there is no dimensional reduction here since $l = n$. In this case, the PCA algorithm amounts simply to a rotation of the original data. However, it still produces $2$ new features which are orthogonal linear combinations of the original features:  petal length and petal width, i.e.
\begin{equation}
    \begin{split}
        w_1 &= 0.922 \times \textrm{petal length} + 0.388 \times \textrm{petal width}, \\
        w_2 &= -0.388 \times \textrm{petal length} + 0.922 \times \textrm{petal width}.
    \end{split}
\end{equation}
We see very clearly that the first new feature $w_1$ has a much larger variance than the second feature $w_2$. In fact, if we are interested in distinguishing the three different species of flowers, as in a classification task, its almost sufficient to use only the data feature with the largest variance, $w_1$. This is the essence of (PCA) dimensional reduction.

Finally, it is important to note that it is not always true that the feature with the largest variance is the most relevant for the task and it is possible to construct counter examples where the feature with the least variance contains all the useful information. However, PCA is often a good guiding principle and can yield interesting insights into the data. Most importantly, it is also \emph{interpretable}\index{interpretability}, i.e., not only does it separate the data, but we also learn \emph{which} linear combination of features can achieve this separation. We will see that for many neural network algorithms, in contrast, a lack of interpretability is a central issue.

\subsection{Kernel PCA}
\begin{figure*}[t]
    \includegraphics{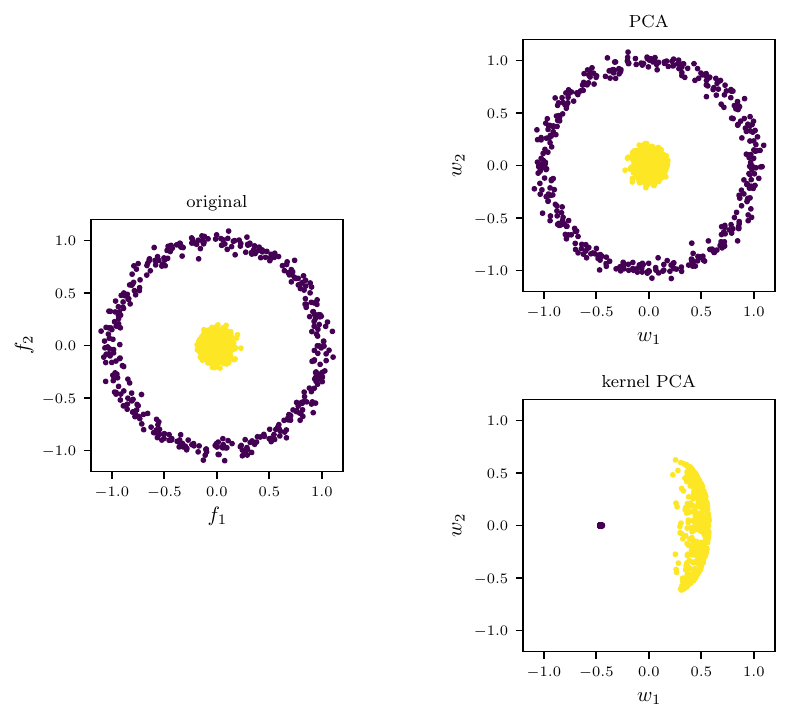}
    \caption{\textbf{Kernel PCA versus PCA.} For kernel PCA, an RBF kernel with $\gamma = 10$ was used.}
\label{fig: Kernel PCA}
\end{figure*}

PCA performs a linear transformation on our data. However, there are cases where such a transformation is unable to produce any meaningful result. Consider for instance the fictitious dataset with $2$ classes and $2$ data features as shown on the left of Fig.~\ref{fig: Kernel PCA}. We see by naked eye that it should be possible to separate this data well, for instance by the distance of the datapoint from the origin, but it is also clear that a linear function cannot be used to compute it. In this case, it can be helpful to consider a non-linear extension of PCA, known as \textit{kernel PCA}\index{kernel principle component analysis}.

The basic idea of this method is to apply to the data $\bm{x} \in \mathbb{R}^{n}$ a chosen non-linear vector-valued transformation function $\bm{\Phi}(\bm{x})$ with
\begin{equation}\label{eqn: kernel pca transformation}
    \bm{\Phi}: \mathbb{R}^{n} \rightarrow \mathbb{R}^{N},
\end{equation}
which is a map from the original $n$-dimensional space (corresponding to the $n$ original data features) to a $N$-dimensional feature space. By embedding the original data in this new space, we hope to be able to separate the data with a linear transformation, in other words, kernel PCA simply involves performing the standard PCA on the transformed data $\bm{\Phi}(\bm{x})$. Here, we will assume that the transformed data is centered, i.e.,
\begin{equation}
\sum_i \Phi(\bm{x}_i) = 0
\end{equation}
to have simpler formulas.

In practice, when $N$ is large, it is not efficient or even possible to explicitly perform the transformation $\bm{\Phi}$. Instead we can make use of a method known as the kernel trick. Recall that in standard PCA, the primary aim is to find the eigenvectors and eigenvalues of the covariance matrix $C$ in Eq.~\eqref{eqn: PCA Covariance Matrix}
. In the case of kernel PCA, this matrix becomes 
\begin{equation}
    C = \sum_{i=1}^{m} \bm{\Phi}(\bm{x}_{i})\bm{\Phi}(\bm{x}_{i})^T,
\end{equation}
with the eigenvalue equation
\begin{equation}
    \sum_{i=1}^{m} \bm{\Phi}(\bm{x}_{i})\bm{\Phi}(\bm{x}_{i})^T \bm{v}_{j} = \lambda_{j}\bm{v}_{j}.
    \label{eqn: pca eigenvalue equation}
\end{equation}
By writing the eigenvectors $\bm{v}_{j}$ as a linear combination of the transformed data features
\begin{equation}
    \bm{v}_{j} = \sum_{i=1}^{m} a_{ji}\bm{\Phi}(\bm{x}_{i}),
\end{equation}
we see that finding the eigenvectors is equivalent to finding the coefficients $a_{ji}$. On substituting this form back into Eq.~\eqref{eqn: pca eigenvalue equation}, we find
\begin{equation}
    \sum_{i=1}^{m} \bm{\Phi}(\bm{x}_{i})\bm{\Phi}(\bm{x}_{i})^T \left[ \sum_{i=1}^{m} a_{ji}\bm{\Phi}(\bm{x}_{j}) \right] = \lambda_{j} \left[ \sum_{i=1}^{m} a_{ji}\bm{\Phi}(\bm{x}_{i}) \right].
\end{equation}
By multiplying both sides of the equation by $\bm{\Phi}(\bm{x}_{k})^{T}$ we arrive at
\begin{equation}\label{eqn: kernel pca eigen equation}
    \begin{split}
        \sum_{i=1}^{m} \bm{\Phi}(\bm{x}_{k})^{T} \bm{\Phi}(\bm{x}_{i})\bm{\Phi}(\bm{x}_{i})^T \left[ \sum_{l=1}^{m} a_{jl}\bm{\Phi}(\bm{x}_{l}) \right] &= \lambda_{j} \bm{\Phi}(\bm{x}_{k})^{T} \left[ \sum_{l=1}^{m} a_{jl} \bm{\Phi}(\bm{x}_{l}) \right] \\
        \sum_{i=1}^{m} \left[ \bm{\Phi}(\bm{x}_{k})^{T} \bm{\Phi}(\bm{x}_{i}) \right]   \sum_{l=1}^{m} a_{jl} \left[ \bm{\Phi}(\bm{x}_{i})^T \bm{\Phi}(\bm{x}_{l}) \right] &= \lambda_{j} \sum_{l=1}^{m} a_{jl} \left[ \bm{\Phi}(\bm{x}_{k})^{T} \bm{\Phi}(\bm{x}_{l}) \right] \\
        \sum_{i=1}^{m} K(\bm{x}_{k},\bm{x}_{i})   \sum_{l=1}^{m} a_{jl} K(\bm{x}_{i},\bm{x}_{l}) &= \lambda_{j} \sum_{l=1}^{m} a_{jl} K(\bm{x}_{k},\bm{x}_{l}), 
    \end{split}
\end{equation}
where $K(\bm{x},\bm{y}) = \bm{\Phi}(\bm{x})^{T} \bm{\Phi}(\bm{y})$ is known as the \textit{kernel}. Thus, we see that if we directly specify the kernels we can avoid explicit performing the transformation $\bm{\Phi}$. In matrix form, we find the eigenvalue equation $K^{2}\bm{a}_{j} = \lambda_{j} K \bm{a}_{j}$, which simplifies to
\begin{equation}
    K \bm{a}_{j} = \lambda_{j} \bm{a}_{j}.
\end{equation}
Note that this simplification requires $\lambda_j \neq 0$, which will be the case for relevant principle components. (If $\lambda_{j} = 0$, then the corresponding eigenvectors would be irrelevant components to be discarded.) 
After solving the above equation and obtaining the coefficients $a_{jl}$, the kernel PCA transformation is then simply given by the overlap with the eigenvectors $\bm{v}_{j}$, namely%
\begin{equation}
    \bm{x} \rightarrow \bm{\Phi}(\bm{x}) \rightarrow y_{j} = \bm{v}_{j}^{T}\bm{\Phi}(\bm{x}) = \sum_{i=1}^{m} a_{ji} \bm{\Phi}(\bm{x}_{i})^{T} \bm{\Phi}(\bm{x}) = \sum_{i=1}^{m} a_{ji} K(\bm{x}_{i},\bm{x}),
\end{equation}
where once again the explicit $\bm{\Phi}$ transformation is avoided.

Common choices for the kernel are the polynomial kernel
\begin{equation}
	K_{\textrm{poly}}(\bm{x},\bm{y}) = (\bm{x}^T \bm{y} + c)^d
\end{equation}
or the Gaussian kernel, also known as the radial basis function (RBF) kernel, defined by
\begin{equation}
    K_{\textrm{RBF}}(\bm{x},\bm{y}) = \exp\left( -\gamma \| \bm{x} - \bm{y}\|^{2} \right),
\end{equation}
where $c$, $d$, and $\gamma$ are tunable parameters and $ \| \bm{x} - \bm{y}  \| = \sqrt{\sum_j (x_j - y_j)^2}$ denotes the Euclidian distance. Using the RBF kernel, we compare the result of kernel PCA with that of standard PCA, as shown on the right of Fig.~(\ref{fig: Kernel PCA}). It is clear that kernel PCA leads to a meaningful separation of the data while standard PCA completely fails.

	\subsection{t-SNE as a nonlinear visualization technique}

%based on 
%https://www.jmlr.org/papers/volume9/vandermaaten08a/vandermaaten08a.pdf

We studied (kernel) PCA as an example for a method that reduces the dimensionality of a dataset and makes features apparent by which data points can be efficiently distinguished. 
Often, it is desirable to more clearly cluster similar data points and visualize this clustering in a low (two- or three-) dimensional space. We focus our attention on a relatively recent algorithm (from 2008) that has proven very performant. It goes by the name \emph{t-distributed stochastic neighborhood embedding} (t-SNE). Unlike PCA, t-SNE is a stochastic algorithm, in other words applying the algorithm twice will lead to different results. 

The basic idea is to think of the data (images, for instance) as objects $\bm{x}_i$ in a very high-dimensional space and characterize their relation by the Euclidean distance $||\bm{x}_i-\bm{x}_j||$ between them. These pairwise distances are mapped to a probability distribution $p_{ij}$. The same is done for the distances $ \| \bm{y}_i-\bm{y}_j\|$ of the images of the data points $\bm{y}_i$ in the low-dimensional target space. Their probability distribution is denoted $q_{ij}$. The mapping is optimized by changing the locations $\bm{y}_i$ so as to minimize the distance between the two probability distributions. Let us substantiate in the following this intuition with concrete formulas. 

In order to introduce the probability distribution in the space of data points, we first introduce the conditional probabilities 
\begin{equation}
p_{j|i}=\frac{\mathrm{exp}\left(-\|\bm{x}_i-\bm{x}_j\|^2/2\sigma_i^2\right)}
{\sum_{k\neq i}\mathrm{exp}\left(-\|\bm{x}_i-\bm{x}_k\|^2/2\sigma_i^2\right)},
	\label{eq:gaussian_distance}
\end{equation}
where the choice of variances $\sigma_i$ will be explained momentarily and $p_{i|i}=0$ by definition.  Distances are thus turned into a Gaussian distribution and the probability measures how likely the point $\bm{x}_j$ would choose $\bm{x}_i$ as its neighbor. For the probability distribution $p_{ij}$ we then use the symmetrized version (joint probability distribution)
\begin{equation}
p_{ij}=\frac{p_{i|j}+p_{j|i}}{2m}.
\end{equation}
Note that $p_{j|i}\neq p_{i|j}$ while $p_{ji}= p_{ij}$. The symmetrized $p_{ij}$ ensures that $\sum_j p_{ij}>1/(2m)$, so that each data point makes a significant contribution and outliers are not simply discarded in the minimization. 

The probability distribution in the target space is chosen to be a so-called Student t-distribution
\begin{equation}
	q_{ij}=\frac{
		(1+\|\bm{y}_i-\bm{y}_j\|^2)^{-1}
			}{
		\sum_{k\neq l}
			(1+\|\bm{y}_k-\bm{y}_l\|^2)^{-1}
			}.
\end{equation}
This choice has several advantages over the Gaussian choice in the space of the original data points: (i) it is symmetric upon interchanging $i$ and $j$, (ii) it is numerically more efficiently evaluated because there are no exponentials, (iii) it has 'fatter' tails which helps to produce more meaningful maps in the lower dimensional space. 

In order to minimize the distance between $p_{ij}$ and $q_{ij}$, we have to introduce a (real-valued) measure for the similarity between the two probability distributions. We will then use this measure as a so-called \emph{loss function}\index{loss function} or \emph{cost function}\index{cost function}, which we aim to minimize by adjusting the $\bm{y}$'s~\footnote{Note that we can also frame PCA as a optimization problem, where we are interested in finding the linear combination of features that maximize the variance. Unlike in t-SNE, this optimization can be done analytically, see exercises}. Here, we choose the \emph{Kullback-Leibler} (KL) divergence
\begin{equation}
L(\{\bm{y}_i\})=\sum_i\sum_jp_{ij}\mathrm{log}\frac{p_{ij}}{q_{ij}}~\footnote{In general, we use $\log$ for the natural logarithm. In case of a different base, this is denoted as $\log_a$.},
\label{eq:KL}
\end{equation}
which we will frequently encounter during this lecture. An important property of the KL divergence is that it is non-negative~\footnote{Importantly, while the KL divergence is a measure for the similarity of two probability distributions, it does not define a metric on the space of probability distributions, but, as the name suggests, a divergence. Amongst other things, the KL divergence is not symmetric.}.

The minimization of $L(\{\bm{y}_i\})$ with respect to the positions $\bm{y}_i$ can be achieved with a variety of methods. In the simplest case it can be gradient descent, which we will discuss in more detail in a later chapter. As the name suggests, gradient descent follows the direction of largest gradient of the cost function to find the minimum. To this end it is useful that these gradients can be calculated in a simple form
\begin{equation}
\frac{\partial L}{\partial \bm{y}_i}
=4\sum_j (p_{ij}-q_{ij})(\bm{y}_i-\bm{y}_j)(1+||\bm{y}_i-\bm{y}_j||^2)^{-1}.
\end{equation} 

\begin{figure*}[t]
    \includegraphics[width=0.5\textwidth]{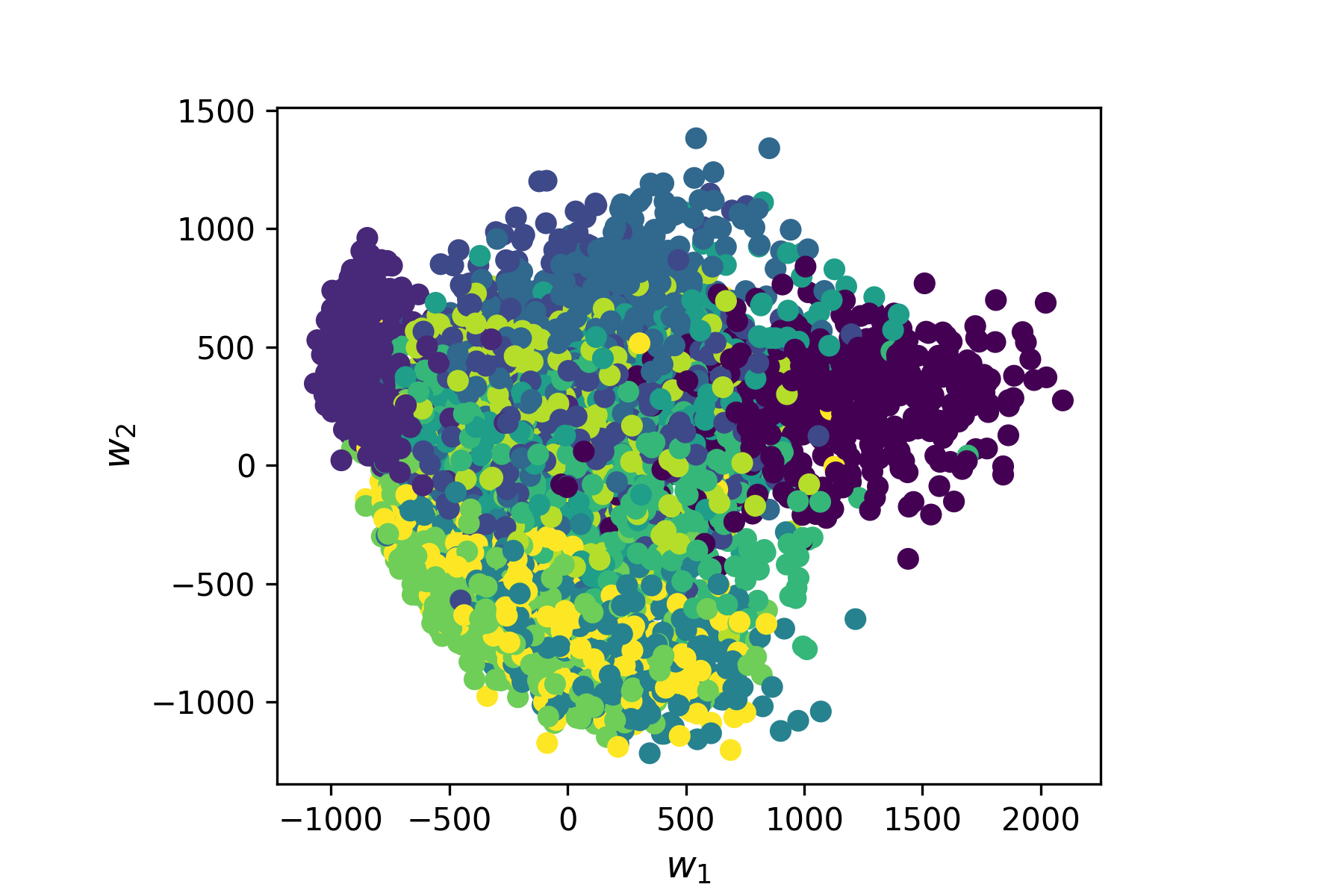}
    \includegraphics[width=0.5\textwidth]{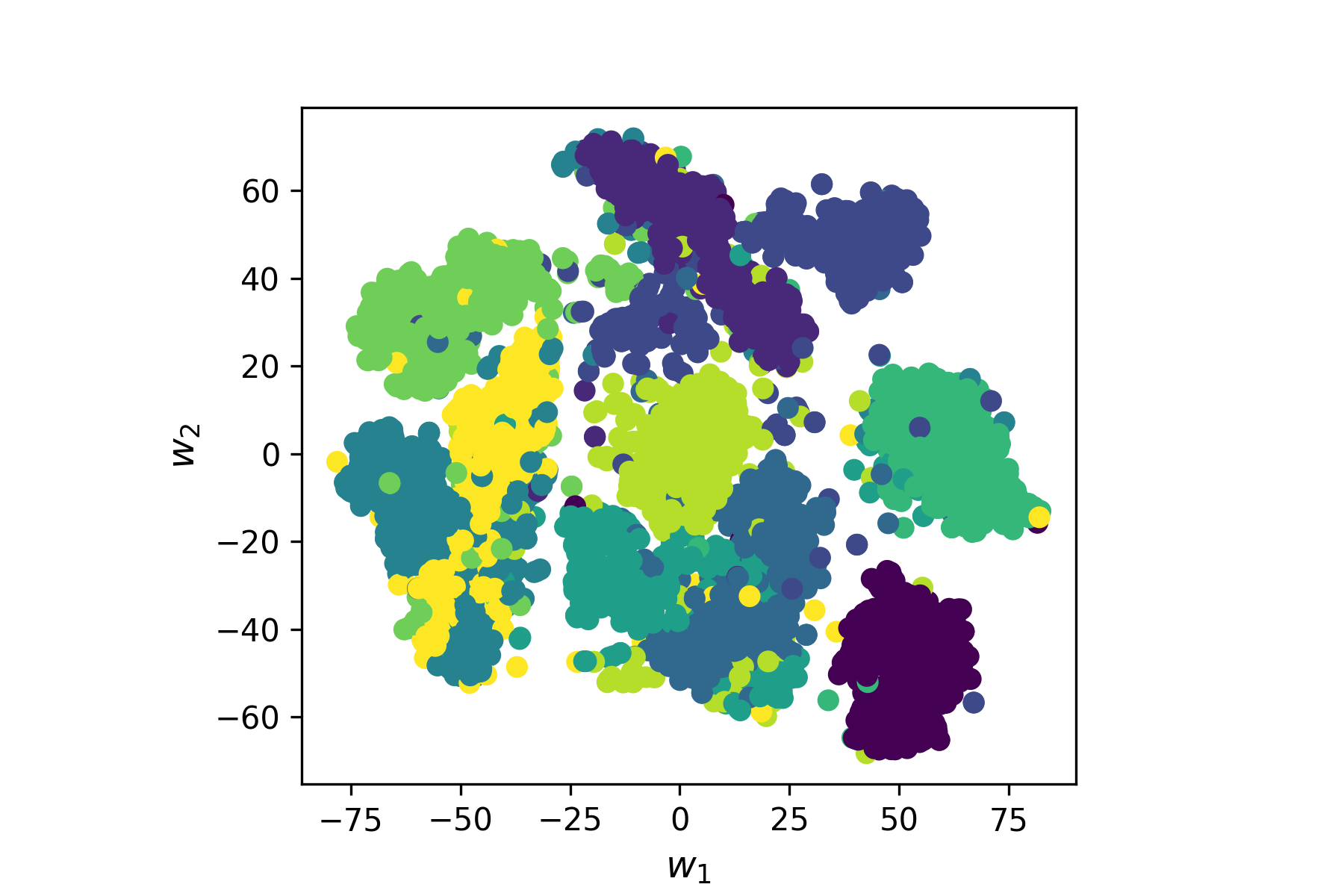}
    \caption{\textbf{PCA vs.\ t-SNE} Application of both methods on 5000 samples from the MNIST handwritten digit dataset. We see that perfect clustering cannot be achieved with either method, but t-SNE delivers the much better result.}
\label{fig: PCA vs tsne}
\end{figure*}

Finally, let us discuss the choice of the $\sigma_i$ in Eq.~\eqref{eq:gaussian_distance}. Intuitively, we want to choose $\sigma_i$ such that distances are equally well resolved in both dense and sparse regions of the dataset. As a consequence, we choose smaller values of $\sigma_i$ in dense regions as compared to sparser regions. More formally, a given value of $\sigma_i$ induces a probability distribution $P_i$ over all the other data points. This distribution has an \emph{entropy}\index{entropy} (here we use the Shannon entropy, in general it is a measure for the ``uncertainty'' represented by the distribution)
\begin{equation}
	H(P_i)=-\sum_j p_{j|i}\, \mathrm{log}_2 \,p_{j|i}.
\end{equation}
The value of $H(P_i)$ increases as $\sigma_i$ increases, in other words the more uncertainty is added to the distances. The algorithm searches for the $\sigma_i$ that result in a $P_i$ with fixed 
\emph{perplexity}\index{perplexity} 
\begin{equation}
\mathrm{Perp}(P_i)=2^{H(P_i)}.
\end{equation}
The target value of the perplexity is chosen a priory and is the main parameter that controls the outcome of the t-SNE algorithm. It can be interpreted as a smooth measure for the effective number of neighbors. Typical values for the perplexity are between 5 and 50.

By now, t-SNE is implemented as standard in many packages. They involve some extra tweaks that force points $\bm{y}_i$ to stay close together at the initial steps of the optimization and create a lot of empty space. This facilitates the moving of larger clusters in early stages of the optimization until a globally good arrangement is found. If the dataset is very high-dimensional it is advisable to perform an initial dimensionality reduction (to somewhere between 10 and 100 dimensions, for instance) with PCA before running t-SNE. 

While t-SNE is a very powerful clustering technique, it has its limitations. (i) The target dimension should be 2 or 3, for much larger dimensions the ansatz for $q_{ij}$ is not suitable. (ii) If the dataset is intrinsically high-dimensional (so that also the PCA pre-processing fails), t-SNE may not be a suitable technique. (iii) Due to the stochastic nature of the optimization, results are not reproducible. The result may end up looking very different when the algorithm is initialized with some slightly different initial values for $\bm{y}_i$.

\subsection{Clustering algorithms: the example of $k$-means}

All of PCA, kernel-PCA and t-SNE may or may not deliver a visualization of the dataset, where clusters emerge. They all leave it to the observer to identify these possible clusters. 
In this section, we want to introduce an algorithm that actually clusters data, in other words, it will assign any data point to one of $k$ clusters. Here, the desired number of clusters $k$ is fixed a priori by us. This is a weakness but may be compensated by running the algorithm with different values of $k$ and assess, where the performance is best. 

We will exemplify a simple iterative clustering algorithm that goes by the name $k$-means. The key idea is that data points are assigned to clusters such that the squared distances between the data points belonging to one cluster and the centroid of the cluster is minimized. The centroid is defined as the arithmetic mean of all data points in a cluster. 

This description already suggests that we will again minimize a loss function (or maximize an expectation function, which just differs in the overall sign from the loss function).
Suppose we are given an assignment of datapoints $\bm{x}_i$ to clusters $j=1,\cdots, k$ that is represented by 
\begin{equation}
w_{ij}=\begin{cases}
1,\qquad \bm{x}_i\text{ in cluster }j,\\
0,\qquad \bm{x}_i\text{ not in cluster }j.
\end{cases}
\end{equation}
Then the loss function is given by
\begin{equation}
L(\{w_{ij}\}; \{\bm{x}_i\})=\sum_{i=1}^m\sum_{j=1}^k w_{ij}||\bm{x}_i-\bm{\mu}_j ||^2,
\end{equation}
with the centroid 
\begin{equation}
\bm{\mu}_j=\frac{\sum_iw_{ij}\bm{x}_i}{\sum_iw_{ij}}.
\label{eq: centroids}
\end{equation}

Naturally, we want to minimize the loss function with respect to the assignment $w_{ij}$. However, a change in this assignment also changes $\bm{\mu}_j$. For this reason, it is natural to use an iterative algorithm and divide each update step into two parts. The first part updates the  $w_{ij}$ according to
\begin{equation}
w_{ij}=\begin{cases}
1,\qquad \text{if } j=\mathrm{argmin}_l ||\bm{x}_i-\bm{\mu}_l||,\\
0,\qquad \text{else },
\end{cases}
\end{equation}
in other words, each data point is assigned to the nearest cluster centroid.
The second part is a recalculation of the centroids according to Eq.~\eqref{eq: centroids}.

\begin{figure}[t]
\centering
    \includegraphics[]{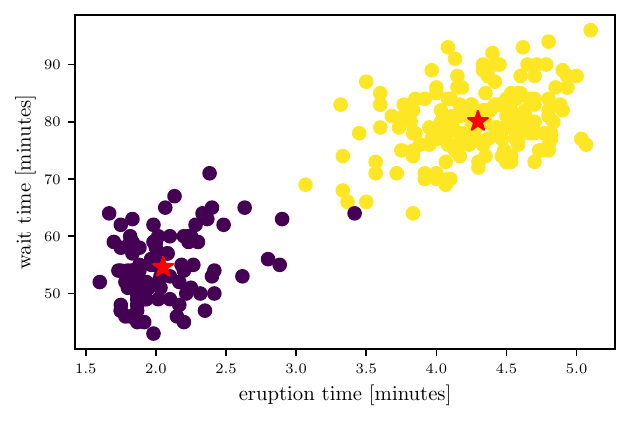}
    \caption{\textbf{K-means} Example of clustering on a dataset including two features on the Old Faithful geyser: duration of eruption and waiting time between eruption suggesting two distinct types of eruptions. The red crosses denotes the centroid location.}
\label{fig:k-means}
\end{figure}

The algorithm is initialized by choosing at random $k$ distinct data points as initial positions of the centroids. Then one repeats the above two-part steps until convergence, meaning until the $w_{ij}$ do not change anymore.
Figure~\ref{fig:k-means} shows an example of clustering on a dataset for the Old Faithful geyser~\footnote{\href{https://www.stat.cmu.edu/~larry/all-of-statistics/=data/faithful.dat}{https://www.stat.cmu.edu/~larry/all-of-statistics/=data/faithful.dat}}.

A few things to keep in mind when using $k$-means: First, 
in this algorithm we use the Euclidean distance measure $\|\cdot \|$. It is advisable to standardize the data such that each feature has mean zero and a standard deviation of one when averaging over all data points. Otherwise---if some features are overall numerically smaller than others---, the differences in various features may be weighted very differently by the algorithm.
Second,  the results of $k$-means may depend on the initialization. One should re-run the algorithm with a few different initializations to avoid running into bad local minima. Finally, $k$-means assumes spherical clusters by construction. If the actual cluster form is very far from such a spherical form, $k$-means fails to assign the points correctly.

Applications of $k$-means are manifold: in economy they include marked segmentation, in science any classification problem such as that of phases of matter, document clustering, image compression (color reduction), etc.. In general, it helps to build intuition about the data at hand. 

\section{Supervised Learning without Neural \\Networks}
\sectionmark{Supervised Learning I}
\label{sec:supervized_noNN}
\textit{Supervised learning}\index{supervised learning} is the term for a machine learning task, where we are given a dataset consisting of input-output pairs $\lbrace(\bm{x}_{1}, y_{1}), \dots, (\bm{x}_{m}, y_{m})\rbrace$ and our task is to ``learn'' a function which maps input to output $f: \bm{x} \mapsto y$. Here we chose a vector-valued input $\bm{x}$ and only a single, real number as output $y$, but in principle also the output can be vector valued. The output data that we have is called the \emph{ground truth} and sometimes also referred to as ``\emph{labels}'' of the input. In contrast to supervised learning, the algorithms presented so far were \emph{unsupervised}, because they relied on input-data $\lbrace\bm{x}_i\rbrace$ only, without any ground truth or output data.
 
Within the scope of supervised learning, there are two main types of tasks: \textit{Classification}\index{classification} and \textit{Regression}\index{regression}. In a classification task, our output $y$ is a discrete variable corresponding to a classification category. An example of such a task would be to distinguish different galaxy types from images or find stars with a planetary system (exoplanets) from those without given time series of images of such objects. On the other hand, in a regression problem, the output $y$ is a continuous number or vector. For example predicting the quantity of rainfall based on meteorological data from the previous days, the area a forest fire destroys, or how long a geyser eruption will last given the wait time. 

In this section, we first familiarize ourselves with linear methods for achieving these tasks. Neural networks, in contrast, are used as a non-linear ansatz for supervised classification and regression tasks.

\subsection{Linear regression}\label{sec:linear regression}
Linear regression, as the name suggests, simply means to fit a linear model to a dataset. Consider a dataset consisting of input-output pairs  $\lbrace(\bm{x}_{1}, y_{1}), \dots, (\bm{x}_{m}, y_{m})\rbrace$, where the inputs are $n$-component vectors $\boldsymbol{x}^{T} = (x_1, x_2, \dots , x_n)$ and the output $y$ is a real-valued number. The linear model then takes the form
\begin{equation} \label{eqn: Univariate Linear Model}
     f(\boldsymbol{x}|\bm{\beta}) = \beta_0 + \sum_{j=1}^{n} \beta_{j}x_{j} 
\end{equation}
or in matrix notation
\begin{equation}\label{eqn: Univariate Linear Model Matrix Form}
    f(\boldsymbol{x}|\bm{\beta}) = \tilde{\bm{x}}^{T}\bm{\beta}
\end{equation}
where $\bm{\tilde{x}}^{T} = (1, x_1, x_2, \dots , x_n)$ 
and $\bm{\beta} = (\beta_0, \dots, \beta_n)^{T}$ are $(n+1)$ dimensional row vectors. 

The aim then is to find parameters $\hat{\bm{\beta}}$ such that $f(\bm{x}|\hat{\bm{\beta}})$ is a good \emph{estimator}\index{estimator} for the output value $y$. In order to quantify what it means to be a ``good'' estimator, we again need to specify a \emph{loss function}\index{loss function} $L(\bm{\beta})$. The good set of parameters $\hat{\bm{\beta}}$ is then the  minimizer of this loss function
\begin{equation}
    \hat{\bm{\beta}} = \mathop{\mathrm{argmin}}_{\bm{\beta}} L(\bm{\beta}).
\end{equation}
There are many, inequivalent, choices for this loss function. For our purpose, we choose the loss function to be the \textit{residual sum of squares} (RSS) defined as
\begin{equation}\label{eqn: RSS}
\begin{split}
    \textrm{RSS}(\bm{\beta}) &= \sum_{i=1}^{m} [y_{i} -  f(\bm{x}_{i}|\bm{\beta})]^{2} \\
    &= \sum_{i=1}^{m} \left(y_{i} -  \beta_0 -\sum_{j=1}^{n} \beta_{j}x_{ij}\right)^{2},
\end{split}
\end{equation}
where the sum runs over the $m$ samples of the dataset. This loss function is sometimes also called the \emph{L2-loss} and can be seen as a measure of the distance between the output values from the dataset $y_i$ and the corresponding predictions $f(\bm{x}_i|\bm{\beta})$.

It is convenient to define the $m$ by $(n+1)$ data matrix $\widetilde{X}$, each row of which corresponds to  an input sample $\bm{\tilde{x}}^{T}_{i}$, as well as the output vector $\bm{Y}^{T} = (y_{1}, \dots, y_{m})$. With this notation, Eq.~\eqref{eqn: RSS} can be expressed succinctly as a matrix equation
\begin{equation}
    \textrm{RSS}(\bm{\beta}) = (\bm{Y} - \widetilde{X}\bm{\beta})^{T}(\bm{Y} - \widetilde{X}\bm{\beta}).
    \end{equation}
The minimum of $\textrm{RSS}(\bm{\beta})$ can be easily solved by considering the partial derivatives with respect to $\bm{\beta}$, i.e.,
\begin{equation}
\begin{split}
  &\frac{\partial \textrm{RSS}}{\partial \bm{\beta}} = -2 \widetilde{X}^{T}(\bm{Y} - \widetilde{X}\bm{\beta}), \\
  &\frac{\partial^{2} \textrm{RSS}}{\partial \bm{\beta}\partial \bm{\beta}^{T}} = 2 \widetilde{X}^{T}\widetilde{X}.
\end{split}
\end{equation}
At the minimum, $\frac{\partial \textrm{RSS}}{\partial \bm{\beta}} = 0$ and $\frac{\partial^{2} \textrm{RSS}}{\partial \bm{\beta}\partial \bm{\beta}^{T}}$ is positive-definite. Assuming $ \widetilde{X}^{T}\widetilde{X}$ is full-rank and hence invertible, we can obtain the solution $\hat{\bm{\beta}}$ as 
\begin{equation}\label{eqn: LG RSS Solution}
\begin{split}
    &\left.\frac{\partial \textrm{RSS}}{\partial \bm{\beta}}\right|_{\bm{\beta}=\hat{\bm{\beta}}} = 0, \\
      \implies &\widetilde{X}^{T}\widetilde{X}\hat{\bm{\beta}} = \widetilde{X}^{T}\bm{Y}, \\
    \implies 
 & \hat{\bm{\beta} }=  (\widetilde{X}^{T}\widetilde{X})^{-1} \widetilde{X}^{T} \bm{Y}.
\end{split}
\end{equation}
If $ \widetilde{X}^{T}\widetilde{X}$ is not full-rank, which can happen if certain data features are perfectly correlated (e.g., $x_1 = 2x_3$), the solution to $\widetilde{X}^{T}\widetilde{X}\bm{\beta} = \widetilde{X}^{T}\bm{Y}$ can still be found, but it would not be unique. Note that the RSS is not the only possible choice for the loss function and a different choice would lead to a different solution.

What we have done so far is uni-variate linear regression, that is linear regression where the output $y$ is a single, real-valued number. The generalization to the multi-variate case, where the output is a $p$-component vector $\bm{y}^{T} = (y_1, \dots y_p)$, is straightforward. The model takes the form
\begin{equation} \label{eqn: Multivariate Linear Model}
    f_{k}(\bm{x}|\beta) = \beta_{0k} + \sum_{j=1}^{n} \beta_{jk}x_{j},
\end{equation}
where the parameters $\beta_{jk}$ now have an additional index $k = 1, \dots, p$. Considering the parameters $\beta$ as a $(n+1)$ by $p$ matrix, we can show that the solution takes the same form as before [Eq.~\eqref{eqn: LG RSS Solution}] with $Y$ as a $m$ by $p$ output matrix.

	\subsubsection{Statistical analysis}
Let us stop here and evaluate the quality of the method we have just introduced. At the same time, we will take the opportunity to introduce some statistics notions, which will be useful throughout the lecture.

Up to now, we have made no assumptions about the dataset we are given, we simply stated that it consisted of input-output pairs, $\{(\bm{x}_{1}, y_{1}), \dots,$ $(\bm{x}_{m}, y_{m})\}$. In order to assess the accuracy of our model in a mathematically clean way, we have to make an additional assumption.
The output data $y_1\ldots, y_m$ may arise from some measurement or observation. Then, each of these values will generically be subject to errors $\epsilon_1,\cdots, \epsilon_m$ by which the values deviate from the ``true'' output without errors,
\begin{equation} \label{eqn: True Linear_b}
\begin{split}
        y_i &= y^{\textrm{true}}_i + \epsilon_i,\qquad i=1,\cdots,m.
\end{split}
\end{equation}
We assume that this error $\epsilon$ is a Gaussian random variable with mean $\mu = 0$ and variance $\sigma^2$, which we denote by $\epsilon \sim \mathcal{N}(0, \sigma^2)$~\footnote{Given a probability distribution $P(x)$, we in general denote by $x\sim P(x)$ that $x$ follows the probability distribution $P(x)$. Concretely, for discrete $x$ this means that we choose $x$ with probability $P(x)$, while for continuous $x$, the probability to lie in the interval $dx$ is giben by $P(x)dx$.}. Assuming that a linear model in Eq.~\eqref{eqn: Univariate Linear Model} 
is a suitable model for our dataset, we are interested in the following question: How does our solution $\hat{\bm{\beta}}$ as given in Eq.~\eqref{eqn: LG RSS Solution} 
compare with the true solution $\bm{\beta}^{\textrm{true}}$ which obeys
\begin{equation} \label{eqn: True Linear}
\begin{split}
        y_i = \beta_0^{\textrm{true}} + \sum_{j=1}^{n} \beta_{j}^{\textrm{true}}x_{ij} + \epsilon_i,\qquad
        i=1,\ldots,m?
\end{split}
\end{equation}

In order to make statistical statements about this question, we have to imagine that we can fix the inputs $\bm{x}_{i}$ of our dataset  and repeatedly draw samples for our outputs $y_i$. Each time we will obtain a different value for $y_i$ following Eq.~\eqref{eqn: True Linear}, in other words the $\epsilon_i$ are uncorrelated random numbers.
This allows us to formalise the notion of an \emph{expectation value}\index{expectation value} $E(\cdots)$ as the average over an infinite number of draws. 
For each draw, we obtain a new dataset, which differs from the other ones by the values of the outputs $y_i$. With each of these datasets, we obtain a different solution $\hat{\bm{\beta}}$ as given by Eq.~\eqref{eqn: LG RSS Solution}
. The expectation value $E(\hat{\bm{\beta}})$ is then simply the average value we obtained across an infinite number of datasets. The deviation of this average value from the ``true'' value given perfect data is called the \emph{bias} of the model,
\begin{equation}\label{eqn: Bias}
    \textrm{Bias}(\hat{\bm{\beta}}) = E(\hat{\bm{\beta}})-\bm{\beta}^{\textrm{true}}.
\end{equation}
%A nonzero value of Eq.~\eqref{eqn: Bias} indicates that ... .

For the linear regression we study here, the bias is exactly zero, because
\begin{equation}\label{eqn: LG RSS Unbiased}
    \begin{split}
      E(\hat{\bm{\beta}}) &= E\left((\widetilde{X}^{T}\widetilde{X})^{-1} \widetilde{X}^{T} (\bm{Y}^{\textrm{true}}+\bm{\epsilon})\right)\\
      &=\bm{\beta}^{\textrm{true}},
    \end{split}
\end{equation}
where the second line follows because $E(\bm{\epsilon}) = \bm{0}$ and $(\widetilde{X}^{T}\widetilde{X})^{-1} \widetilde{X}^{T} \bm{Y}^{\textrm{true}} = \bm{\beta}^{\textrm{true}}$. Equation~\eqref{eqn: LG RSS Unbiased} implies that linear regression is unbiased. Note that other machine learning algorithms will in general be biased.

What about the standard error or uncertainty of our solution? This information is contained in the \emph{covariance matrix}\index{covariance matrix}
\begin{equation}
\begin{split}
    \textrm{Var}(\hat{\bm{\beta}}) &= E\left([\hat{\bm{\beta}} - E(\hat{\bm{\beta}})][\hat{\bm{\beta}} - E(\hat{\bm{\beta}})]^{T} \right).
\end{split}
\end{equation}
The covariance matrix can be computed for the case of linear regression using the solution in Eq.~\eqref{eqn: LG RSS Solution}, the expectation value in Eq.~\eqref{eqn: LG RSS Unbiased} and the assumption in Eq.~\eqref{eqn: True Linear} that $Y = Y^{\textrm{true}} + \bm{\epsilon}$ yielding
\begin{equation}
\begin{split}
    \textrm{Var}(\hat{\bm{\beta}}) &= E\left([\hat{\bm{\beta}} - E(\hat{\bm{\beta}})][\hat{\bm{\beta}} - E(\hat{\bm{\beta}})]^{T} \right)\\
     &= E\left( \left[ (\widetilde{X}^{T}\widetilde{X})^{-1} \widetilde{X}^{T} \bm{\epsilon} \right] \left[ (\widetilde{X}^{T}\widetilde{X})^{-1} \widetilde{X}^{T} \bm{\epsilon}\right]^{T} \right) \\
    &= E\left( (\widetilde{X}^{T}\widetilde{X})^{-1} \widetilde{X}^{T} \bm{\epsilon} \bm{\epsilon}^{T} \widetilde{X} (\widetilde{X}^{T}\widetilde{X})^{-1}  \right).
\end{split}
\end{equation}
This expression can be simplified by using the fact that our input matrices $\widetilde{X}$ are independent of the draw such that
\begin{equation}
\begin{split}
\textrm{Var}(\hat{\bm{\beta}})
    &= (\widetilde{X}^{T}\widetilde{X})^{-1} \widetilde{X}^{T} E(\bm{\epsilon} \bm{\epsilon}^{T}) \widetilde{X} (\widetilde{X}^{T}\widetilde{X})^{-1} \\
    &= (\widetilde{X}^{T}\widetilde{X})^{-1} \widetilde{X}^{T} \sigma^2 I \widetilde{X} (\widetilde{X}^{T}\widetilde{X})^{-1} \\
    &= \sigma^2 (\widetilde{X}^{T}\widetilde{X})^{-1}.
\end{split}
\end{equation}
Here, the second line follows from the fact that different samples are uncorrelated, which implies that $E(\bm{\epsilon} \bm{\epsilon}^{T}) = \sigma^2 I$ with $I$ the identity matrix. The diagonal elements of $\sigma^2 (\widetilde{X}^{T}\widetilde{X})^{-1}$ then correspond to the variance\index{variance}
\begin{equation}
\begin{split}
    \textrm{Var}(\hat{\bm{\beta}}) &= E\left([\hat{\bm{\beta}} - E(\hat{\bm{\beta}})][\hat{\bm{\beta}} - E(\hat{\bm{\beta}})]^{T} \right)\\
     &= \sigma^2 (\widetilde{X}^{T}\widetilde{X})^{-1}
\end{split}
\end{equation}
of the individual parameters $\beta_i$. The standard error or uncertainty is then $\sqrt{\textrm{Var}(\hat{\beta}_{i})}$.

There is one more  missing element: we have not explained how to obtain the variances $\sigma^2$ of the outputs $y$. In an actual machine learning task, we would not know anything about the true relation, as given in Eq.~\eqref{eqn: True Linear}, governing our dataset. The only information we have access to is a single dataset. Therefore, we have to estimate the variance using the samples in our dataset, which is given by
\begin{equation}
    \hat{\sigma}^2 = \frac{1}{m - n - 1}\sum_{i=1}^{m} (y_{i} - f(\bm{x}_i|\hat{\bm{\beta}}))^2,
\end{equation}
where $y_i$ are the output values from our dataset and $f(\bm{x}_i|\hat{\bm{\beta}})$ is the corresponding prediction. Note that we normalized the above expression by $(m - n - 1)$ instead of $m$ to ensure that $E(\hat{\sigma}^2) = \sigma^2$, meaning that $\hat{\sigma}^2$ is an unbiased estimator of $\sigma^2$.

Our ultimate goal is not simply to fit a model to the dataset. We want our model to generalize\index{generalization} to inputs not within the dataset. To assess how well this is achieved, let us consider the prediction $\tilde{\bm{a}}^{T} \hat{\bm{\beta}}$  on a new random input-output pair $(\bm{a},y_{0})$. The output is again subject to an error $y_{0} = \tilde{\bm{a}}^{T}\bm{\beta}^{\textrm{true}} + \epsilon$. 
In order to compute the expected error of the prediction, we  compute the expectation value of the loss function over these previously unseen data. This is also known as the \emph{test or generalization error}\index{text error}\index{generalization error}. For the square-distance loss function, this is the \emph{mean square error} (MSE)\index{mean square error}
\begin{equation}\label{eqn: MSE Generalization Error}
    \begin{split}
        \textrm{MSE}(\hat{\bm{\beta}}) =&E\left((y_{0} - \tilde{\bm{a}}^{T} \hat{\bm{\beta}})^2\right) \\
        = &E\left((\epsilon + \tilde{\bm{a}}^{T}\bm{\beta}^{\textrm{true}} - \tilde{\bm{a}}^{T}\hat{\bm{\beta}})^2\right) \\
        = &E(\epsilon^2) +  [\tilde{\bm{a}}^{T}(\bm{\beta}^{\textrm{true}} - E(\hat{\bm{\beta}}))]^2 + E\left( [\tilde{\bm{a}}^{T}(\hat{\bm{\beta}} - E(\hat{\bm{\beta}}))]^2\right) \\
        = &\sigma^2 + [\tilde{\bm{a}}^{T}\textrm{Bias}(\hat{\bm{\beta}})]^2 + \tilde{\bm{a}}^{T} \textrm{Var}(\hat{\bm{\beta}}) \tilde{\bm{a}}.
    \end{split}
\end{equation}
There are three terms in the expression. The first term is the irreducible or intrinsic uncertainty of the dataset. The second term represents the bias and the third term is the variance of the model. For RSS linear regression, the estimate is unbiased so that
\begin{equation}
        \textrm{MSE}(\hat{\bm{\beta}}) = \sigma^2 + \tilde{\bm{a}}^{T} \textrm{Var}(\hat{\bm{\beta}}) \tilde{\bm{a}}.
\end{equation}
Based on the assumption that the dataset indeed derives from a linear model as given by Eq.~\eqref{eqn: True Linear} with a Gaussian error, it can be shown that the RSS solution, Eq.~\eqref{eqn: LG RSS Solution}, gives the minimum error among all unbiased linear estimators, Eq.~\eqref{eqn: Univariate Linear Model}.
This is known as the Gauss-Markov theorem.

This completes our error analysis of the method.

\subsubsection{Regularization and the bias-variance tradeoff}
Although the RSS solution has the minimum error among unbiased linear estimators, the expression for the generalization error, Eq.~\eqref{eqn: MSE Generalization Error}, suggests that we can actually still reduce the error by sacrificing some bias in our estimate. 

A possible way to reduce the generalization error is to simply drop some data features. From the $n$ data features $\lbrace x_{1}, \dots x_{n} \rbrace$, we can pick a reduced set $\mathcal{M}$. For example, we can choose $\mathcal{M} = \lbrace x_{1}, x_{3}, x_{7} \rbrace$, and define our new linear model as
\begin{equation}\label{eqn: Univariate Subset Linear Model}
    f(\boldsymbol{x}|\bm{\beta}) = \beta_0 + \sum_{x_j \in \mathcal{M}} \beta_{j}x_{j}.
\end{equation}
This is equivalent to fixing some parameters to zero, i.e., $\beta_k = 0$ if $x_{k} \notin \mathcal{M}$. Minimizing the RSS with this constraint results in a biased estimator but the reduction in model variance can sometimes help to reduce the overall generalization error. For a small number of features, $n \sim 20$, one can search exhaustively for the best subset of features that minimizes the error, but beyond that the search becomes computationally unfeasible.

A common alternative is called \emph{ridge regression}\index{ridge regression}. In this method, we consider the same linear model given in Eq.~\eqref{eqn: Univariate Linear Model} but with a modified loss function
\begin{equation}\label{eqn: Ridge}
    L_{\textrm{ridge}}(\bm{\beta}) = \sum_{i=1}^{m} \left[y_{i} -   f(\boldsymbol{x}_i|\bm{\beta})\right]^{2} + \lambda \sum_{j=0}^{n} \beta_{j}^{2},
\end{equation}
where $\lambda > 0$ is a positive parameter. 
This is almost the same as the RSS apart from the term proportional to $\lambda$ [c.f. Eq.~\eqref{eqn: RSS}]. The effect of this new term is to penalize large parameters $\beta_j$ and bias the model towards smaller absolute values. The parameter $\lambda$ is an example of a \emph{hyper-parameter}\index{hyper-parameter}, which is kept fixed during the training. On fixing $\lambda$ and minimizing the loss function, we obtain the solution
\begin{equation} \label{eqn: Ridge Solution}
    \hat{\bm{\beta}}_{\textrm{ridge}} = (\widetilde{X}^{T}\widetilde{X} + \lambda I)^{-1}\widetilde{X}^{T}\bm{Y},
\end{equation}
from which we can see that as $\lambda \rightarrow \infty$, $\hat{\bm{\beta}}_{\textrm{ridge}} \rightarrow \bm{0}$.
By computing the bias and variance,
\begin{equation}\label{eqn: Ridge Bias-Variance}
    \begin{split}
        \textrm{Bias}(\hat{\bm{\beta}}_{\textrm{ridge}}) &= -\lambda (\widetilde{X}^{T}\widetilde{X} + \lambda I)^{-1} \bm{\beta}^{\textrm{true}}\\
        \textrm{Var}(\hat{\bm{\beta}}_{\textrm{ridge}}) &= \sigma^2 (\widetilde{X}^{T}\widetilde{X} + \lambda I)^{-1} \widetilde{X}^{T}\widetilde{X}(\widetilde{X}^{T}\widetilde{X} + \lambda I)^{-1},
    \end{split}
\end{equation}
it is also obvious that increasing $\lambda$ increases the bias, while reducing the variance. This is the tradeoff between bias and variance. By appropriately choosing $\lambda$ it is possible that the generalization error can be reduced. We will introduce in the next section a common strategy how to find the optimal value for $\lambda$.
\begin{figure}[t]
\centering
    \includegraphics[width=0.5\textwidth]{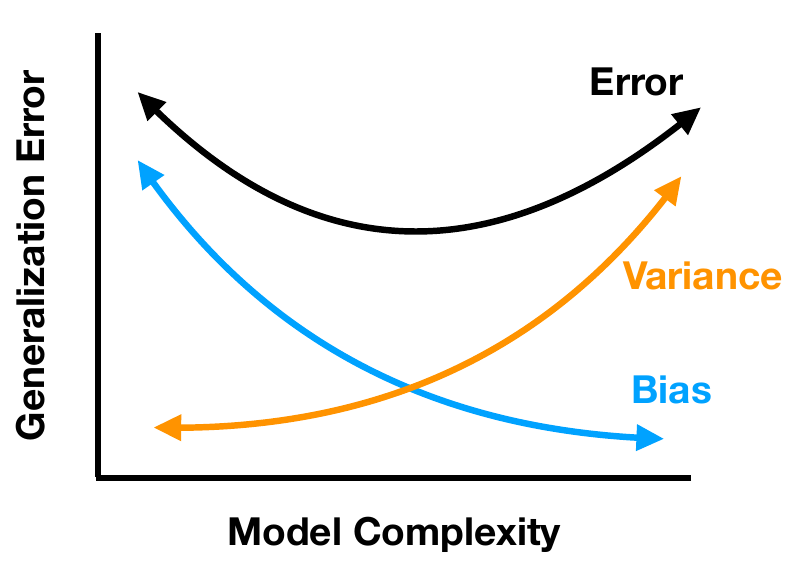}
    \caption{\textbf{Schematic depiction of the bias-variance tradeoff.}
    }
\label{fig: Bias-Variance Tradeoff}
\end{figure}

The techniques presented here to reduce the generalization error, namely dropping of features and biasing the model to small parameters, are part of a large class of methods known as \emph{regularization}\index{regularization}. Comparing the two methods, we can see a similarity. Both methods actually reduce the complexity of our model. In the former, some parameters are set to zero, while in the latter, there is a constraint which effectively reduces the magnitude of all parameters. A less complex model has a smaller variance but larger bias. By balancing these competing effects, generalization can be improved, as illustrated schematically in Fig.~\ref{fig: Bias-Variance Tradeoff}.

In the next chapter, we will see that these techniques are useful beyond applications to linear methods. We illustrate the different concepts in the following example.

\subsubsection{Example}
\begin{figure}[t]
\centering
    \includegraphics[width=0.7\textwidth]{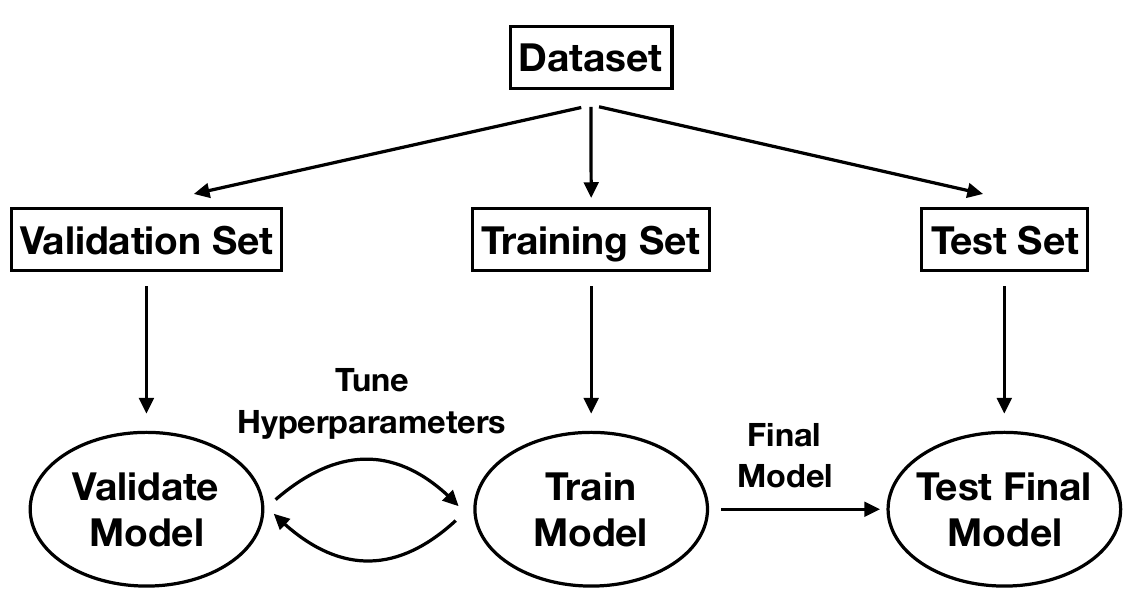}
    \caption{\textbf{Machine Learning Workflow.}}
\label{fig: ML Workflow}
\end{figure}

We illustrate the concepts of linear regression using a medical dataset. In the process, we will also familiarize ourselves with the standard machine learning workflow [see Fig.~\ref{fig: ML Workflow}]. For this example, we are given $10$ data features, namely age, sex, body mass index, average blood pressure, and six blood serum measurements from $442$ diabetes patients, and our task is to train a model $f(\bm{x}|\bm{\beta})$ [Eq.~\eqref{eqn: Univariate Linear Model}] to predict a quantitative measure of the disease progression after one year. 

Recall that the final aim of a machine-learning task is not to obtain the smallest possible value for the loss function such as the RSS, but to minimize the generalization error on unseen data [c.f. Eq.~\eqref{eqn: MSE Generalization Error}]. The standard approach relies on a division of the dataset into three subsets: training set, validation set and test set. The standard workflow is summarized in Box \ref{box: ML Workflow}.
\begin{mybox}[ML Workflow]{box: ML Workflow}
\begin{enumerate}
    \item Divide the dataset into training set $\mathcal{T}$, validation set $\mathcal{V}$ and test set $\mathcal{S}$. A common ratio for the split is $70 : 15 : 15$.
    \item Pick the hyperparameters, e.g., $\lambda$ in Eq.~\eqref{eqn: Ridge}.
    \item Train the model with only the training set, in other words minimize the loss function on the training set. [This corresponds to Eq.~\eqref{eqn: LG RSS Solution} or \eqref{eqn: Ridge Solution} for the linear regression, where $\widetilde{X}$ only contains the training set.]
    \item Evaluate the MSE (or any other chosen metric) on the validation set, [c.f. Eq.~\eqref{eqn: MSE Generalization Error}]
    \begin{equation}
        \textrm{MSE}_{\textrm{validation}}(\hat{\bm{\beta}}) = \frac{1}{|\mathcal{V}|}\sum_{j\in\mathcal{V}} (y_j - f(\bm{x}_j|\hat{\bm{\beta}}))^2.
    \end{equation}
This is known as the \emph{validation error}\index{validation error}.
    \item Pick a different value for the hyperparameters and repeat steps $3$ and $4$, until validation error is minimized.
    \item Evaluate the final model on the test set
    \begin{equation}
        \textrm{MSE}_{\textrm{test}}(\hat{\bm{\beta}}) = \frac{1}{|\mathcal{S}|}\sum_{j\in\mathcal{S}} (y_j - f(\bm{x}_j|\hat{\bm{\beta}}))^2.
    \end{equation}
\end{enumerate}
\end{mybox}

It is important to note that the test set $\mathcal{S}$ was not involved in optimizing either parameters $\bm{\beta}$ or the hyperparameters such as $\lambda$.

\begin{figure}[t]
\centering
    \includegraphics[width=1.0\textwidth]{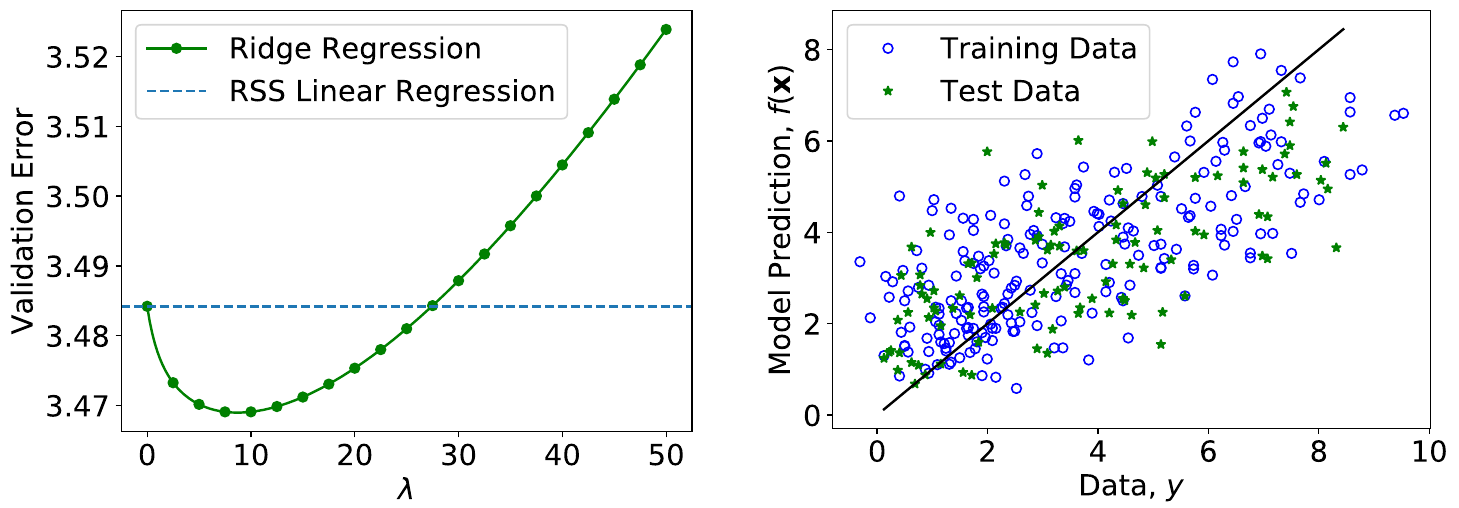}
    \caption{\textbf{Ridge Regression on Diabetes patients dataset.} Left: Validation error versus $\lambda$. Right: Test data versus the prediction from the trained model. If 
    the prediction were free of any error, all the points would fall on the black line.
    }
\label{fig: Regression on Diabetes Dataset}
\end{figure}

Applying this procedure to the diabetes dataset\footnote{Source: \href{https://www4.stat.ncsu.edu/~boos/var.select/diabetes.html}{https://www4.stat.ncsu.edu/$\small{\sim}$boos/var.select/diabetes.html}}, we obtain the results in Fig.~\ref{fig: Regression on Diabetes Dataset}. We compare RSS linear regression with the ridge regression, and indeed we see that by appropriately choosing the regularization hyperparameter $\lambda$, the generalization error can be minimized.

\begin{figure}[bh]
\centering
    \includegraphics[width=1.0\textwidth]{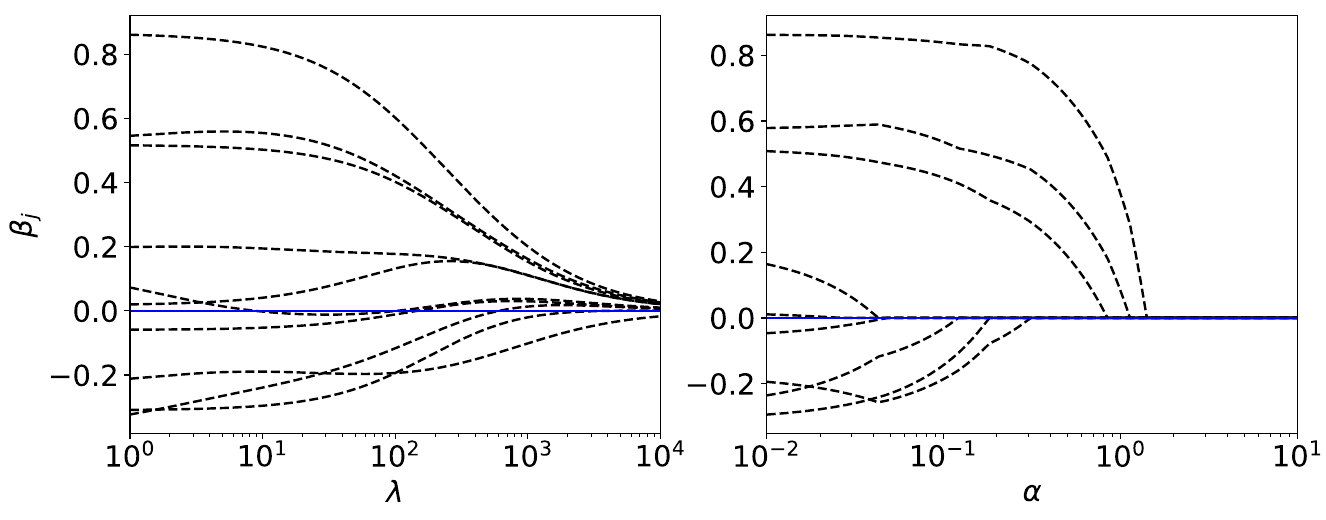}
    \caption{
    \textbf{Evolution of the model parameters.} Increasing the hyperparameter $\lambda$ or $\alpha$ leads to a reduction of the absolute value of the model parameters, here shown for the ridge (left) and Lasso (right) regression for the Diabetes dataset.
    }
\label{fig: Ridge Parameters}
\end{figure}

As side remark regarding the ridge regression, we can see on the left of Fig.~\ref{fig: Ridge Parameters}, that as $\lambda$ increases, the magnitude of the parameters, Eq.~\eqref{eqn: Ridge Solution}, $\hat{\bm{\beta}}_{\textrm{ridge}}$ decreases. Consider on the other hand, a different form of regularization, which goes by the name \emph{lasso regression}\index{lasso regression}, where the loss function is given by
\begin{equation}\label{eqn: Lasso}
    L_{\textrm{lasso}}(\bm{\beta}) = \sum_{i=1}^{m} \left[y_{i} -  f(\boldsymbol{x}_i|\bm{\beta})\right]^{2} + \alpha \sum_{j=0}^{n} |\beta_{j}|.
\end{equation}
Despite the similarities, lasso regression has a very different behavior as depicted on the right side of Fig.~\ref{fig: Ridge Parameters}. As $\alpha$ increases some parameters actually vanish and can be ignored completely. This actually corresponds to dropping certain data features completely and can be useful if we are interested in selecting the most important features in a dataset. 

	\subsection{Linear classifiers and their extensions}
\subsubsection{Binary classification and support vector machines}
In a classification problem, the aim is to categorize the inputs into one of a finite set of classes. Formulated as a supervised learning task, the dataset again consists of input-output pairs, i.e. $\lbrace(\bm{x}_{1}, y_{1}), \dots, (\bm{x}_{m}, y_{m})\rbrace$ with $\bm{x}\in \mathbb{R}^n$. However, unlike regression problems, the output $y$ is a discrete integer number representing one of the classes. In a binary classification problem, in other words a problem with only two classes, it is natural to choose $y\in\{-1, 1\}$.  

We have introduced linear regression in the previous section as a method for supervised learning when the output is a real number. Here, we will see how we can use the same model for a binary classification task.
If we look at the regression problem, we first note that geometrically 
\begin{equation} \label{eqn: Univariate Linear Model B}
     f(\boldsymbol{x}|\bm{\beta}) = \beta_0 + \sum_{j=1}^{n} \beta_{j}x_{j} = 0 
\end{equation}
defines a hyperplane perpendicular to the vector with elements $\beta_{j\geq1}$. If we fix the length $\sum_{j=1}^n \beta_j^2=1$, then $f(\bm{x}|\bm{\beta})$ measures the (signed) distance of $\bm{x}$ to the hyperplane with a sign depending on which side of the plane the point $\bm{x}_i$ lies. To use this model as a classifier, we thus define
\begin{equation}
  F(\bm{x}|\bm{\beta}) = \sign f(\bm{x}|\bm{\beta}),
  \label{eq:binaryclassifier}
\end{equation}
which yields $\{-1, +1\}$.
If the two classes are (completely) linearly separable, then the goal of the classification is to find a hyperplane that separates the two classes in feature space. 
Specifically, we look for parameters $\bm{\beta}$, such that
\begin{equation}
  y_i \tilde{\bm{x}}_i^T\bm{\beta} > M, \quad \forall i,
  \label{eq:separable}
\end{equation}
where $M$ is called the \emph{margin}.
The optimal solution $\hat{\bm{\beta}}$ then maximizes this margin. Note that instead of fixing the norm of $\beta_{j\geq1}$ and maximizing $M$, it is customary to minimize $\sum_{j=1}^n \beta_j^2$ setting $M=1$ in Eq.~\eqref{eq:separable}.

In most cases, the two classes are not completely separable. In order to still find a good classifier, we allow some of the points $\bm{x}_i$ to lie within the margin or even on the wrong side of the hyperplane. For this purpose, we rewrite the optimization constraint Eq.~\eqref{eq:separable} to  
\begin{equation}
  y_i \tilde{\bm{x}}_i^T\bm{\beta} > (1-\xi_i), \textrm{with } \xi_i \geq 0, \quad \forall i.
  \label{eq:notseparable}
\end{equation}
We can now define the optimization problem as finding 
\begin{equation}
  \min_{\bm{\beta},\{\xi_i\}} \frac12 \sum_{j=1}^{n} \beta_j^2 + C\sum_i \xi_i
  \label{eq:optimalclassifierbeta}
\end{equation}
subject to the constraint Eq.~\eqref{eq:notseparable}. Note that the second term with hyperparameter $C$ acts like a regularizer, in particular a lasso regularizer. As we have seen in the example of the previous section, such a regularizer tries to set as many $\xi_i$ to zero as possible.

\begin{figure}[bt]
  \centering
  \includegraphics{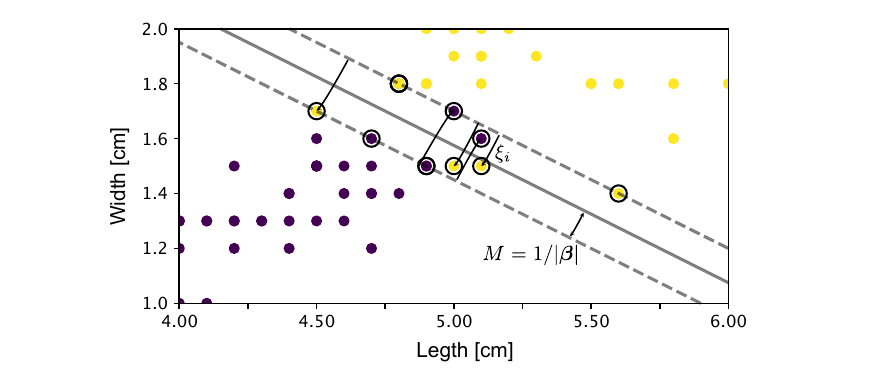}
  \caption{{\bf Binary classification.} Hyperplane separating the two classes and margin $M$ of the linear binary classifier. The support vectors are denoted by a circle around them.}
  \label{fig:svm}
\end{figure}
We can solve this constrained minimization problem by introducing Lagrange multipliers $\alpha_i$ and $\mu_i$ and solving 
\begin{equation}
  \min_{\beta, \{\xi_i\}} \frac12 \sum_{j=1}^{n} \beta_j^2 + C\sum_i \xi_i - \sum_i \alpha_i [y_i \tilde{\bm{x}}_i^T\bm{\beta} - (1-\xi_i)] - \sum_i\mu_i\xi_i,
  \label{eq:svm_lagrange}
\end{equation}
which yields the conditions
\begin{eqnarray}
  \beta_j &=& \sum_i \alpha_i y_i x_{ij},\label{eq:svm_beta}\\
  0 &=& \sum_i \alpha_i y_i\\
  \alpha_i &=& C-\mu_i, \quad \forall i.
\label{eq:svm_derivatives}
\end{eqnarray}
It is numerically simpler to solve the dual problem
\begin{equation}
  \min_{\{\alpha_i\}} \frac12 \sum_{i,i'} \alpha_i \alpha_{i'} y_i y_{i'} \bm{x}_i^T \bm{x}_{i'} - \sum_i \alpha_i
  \label{eq:svm_dual}
\end{equation}
subject to $\sum_i \alpha_i y_i =0$ and $0\leq \alpha_i \leq C$~\footnote{Note that the constraints for the minimization are not equalities, but actually inequalities. A solution thus has to fulfil the additional Karush-Kuhn-Tucker constraints
  \begin{eqnarray}
    \alpha_i [y_i \tilde{\bm{x}}_i^T\bm{\beta} - (1-\xi_i)]&=&0,\label{eq:firstKKT}\\
    \mu_i\xi_i &=& 0,\\
    y_i \tilde{\bm{x}}_i^T\bm{\beta} - (1-\xi_i)&>& 0.
  \end{eqnarray}
}.
Using Eq.~\eqref{eq:svm_beta}, we can reexpress $\beta_j$ to find
\begin{equation}
  f(\bm{x}|\{\alpha_i\}) = \sum_i{}' \alpha_i y_i \bm{x}^T \bm{x}_i + \beta_0,
  \label{eq:svm_f}
\end{equation}
where the sum only runs over the points $\bm{x}_i$, which lie within the margin, as all other points have $\alpha_i\equiv0$ [see Eq.~\eqref{eq:firstKKT}]. These points are thus called the \emph{support vectors} and are denoted in Fig.~\ref{fig:svm} with a circle around them. Finally, note that we can use Eq.~\eqref{eq:firstKKT} again to find $\beta_0$.
\vspace{11pt}

\noindent
\textbf{The Kernel trick and support vector machines}\\
We have seen in our discussion of PCA that most data is not separable linearly. However, we have also seen how the kernel trick can help us in such situations. In particular, we have seen how a non-linear function $\bm{\Phi}(\bm{x})$, which we first apply to the data $\bm{x}$, can help us separate data that is not linearly separable. Importantly, we never actually use the non-linear function $\bm{\Phi}(\bm{x})$, but only the kernel.
Looking at the dual optimization problem Eq.~\eqref{eq:svm_dual} and the resulting classifier Eq.~\eqref{eq:svm_f}, we see that, as in the case of Kernel PCA, only the kernel $K(\bm{x}, \bm{y}) = \bm{\Phi}(\bm{x})^T\bm{\Phi}(\bm{y})$ enters, simplifying the problem. This non-linear extension of the binary classifier is called a \emph{support vector machine}.

\subsubsection{More than two classes: logistic regression}
In the following, we are interested in the case of $p$ classes with $p>2$.
After the previous discussion, it seems natural for the output to take the integer values $y = 1, \dots, p$. However, it turns out to be helpful to use a different, so-called \emph{one-hot encoding}\index{one-hot encoding}. In this encoding, the output $y$ is instead represented by the $p$-dimensional unit vector in $y$ direction $\bm{e}^{(y)}$,
\begin{equation} \label{eqn: One-Hot Encoding}
    y \longrightarrow \bm{e}^{(y)} =
    \begin{bmatrix}
        e^{(y)}_1 \\
        \vdots \\
        e^{(y)}_y \\
        \vdots \\
        e^{(y)}_{p}
    \end{bmatrix}
    =
    \begin{bmatrix}
        0 \\
        \vdots \\
        1 \\
        \vdots \\
        0
    \end{bmatrix},
\end{equation}
where $e^{(y)}_l = 1$ if $l = y$  and zero for all other $l=1,\ldots, p$. A main advantage of this encoding is that we are not forced to choose a potentially biasing ordering of the classes as we would when arranging them along the ray of integers.

A linear approach to this problem then again mirrors the case for linear regression.
We fit a multi-variate linear model, Eq.~\eqref{eqn: Multivariate Linear Model}, to the one-hot encoded dataset \allowbreak$\lbrace(\bm{x}_{1}, \bm{e}^{(y_1)}), \dots, (\bm{x}_{m}, \bm{e}^{(y_m)})\rbrace$. By minimising the RSS, Eq.~\eqref{eqn: RSS}, we obtain the solution
\begin{equation}
    \hat{\beta} = (\widetilde{X}^{T}\widetilde{X})^{-1} \widetilde{X}^{T} Y,
\end{equation}
where $Y$ is the $m$ by $p$ output matrix. The prediction given an input $\bm{x}$ is then a $p$-dimensional vector $\bm{f}(\bm{x}|\hat{\beta}) = \tilde{\bm{x}}^{T} \hat{\beta}$. On a generic input $\bm{x}$, it is obvious that the components of this prediction vector would be real valued, rather than being one of the one-hot basis vectors. To obtain a class prediction $F(\bm{x}|\hat{\beta}) = 1, \dots, p$, we simply take the index of the largest component of that vector, i.e.,
\begin{equation}
    F(\bm{x}|\hat{\beta}) = \underset{k}{\textrm{argmax}} \,f_{k}(\bm{x}|\hat{\beta}).
\end{equation}
The $\textrm{argmax}$ function is a non-linear function and is a first example of what is referred to as \emph{activation function}\index{activation function}. 

For numerical minimization, it is better to use a smooth generalization of the $\textrm{argmax}$ function as activation function. Such an activation function is given by the \emph{softmax} function
\begin{equation}
  F_k(\bm{x}|\hat{\beta})= \frac{e^{f_k(\bm{x}|\hat{\beta})}}{\sum_{k'=1}^pe^{f_{k'}(\bm{x}|\hat{\beta})}}.
\end{equation}
Importantly, the output of the softmax function is now a probability $P(y = k|\bm{x})$, since $\sum_k F_k(\bm{x}|\hat{\beta}) = 1$.
This extended linear model is referred to as \emph{logistic regression}~\footnote{Note that the softmax function for two classes is the logistic function.}.

The current linear approach based on classification of one-hot encoded data generally works poorly when there are more than two classes. We will see in the next chapter that relatively straightforward non-linear extensions of this approach can lead to much better results.

%-----ML with neural networks-----------------------------%
	\section{Supervised Learning with Neural Networks}
\sectionmark{Supervised Learning II}
\label{sec:supervized}
In the previous chapter, we covered the basics of machine learning using conventional methods such as linear regression and principle component analysis. In the present chapter, we move towards a more complex class of machine learning models: \emph{neural networks}\index{neural networks}. Neural networks have been central to the recent vast success of machine learning in many practical applications.

\begin{figure}[b]
\centering
\includegraphics{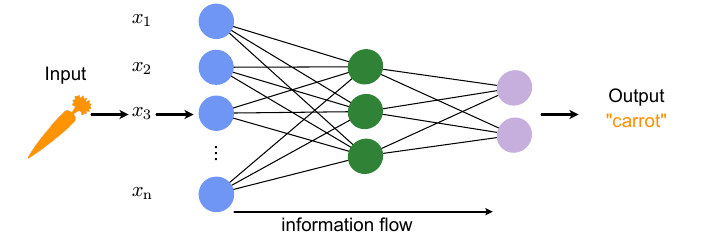}
\caption{{\bf Neural Network.} Graphical representation and basic architecture.}
 \label{fig:NN_carrot}
\end{figure}

The idea for the design of a neural network model is an analogy to how biological organisms process information. Biological brains contain neurons, electrically activated nerve cells, connected by synapses that facilitate information transfer between neurons. The machine learning equivalent of this structure, the so-called artificial neural network or neural network in short, is a mathematical function developed with the same principles in mind. It is composed from elementary functions, the \emph{neurons}\index{neuron}, which are organized in \emph{layers} that are connected to each other. To simplify the notation, a graphical representation of the neurons and network is used, see Fig.~\ref{fig:NN_carrot}. The connections in the graphical representation means that the output from one set of neurons (forming one layer) serves as the input for the next set of neurons (the next layer). This defines a sense of direction in which information is handed over from layer to layer, and thus the architecture is referred to as a \emph{feed-forward} neural network.

In general, an artificial neural network is simply an example of a variational non-linear function that maps some (potentially high-dimensional) input data to a desired output.
Neural networks are remarkably powerful and it has been  proven that under some  mild structure assumptions they can approximate any smooth function arbitrarily well as the number of neurons tends to infinity.
A drawback is that neural networks typically depend on a large amount of parameters. In the following, we will learn how to construct these neural networks and find optimal values for the variational parameters.

In this chapter, we are going to discuss one option for optimizing neural networks: the so-called \index{supervised learning}\emph{supervised learning}. A machine learning process is called supervised whenever we use training data comprising input-output pairs, in other words input with known correct answer (the label), to teach the network-required task.

\subsection{Computational neurons}
The basic building block of a neural network is the neuron. Let us consider a single neuron which we assume to be connected to $k$ neurons in the preceding layer, see Fig.~\ref{fig:NN_act} left side. The neuron corresponds to a function $f:\mathbb{R}^k\to \mathbb{R}$ which is a composition of a linear function $q:\mathbb{R}^k\to \mathbb{R}$  and a non-linear (so-called \emph{activation function}) $g:\mathbb{R}\to \mathbb{R}$. Specifically,
\begin{equation}
    f(z_1,\ldots,z_k)
    =
    g(q(z_1,\ldots,z_k))
\end{equation}
where  $z_1, z_2, \dots, z_k$ are the outputs of the neurons from the preceding layer to which the neuron is connected.

The linear function is parametrized as
\begin{equation}
q(z_1,\ldots,z_k) = \sum_{j=1}^k w_jz_j + b.
\end{equation}
Here, the real numbers $w_1, w_2, \dots, w_k$ are called \emph{weights} and can be thought of as the ``strength'' of each respective connection between neurons in the preceding layer and this neuron.
The real parameter $b$ is known as the \emph{bias} and is simply a constant offset~\footnote{Note that this bias is unrelated to the bias we learned about in regression.}. The weights and biases are the variational parameters we will need to optimize when we train the network.

The activation function $g$ is crucial for the neural network to be able to approximate any smooth function, since so far we merely performed a linear transformation. For this reason, $g$ has to be nonlinear. 
In analogy to biological neurons, $g$ represents the property of the neuron that it ``spikes'', in other words it produces a noticeable output only when the input potential grows beyond a certain threshold value. The most common choices for activation functions, shown in Fig.~\ref{fig:NN_act}, include:
\begin{figure}
  \centering
  \includegraphics{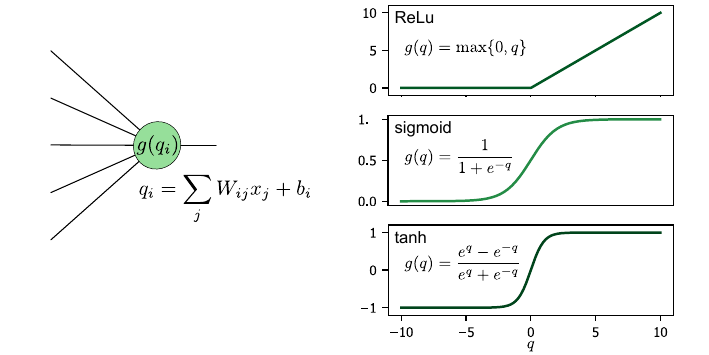}
  \caption{{\bf The artificial neuron.} Left: schematic of a single neuron and its functional form. Right: examples of the commonly used activation functions: ReLU, sigmoid function and hyperbolic tangent.}
  \label{fig:NN_act}
\end{figure}

\begin{itemize}
\item \emph{ReLU}\index{ReLU}: 
ReLU stands for rectified linear unit and is zero for all numbers smaller than zero, while a linear function for all positive numbers.
\item \emph{Sigmoid}\index{Sigmoid}:
  The sigmoid function, usually taken as the logistic function, is a smoothed version of the step function.
\item \emph{Hyperbolic tangent}\index{Hyperbolic tangent}:
The hyperbolic tangent function has a similar behavior as sigmoid but has both positive and negative values.
\item \emph{Softmax}\index{Softmax}: The softmax function is a common activation function for the last layer in a classification problem (see below).
\end{itemize}

The choice of activation function is part of the neural network architecture and is therefore not changed during training (in contrast to the variational parameters weights and bias, which are adjusted during training). Typically, the same activation function is used for all neurons in a layer, while the activation function may vary from layer to layer. Determining what a good activation function is for a given layer of a neural network is typically a heuristic rather than systematic task.

Note that the softmax provides a special case of an activation function as it explicitly depends on the output of the $q$ functions in the other neurons of the same layer. Let us label by $l=1,\ldots,n $ the $n$ neurons in a given layer and by $q_l$ the output of their respective linear transformation. Then, the \emph{softmax} is defined as
\begin{equation}
    g_l(q_1,\ldots, q_n)= \frac{e^{q_{l}}}{\sum_{l'=1}^ne^{q_{l'}}}
\end{equation}
for the output of neuron $l$. A useful property of softmax is that
\begin{equation}
    \sum_l g_l(q_1,\ldots, q_n)=1,
\end{equation}
so that the layer output can be interpreted as a probability distribution. As we saw in the previous chapter, this is particularly useful for classification tasks.

\subsection{A simple network structure}
\begin{figure}
    \centering
    \includegraphics{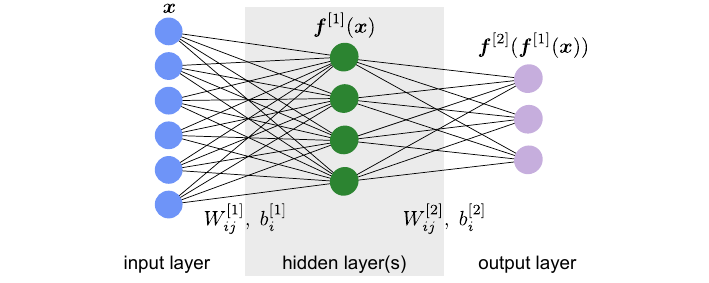}
    \caption{{\bf Simple neural network.} Architecture and variational parameters. }
    \label{fig:simple_network}
\end{figure}

Now that we understand how a single neuron works, we can connect many of them together and create an artificial neural network.  The general structure of a simple (feed-forward) neural network\index{feed-forward network} is shown in Fig.~\ref{fig:simple_network}. The first and last layers\index{layer} are the input and output layers (blue and violet, respectively, in Fig.~\ref{fig:simple_network}) and are called \emph{visible layers}\index{visible layer} as they are directly accessed. All the other layers in between them are neither accessible for input nor providing any direct output, and thus are called \emph{hidden layers}\index{hidden layer} (green layer in Fig.~\ref{fig:simple_network}).

Assuming we can feed the input to the network as a vector, we denote the input data with $\bm x$. The network then transforms this input into the output $\bm{F}(\bm{x})$, which in general is also a vector.
As a simple and concrete example, we write the complete functional form of a neural network with one hidden layer as shown in Fig.~\ref{fig:simple_network},
\begin{equation}
    \bm{F}(\bm{x})
    =
    \bm{g}^{[2]}\left(
    W^{[2]}\bm{g}^{[1]}
    \left(W^{[1]}\bm{x}+\bm{b}^{[1]}\right)+\bm{b}^{[2]}
    \right).
    \label{eq: 2-layer NN}
\end{equation}
Here, $W^{[n]}$ and $\bm{b}^{[n]}$ are the weight matrix and bias vectors of the $n$-th layer. Specifically, $W^{[1]}$ is the $k\times l$ weight matrix of the hidden layer with $k$ and $l$ the number of neurons in the input and hidden layer, respectively. $W_{ij}^{[1]}$ is the $j$-the entry of the weight vector of the $i$-th neuron in the hidden layer, while $b_i^{[1]}$ is the bias of this neuron. The $W_{ij}^{[2]}$ and $\bm{b}_i^{[2]}$ are the respective quantities for the output layer.  
This network is called \emph{fully connected}\index{fully connected} or \emph{dense}, because each neuron in a given layer takes as input the output from all the neurons in the previous layer, in other words all weights are allowed to be non-zero.

Note that for the evaluation of such a network, we first calculate all the neurons' values of the first hidden layer, which feed into the neurons of the second hidden layer and so on until we reach the output layer. This procedure, which is possible only for feed-forward neural networks, is obviously much more efficient than evaluating the nested function of each output neuron independently.

	\subsection{Training}\label{sec:training}
Adjusting all the weights and biases  to achieve the task given using data samples $\mathcal{D}= \{(\bm{x}_1,\bm{y}_1),\dots, (\bm{x}_m,\bm{y}_m)\}$ constitutes the \index{training}\emph{training} of the network. In other words, the training is the process that makes the network an approximation $\bm{F}_{\theta}(\bm{x})\approx \mathcal{F}(\bm{x})$ to the mathematical function $ \mathcal{F}(\bm{x}) = \bm{y}$ that we want it to represent. Each neuron has its own bias and weights, a potentially huge number of variational parameters that we collectively denote by $\theta = \{W,B\}$, and we will need to adjust all of them.

We have already seen in the previous chapter how one in principle trains a variational function: We introduce a \index{loss function}\emph{loss function} $L(\theta)$, which characterizes how well the network is doing at predicting the correct output for each input. The loss function now depends, through the neural network, on all the weights and biases. Once we have defined a loss function, we also already understand how to train the network: we need to minimize $L(\theta)$ with respect to $W$ and $B$. However, $L$ is typically a high-dimensional function and may have many nearly degenerate minima. 
Unlike in the previous chapter, finding the loss function's absolute minimum exactly is typically intractable analytically and may come at prohibitive costs computationally.
The practical goal is therefore rather to find a ``good'' instead than the absolute minimum through training. 
Having found such ``good'' values for $W,B$, the network can then be applied on previously unseen data.

In order to minimize the loss function numerically, we employ an iterative method called \emph{gradient descent}\index{gradient descent}~\footnote{We will discuss more elaborate variations of this method in the exercises.}. Intuitively, the method corresponds to ``walking down the hill'' in our many parameter landscape until we reach a (local) minimum. For this purpose, we use the derivatives of the cost function to update all the weights and biases incrementally and search for the minimum of the function via tiny steps on the many-dimensional surface. More specifically, we can update all weights and biases in each step as
 \begin{equation}
 \theta_\alpha \rightarrow  \theta_\alpha - \eta \frac{\partial L(\theta)}{\partial  \theta_\alpha}.
 \label{eq:gradient descent}
 \end{equation}
The variable $\eta$, also referred to as \emph{learning rate}\index{learning rate}, specifies the size of the step we use to walk the landscape. If the learning rate is too small in the beginning, we might get stuck in a local minimum early on, while for too large $\eta$ we might never find a minimum.  The learning rate is a hyperparameter\index{hyperparameter} of the training algorithm.
Note that gradient descent is just a discrete many-variable version of the analytical search for extrema which we know from calculus: An extremum is characterized by vanishing derivatives in all directions, which results in convergence in the gradient descent algorithm outlined above.

The choice of loss function may strongly impact the efficiency of the training and is based on heuristics (as was the case with the choice of activation functions). In the previous chapter, we already encountered one loss function, the \index{mean square error}mean square error
\begin{equation}
    L(\theta) = \frac{1}{2m}\sum_{i=1}^m||\bm{F}_{\theta}(\bm{x}_i) - \bm{y}_i ||_2^2.
    \label{eq:MSE}
\end{equation}
Here, $||\bm{a}||_2=\sqrt{\sum_i a_i^2}$ is the $L2$ norm and thus, this loss function is also referred to as \emph{$L2$ loss}\index{$L2$ loss|see{mean square error}}. An advantage of the L2 loss is that it is a smooth function of the variational parameters.
Another natural loss function is the \index{mean absolute error}\emph{mean absolute error}, which is given by
\begin{equation}
    L(\theta) = \frac{1}{2m}\sum_{i=1}^m||\bm{F}_{\theta}(\bm{x}_i) - \bm{y}_i ||_1,
    \label{eq:MAE}
\end{equation}
where $||\bm{a}||_1 = \sum_i |a_i|$ denotes the $L1$ norm. This loss function is thus also called the \emph{$L1$ loss}\index{$L1$ loss|see{mean absolute error}}. Note that the $L2$ norm, given the squares, puts more weight on outliers than the $L1$ loss. 

The two loss functions introduced so far are the most common loss functions for regression tasks, where the network provides a continuous output. For tasks, where each neuron outputs a probability, for example in classification problems, a great choice is the \index{cross-entropy}\emph{cross-entropy} between true label, $\bm{y}_i$ and the network output, $\bm{F}_{\theta}(\bm{x}_i)$ defined as
\begin{equation}
    L_{\mathrm{ent}}{}(\theta)
    =-\sum_{i=1}^m 
    \left[
    \bm{y}_i\cdot
    \log \left(
    \bm{F}_{\theta}(\bm{x}_i)
    \right)
+
   (1- \bm{y}_i)\cdot
    \log \left(1-
    \bm{F}_{\theta}(\bm{x}_i)
    \right)
    \right]
    ,
    \label{eq:cross-entropy}
\end{equation}
where the logarithm is taken element-wise. This loss function is also called \emph{negative log-likelihood}\index{negative log-likelihood|see{cross-entropy}}. 
It is here written for outputs that lie between 0 and 1, as is the case when the activation function of the last layer of the network is sigmoid $\sigma(z)=1/(1+e^{-z})$.

Before continuing with the intricacies of implementing such a gradient descent algorithm for (deep) neural networks, we can try to better understand the loss functions. While the L1 and L2 loss are obvious measures, the choice of cross entropy is less intuitive. The different cost functions actually differ by the speed of the learning process. The learning rate is largely determined by the partial derivatives of the cost function $\partial L/\partial \theta$, see Eq.~\eqref{eq:gradient descent} and slow learning appears when these derivatives become small. Let us consider the toy example of a single neuron with sigmoid activation $F(x)=\sigma(wx+b)$. Then, the L2 cost function has derivatives
$ \partial_w L$ and $\partial_b L$ that are proportional to $\sigma'(w+b)$.
We observe that these derivatives become very small for $\sigma(w+b)\to 1$, because $\sigma'$ is very small in that limit, leading to a slowdown of learning. This slowdown is also observed in more complex neural networks with L2 loss; we considered the simple case here only for analytical simplicity. 

Given this observation, we want to see whether the cross entropy can improve the situation. 
We again compute the derivative of the cost function with respect to the weights for a single term in the sum and a network that is composed of a single sigmoid and a general input-output pair $\{x,y\}$
\begin{equation}
\begin{split}
\frac{\partial L_{\mathrm{ent}}}{\partial w}
&=-\left(
\frac{y}{\sigma(wx+b)}-\frac{1-y}{1-\sigma(wx+b)}\right)\sigma'(wx+b)x
\\
&=\frac{\sigma'(wx+b) x}{\sigma(wx+b)[1-\sigma(wx+b)]}[\sigma(wx+b)-y]
\\
&=x[\sigma(wx+b)-y],
\end{split}
\label{eq: cost derivative w}
\end{equation}
where in the last step we used that $\sigma'(z)=\sigma(z)[1-\sigma(z)]$. This is a much better result than what we got for the L2 loss. The learning rate is here directly proportional to the error between data point and prediction $[\sigma(wx+b)-y]$. The mathematical reason for this change is that $\sigma'(z)$ cancels out due to the specific form of the cross entropy. A similar expression holds true for the derivative with respect to $b$, 
\begin{equation}
\frac{\partial L_{\mathrm{ent}}}{\partial b}
=[\sigma(wx+b)-y].
\label{eq: cost derivative b}
\end{equation}
In fact, if we insisted that we want the very intuitive form of Eqs.~\eqref{eq: cost derivative w} and~\eqref{eq: cost derivative b} for the gradients, we can derive the cost function for the sigmoid activation function to be the cross-entropy. To see this, we start from
\begin{equation}
\frac{\partial L}{\partial b}=\frac{\partial L}{\partial F}F'
\end{equation}
and $F'=F(1-F)$ for the sigmoid activation, which, comparing to the desired form of Eq.~\eqref{eq: cost derivative b}, yields
\begin{equation}
\frac{\partial L}{\partial F}=\frac{F-y}{F(1-F)}.
\label{eq:partialLF}
\end{equation}
When integrated with respect to $F$, Eq.~\eqref{eq:partialLF} gives exactly the cross-entropy (up to a constant). Starting from Eqs.~\eqref{eq: cost derivative w} and~\eqref{eq: cost derivative b}, we can thus think of choosing the cost function using backward engineering. Following this logic, we can think of other pairs of final-layer activations and cost functions that may work well together. 

What happens if we change the activation function in the last layer from sigmoid to softmax, which is usually used for the case of a classification task? For the loss function, we consider just the first term in the cross entropy---for softmax, this form is appropriate---
\begin{equation}
    L(\theta)
    =-\sum_{i=1}^m 
    \bm{y}_i\cdot
    \log \left(
    \bm{F}_{\theta}(\bm{x}_i)
    \right)
    ,
    \label{eq:cross-entropy 2}
\end{equation}
where, again, the logarithm is taken element-wise.
For concreteness, let us look at one-hot encoded classification problem. Then, all $\bm{y}_i$ labels are vectors with exactly one entry ``1''. Let that entry have index $n_i$ in the vector, such that $(\bm{y}_i)_j = \delta_{n_i,j}$.
The loss function then reads
\begin{equation}
    L(\theta)
    =-\sum_{i=1}^m 
    \log \left(
    F_{n_i}(\bm{x}_i)
    \right)
    .
    \label{eq:cross-entropy 3}
\end{equation}
Due to the properties of the softmax, $ F_{n_i}(\bm{x}_i)$ is always $\leq 1$, so that the loss function is minimized if it approaches 1, the value of the label. 
For the gradients, we obtain
\begin{equation}
\begin{split}
\frac{\partial L}{\partial b_{j}}=&
-\sum_{i=1}^m\frac{1}{F_{n_i}(\bm{x}_i)}\frac{\partial F_{n_i}(\bm{x}_i)}{\partial b_j}
\\
=&
-\sum_{i=1}^m\frac{1}{F_{n_i}(\bm{x}_i)}
\left[
F_{n_i}(\bm{x}_i)\delta_{n_i,j}
-F_{n_i}(\bm{x}_i)F_j(\bm{x}_i)
\right]
\\
=&
\sum_{i=1}^m
\left[
F_{j}(\bm{x}_i)
-\delta_{n_i,j}
\right]
\end{split}
\end{equation}
and a similar result for the derivatives with respect to the weights.
We observe that, again, the gradient has a similar favorable structure to the previous case, in that it is linearly dependent on the error that the network makes.
\vspace{5pt}

\noindent{\bf Backpropagation}\\
\noindent While the process of optimizing the many variables of the loss function is mathematically straightforward to understand, it presents a significant numerical challenge: For each variational parameter, for instance a weight in the $k$-th layer $W_{ij}^{[k]}$, the partial derivative $\partial L/ \partial W_{ij}^{[k]}$ has to be computed. And this has to be done each time the network is evaluated for a new dataset during training. Naively, one could assume that the whole network has to be evaluated for each derivative.  
Luckily there is an algorithm that allows for an efficient and parallel computation of all derivatives---this algorithm is known as  \index{backpropagation}\emph{backpropagation}. The algorithm derives directly from the chain rule of differentiation for nested functions and is based on two observations: 
\begin{itemize}
    \item [(1)] The loss function is a function of the neural network $\bm{F}_{\theta}(\bm{x})$, that is $L \equiv L(\bm{F})$. 
    \item [(2)] To determine the derivatives in layer $k$ only the derivatives of the subsequent layers, given as Jacobi matrix 
    \begin{equation} 
        D\bm{f}^{[l]}(\bm{z}^{[l-1]}) = \partial \bm{f}^{[l]}/\partial \bm{z}^{[l-1]},
    \end{equation}
    with $l>k$ and $z^{[l-1]}$ the output of the previous layer, as well as 
    \begin{equation}
        \frac{\partial \bm{z}^{[k]} }{ \partial \theta_\alpha^{[k]}} =
        \frac{\partial \bm{g}^{[k]}}{\partial q_i^{[k]}}
        \frac{{\partial q_i^{[k]}}}{\partial\theta_\alpha}
        =
        \begin{cases}
        \frac{\partial \bm{g}^{[k]}}{\partial q_i^{[k]}} z^{[k-1]}_j&\theta_\alpha=W^{[k]}_{ij}
        \\
        \frac{\partial \bm{g}^{[k]}}{\partial q_i^{[k]}} &\theta_\alpha=b^{[k]}_{i}
        \end{cases}
    \end{equation} 
    are required.
\end{itemize}
The calculation of the Jacobi matrix thus has to be performed only once for every update. In contrast to the evaluation of the network itself, which is propagating forward, (output of layer $k$ is input to layer $k+1$), we find that a change in the Output propagates  backwards through the network, hence the name\footnote{Backpropagation is actually a special case of a set of techniques known as \emph{automatic differentiation}\index{automatic differentiation} (AD). AD makes use of the fact that any computer program can be composed of elementary operations (addition, subtraction, multiplication, division) and elementary functions ($\sin, \exp, \dots$). By repeated application of the chain rule, derivatives of arbitrary order can be computed automatically.}.

The full algorithm looks then as follows:
\begin{algbox}[Backpropagation]{alg: Backpropagation}
\begin{algorithm}[H]
\label{alg: backpropagation}
\SetAlgoLined
\KwIn{Loss function $L$ that in turn depends on the neural network, which is parametrized by weights and biases, summarized as $\theta=\{W,B\}$.
}
\KwOut{Partial derivatives $\partial L / \partial \theta^{[k]}_{\alpha}$ with respect to all parameters  $\theta^{[k]}$ of all layers $k=1\dots n$.}
Calculate the derivatives with respect to the parameters of the output layer:
$\partial L / \partial W^{[n]}_{ij} = (\bm{\nabla} L)^T 
 \frac{\partial \bm{g}^{[n]}}{\partial q_i^{[n]}} z^{[n-1]}_j
$, 
$\quad\partial L / \partial b^{[n]}_{i} = (\bm{\nabla} L)^T \frac{\partial \bm{g}^{[n]}}{\partial q_i^{[n]}}$\\
 \For{$k=n-1$\dots $1$}{
Calculate the Jacobi matrices for layer $k$: $D\bm{g}^{[k]}=(\partial \bm{g}^{[k]}/\partial \bm{q}^{[k]})$ and $D\bm{f}^{[k]}=(\partial \bm{f}^{[k]}/\partial \bm{z}^{[k-1]})$\;
 Multiply all following Jacobi matrices to obtain the derivatives of layer $k$:
  $\partial L / \partial \theta^{[k]}_{\alpha} = (\nabla L)^T D\bm{f}^{[n]}\cdots D\bm{f}^{[k+1]}D\bm{g}^{[k]} (\partial \bm{q}^{[k]}/\partial \theta^{[k]}_\alpha)$
\;
 }
\end{algorithm}
\end{algbox}

A remaining question is when to actually perform updates to the network parameters. One possibility would be to perform the above procedure for each training data individually. Another extreme is to use all the training data available and perform the update with an averaged derivative. Not surprisingly, the answer lies somewhere in the middle: Often, we do not present training data to the network one item at the time, but the full training data is divided into co-called \emph{mini-batches}\index{mini-batch}, a group of training data that is fed into the network together. Chances are the weights and biases can be adjusted better if the network is presented with more information in each training step. However, the price to pay for larger mini-batches is a higher computational cost. Therefore, the mini-batch size (often simply called batch size) can greatly impact the efficiency of training. The random partitioning of the training data into batches is kept for a certain number of iterations, before a new partitioning is chosen. The consecutive iterations carried out with a chosen set of batches constitute a training \index{epoch}\emph{epoch}. 

\subsection{Simple example: MNIST}\label{sec:simple MNIST}
As we discussed in the introduction, the recognition of hand-written digits $0$, $1$, $\ldots 9$ is the ``Drosophila'' of machine learning with neural networks. There is a dataset with tens of thousands of examples of hand-written digits, the so-called MNIST data set.
Each data sample in the MNIST dataset, a $28\times28$ grayscale image, comes with a \emph{label}, which holds the information which digit is stored in the image. The difficulty of learning to recognize the digits is that handwriting styles are incredibly personal and different people will write the digit ``4'' slightly differently. It would be very challenging to hardcode all the criteria to recognize ``4'' and not confuse it with, say, a ``9''.

We can use a simple neural network as introduced earlier in the chapter to tackle this complex task. We will use a network as shown in Fig.~\ref{fig:simple_network} and given in Eq.~\eqref{eq: 2-layer NN} to do just that. The input is the image of the handwritten digit, transformed into a $k=28^2$ long vector,
the hidden layer contains $l$ neurons and the output layer has  $p=10$ neurons, each corresponding to one digit in the one-hot encoding. The output is then a probability distribution over these 10 neurons that will determine which digit the network identifies. 

As an exercise, we build a neural network according to these guidelines and train it. How exactly one writes the code depends on the library of choice, but the generic structure will be the following:

\begin{exbox}[MNIST]{ex: MNIST}
\begin{enumerate}
\item
\emph{Import the data}:
The MNIST database is available for download at
\url{http://yann.lecun.com/exdb/mnist/}
\item
\emph{Define the model}:
\begin{itemize}
\item
\emph{Input layer}: $28^2=784$ neurons (the greyscale value of each pixel of the image, normalized to a value in $[0,1)$, is one component of the input vector).
\item
\emph{Fully connected hidden layer}: Here one can experiment, starting from as few as 10 neurons. The use of a sigmoid activation function is recommended, but others can in principle be used.
\item
\emph{Output layer}: Use 10 neurons, one for each digit. The proper activation function for this classification task is, as discussed, a softmax function.
\end{itemize}

\item
\emph{Choose the loss function}: Since we are dealing with a classification task with a softmax output activation layer, we use the cross-entropy, Eq.~\eqref{eq:cross-entropy 2}.
\item
\emph{Train and evaluate the model}: Follow the standard machine-learning workflow to train\footnote{Most ML packages have some type of 'train' function built in, so no need to worry about implementing back-propagation by hand. All that is needed here is to call the 'train' function} and evaluate the model. However, unlike in the regression example of the previous chapter, where we evaluated the model using the mean square error, here we are rather interested in the accuracy of our prediction.
\end{enumerate}
\end{exbox}

With the training completed, we want to understand how well the final model performs in recognizing handwritten digits. For that, we introduce the \emph{accuracy}\index{accuracy} defined by
\begin{equation}
    \text{accuracy} = \frac{\text{correct predictions}}{\text{total predictions}}.
    \label{eq:accuracy}
\end{equation}

Running 20 trainings with 30 hidden neurons with a learning rate $\eta=0.5$ (a mini-batch size of 10 and train for 30 epochs), we obtain an average accuracy of 96.11 \%. With a L2 loss, we obtain only slightly worse results of 95.95\%. For 100 hidden neurons, we obtain 97.99\%. That is a considerable improvement over a quadratic cost, where we obtain 96.73\%---the more complex the network, the more important it is to choose a good loss function.
Still, these numbers are not even close to state of the art neural network performances. The reason is that we have used the simplest possible all-to-all connected architecture with only one hidden layer. In the following, we will introduce more advanced neural-network architectures and show how to increase the performance.

Before doing so, we briefly introduce other important measures used to characterize the performance of specifically \textbf{binary-classification} models in statistics: \emph{precision}, \emph{specificity} and \emph{recall}\index{precision}\index{specificity}\index{recall}.
In the language of true/false positives/negatives, the precision is defined as
\begin{equation}
    \text{precision} = \frac{\text{true positives}}{\text{true positives}+\text{false positives}}.
\end{equation}
Recall (also referred to as sensitivity) is defined as 
\begin{equation}
    \text{recall} = \frac{\text{true positives}}{\text{true positives}+\text{false negatives}}.
\end{equation}
While recall can be interpreted as true positive rate as it represents the ratio between outcomes identified as positive and actual positives, the specificity is an analogous measures for negatives
\begin{equation}
    \text{specificity} = \frac{\text{true negatives}}{\text{true negatives}+\text{false positives}}.
\end{equation}
Note, however, that these measures can be misleading, in particular when dealing with very unbalanced data sets.

	\subsection{Regularization}
In the previous sections, we have illustrated an artificial neural network that is constructed analogous to neuronal networks in the brain. A priori, a model is only given a rough structure, within which it has a huge number of parameters to adjust by learning from the training set. While we already understand that this is an extraordinarily powerful approach, this method of learning comes with its own set of challenges. The most prominent of them is the generalization\index{generalization} of the rules learned from training data to unseen data.

We have already encountered in the previous chapter how the naive optimization of a linear model reduces its generalization. However, we have also seen how the generalization error can be improved by reducing the variance at the cost of a biased model using regularization\index{regularization}.
Training neural networks comes with the same issue: we are always showing the model we built a training set that is limited in one way or another and we need to make sure that the neural network does not learn particularities of that given training set, but actually extracts a general knowledge. As we will discuss in the following, the mitigation strategies of regularization encountered for linear models can be similarly used for neural networks.

\begin{wrapfigure}{r}{0.2\textwidth}
  \centering
    \includegraphics{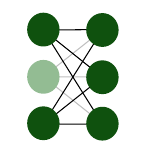}
   \vspace{-0.3cm}
   \caption{{\bf Dropout layer}}
  \label{fig:dropout}
\end{wrapfigure}

Step zero to avoid over-fitting is to create a sufficiently representative and diverse training set. Once this is taken care of, we can take several steps for the  regularization of the network. The simplest, but at the same time most powerful option is introducing \emph{dropout layers}\index{dropout}. This regularization is very similar to dropping features that we discussed for linear regression. However, the dropout layer ignores a randomly selected subset of neuron outputs in the network only during training. Which neurons are dropped is chosen at random for each training step. This regularization procedure is illustrated in Fig.~\ref{fig:dropout}.
By randomly discarding a certain fraction of neurons we add noise to the hidden layers that ensures that the network does not get fixed at small particular features of the training set and is better equipped to recognize the more general features. Another way of looking at it is that this procedure corresponds to training a large number of neural networks with different neuron connections in parallel. The fraction of neurons that are ignored in a dropout layer is a hyperparameter that is fixed a priori. Maybe it is counter-intuitive but the best performance is often achieved when this number is sizable, between 20\% and 50\%. It shows the remarkable resilience of the network against fluctuations.

As for the linear models, regularization can also be achieved by adding regularization terms $R(\theta)$ to the loss function\index{loss function} $L(\theta)$, $L(\theta) \rightarrow L(\theta) + \lambda R(\theta)$ with $\lambda$ a hyperparameter.
We have already encountered the two most common regularization terms: These are 
the  $L2$ regularization\index{$L2$ regression}~\footnote{L2 regularization is also known as Tikhonov regularization} that we encountered in ridge regression\index{ridge regression} with
    \begin{equation}
        R_{L2} = \frac{\lambda}{2} \sum_j W_j^2
    \end{equation} 
and $L1$ regularization that we saw in Lasso regression\index{lasso regression}\index{$L1$ regression}, where
    \begin{equation}
        R_{L1} = \frac{\lambda}{2} \sum_j |W_j|.
    \end{equation}
Note that, as is usually done, we penalize only the weights of the neural network, such that the sums above run over all weights $W_j$ of the network. Finally, while we can in principle set different $\lambda$ for each layer, it is common to use one for the whole network.

As for the linear models, $L2$ regularization shrinks all parameters symmetrically, whereas $L1$ regularization usually causes a subset of parameters to vanish, a property often used for feature selection. For this reason, both methods are also referred to as \emph{weight decay}\index{weight decay}. Either way, both $L1$ and $L2$ regularizations restrict the expressiveness of the neural network, thus encouraging it to learn generalizable features rather than overfitting\index{overfitting} specific features of the data set. 

The weights are commonly initialized with small values and increase during training. This naturally limits the capacity of the network, because for very small weights it effectively acts as a linear model (when one approximates the activation function by a linear function). Only once the weights become bigger, the network explores its nonlinearity. 

Another regularization technique consists in artificially enlarging the data set. Data is often costly, but we have extra knowledge about what further data might look like and feed this information in the machine learning workflow. For instance, going back to the MNIST example, we may shift or tilt the existing images or apply small transformations to them. By doing that, researchers were able to improve MNIST performance by almost 1 percent \footnote{ See Simard et al., \url{http://dx.doi.org/10.1109/ICDAR.2003.1227801}}. In particular if we know symmetries of the problem from which the data originates (such as time translation invariance, invariance under spatial translations or rotations), effective generation of augmented datasets is possible. For classification problems in particular, data may not be distributed between categories equally. To avoid a bias, it is the desirable to enhance the data in the underrepresented categories. Finally, adding various forms of noise to the data set can help avoid overfitting of the existing noise and increase the resilience of the neural network to noise. Note that this noise can be added to both, the input data and the labels.

\subsection{Convolutional neural networks}

The fully-connected simple single-layer architecture for a neural network is in principle universally applicable. However, this architecture is often inefficient and hard to train. In this section, we introduce more advanced neural-network layers and examples of the types of problems for which they are suitable.

\subsubsection{Convolutional layers}

\begin{figure}
\centering
\includegraphics{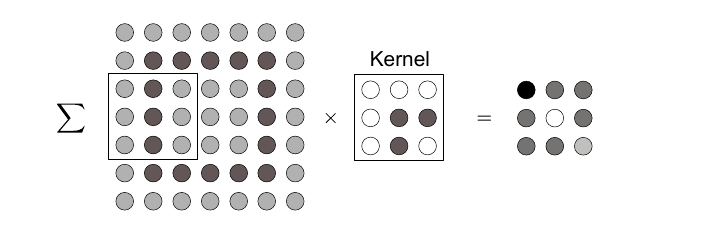}
\caption{{\bf Convolutional layer in 2D.} Here with filter size $k=3$ and stride $s=2$. The filter is first applied to the $3\times 3$ sub-image in the top left of the input, which yields the first pixel in the feature map. The filter then moves $s$ neurons to the right, which yields the next pixel and so on. After moving all the way to the right, the filter moves $s$ pixels down and starts from the left again until reaching the bottom right.}\label{fig:conv_2D}
\end{figure}

The achieved accuracy in the MNIST example above was not as high as one may have hoped, being much worse than the performance of a human. A main reason was that, using a dense network structure, we discarded all local information contained in the pictures. In other words, connecting every input neuron with every neuron in the next layer, the information whether two neurons are close to each other is lost. This information is, however, not only crucial for pictures, but quite often for input data with an underlying geometric structure or natural notion of `distance' in its correlations. 
To use this local information, so-called \emph{convolutional layers}\index{convolutional layers} were introduced.
The neural networks that contain such layers are called \emph{convolutional neural networks}\index{convolutional neural network} (CNNs).

%https://www.cs.ryerson.ca/~aharley/vis/conv/flat.html
%explain padding: original picture is  32x32, k=5, s=1, makes 28x28 pixel in first layer
%Draw K to show function of the filters
%Show Pooling layers at work: 2x2 pooling layer; average pooling; second pooling is max pooling
%Draw line on the bottom to show that spatial structure is kept through all convolutional layers

The key idea behind convolutional layers is to identify certain (local) patterns in the data. In the example of the MNIST images, such patterns could be straight and curved lines, or corners. A pattern is then encoded in a \emph{kernel}\index{kernel} or \index{filter}\emph{filter} in the form of weights, which are again part of the training. The convolutional layer than compares these patterns with a local patch of the input data. Mathematically, identifying the features in the data corresponds to a convolution $(f * x)(t)=\sum_{\tau}f(\tau)x(t-\tau)$ of the kernel $f$ with the original data $x$. 

For two-dimensional data, such as shown in the example in Fig.~\ref{fig:conv_2D}, we write the discrete convolution explicitly as
\begin{equation}
    q_{i,j} = \sum_{m=1}^{k} \sum_{n=1}^{k} f_{n,m} x_{si-m,sj-n} + b_0,
\end{equation}
where $f_{n,m}$ are the weights of the kernel, which has linear size $k$, and $b_0$ is a bias. Finally, $s$ is called \emph{stride}\index{stride} and refers to the number of pixels the filter moves per application. The output, $q$, is called \index{feature map}\emph{feature map}.
Note that the dimension of the feature map is $n_q\times n_q$ with $n_q = \lfloor (n_{in} - k)/s + 1 \rfloor $ when the input image is of dimensions $n_{in} \times n_{in}$: application of a convolutional layer thus reduces the image size, an effect not always intended. To avoid this reduction, the original data can be \index{padding}\emph{padded}, for example by adding zeros around the border of the data to ensure the feature map has the same dimension as the input data.

For typical convolutional networks, one applies a number of filters for each layer in parallel, where each filter is trained to recognize different features. For instance, one filter could start to be sensitive to contours in an image, while another filter recognizes the brightness of a region. Further, while filters in the first layers may be sensitive to local patterns, the ones in the later layers recognize larger structures. This distribution of functionalities between filters happens automatically, it is not preconceived when building the neural network.

\subsubsection{Pooling}

\begin{figure}
\centering
    \includegraphics{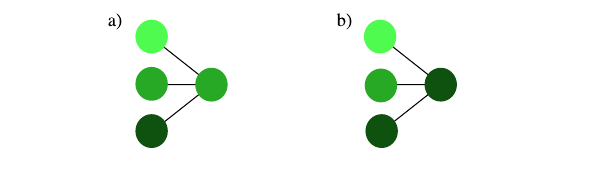}
    \caption{{\bf Pooling layer} (a) an average pooling and (b) a max pooling layer (both $n=3$).}
    \label{fig:pooling}
\end{figure}

Another very useful layer, in particular in combination with convolutional layers, 
is the \index{pooling layer}\emph{pooling layer}. Each neuron in the pooling layer takes input from $n$ (neighboring) neurons in the previous layer---in the case of a convolutional network for each feature map individually---and only retains the most significant information. Thus, the pooling layer helps to reduce the spatial dimension of the data. What is considered significant depends on the particular circumstances: Picking the neuron with the maximum input among the $n$, called \emph{max pooling}\index{max pooling}, detects whether a given feature is present in the window.  Furthermore, max pooling is useful to avoid \emph{dead neurons}\index{dead neurons}, in other words neurons that are stuck with a value near 0 irrespective of the input and such a small gradient for its weights and biases that this is unlikely to change with further training. This is a scenario that can often happen especially when using the ReLU\index{ReLU} activation function. \emph{Average pooling}\index{Average pooling}, in other words taking the average value of the $n$ inputs is a straight forward compression. 
Note that unlike other layers, the pooling layer has just a small set of $n$ connections with no adjustable weights. The functionality of the pooling layer is shown in Fig.~\ref{fig:pooling}~(a) and~(b).

An extreme case of pooling is global pooling, where the full input is converted to a single output. Using a max pooling, this would then immediately tell us, whether a given feature is present in the data.

\subsubsection{Example: DNA sequencing}
With lowering costs and expanding applications, DNA sequencing has become a widespread tool in biological research. Especially the introduction of high-throughput sequencing methods and the related increase of data has required the introduction of data science methods into biology. 
Sequenced data is extremely complex and thus a great playground for machine learning applications. Here, we consider a simple classification as an example. The primary structure of DNA consists of a linear sequence of basic building blocks called nucleotides. The key component of nucleotides are nitrogen bases: Adenine (A), Guanine (G), Cytosine (C), and Thymine (T). The order of the bases in the linear chains defines the  DNA sequence. Which sequences are meaningful is determined by a set of complex specific rules. In other words, there are series of letters A, G, C, and T that correspond to DNA and while many other sequences do not resemble DNA. Trying to distinguish between strings of nitrogen bases that correspond to human DNA and those that don not is a simple example of a classification task that is at the same time not so easy for an untrained human eye.

\begin{figure}
\centering
\includegraphics{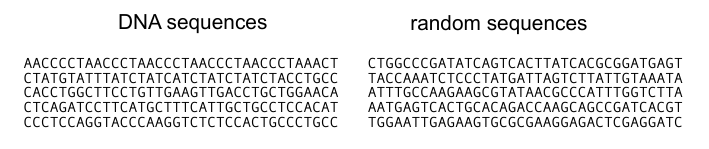}
\caption{{\bf Comparison of DNA and random sequences.}}
\label{fig:DNAcompare}
\end{figure}
In Fig.~\ref{fig:DNAcompare}, we show a comparison of five strings of human DNA and five strings of 36 randomly generated letters A, G, C, and T. Without deeper knowledge it is hard to distinguish the two classes and even harder to find the set of empirical rules that quantify their distinction. We can let a neural network  have a go and see if it performs any better than us studying these sequences by visual analysis. 

\begin{figure}
\centering
\includegraphics{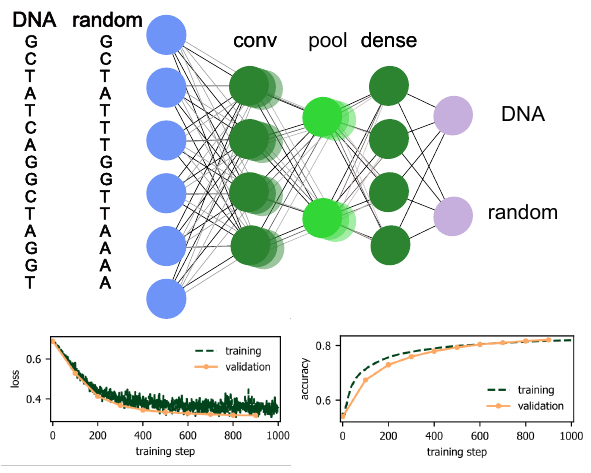}
\caption{{\bf Neural network classification of DNA sequences.} The two lower panels show loss function and accuracy on the training (evaluation) data in green (orange) as a function of the training step respectively.}
\label{fig:DNA}
\end{figure}

We have all ingredients to build a binary classifier that will be able to distinguish between DNA and non-DNA sequences. First, we download a freely available database of the human genome from \url{https://genome.ucsc.edu}\footnote{\url{http://hgdownload.cse.ucsc.edu/goldenpath/hg19/encodeDCC/wgEncodeUwRepliSeq/wgEncodeUwRepliSeqBg02esG1bAlnRep1.bam}}. Here, we downloaded a database of encoding genomes that contains $100 000$ sequences of human DNA (each is 36 letters long). Additionally, we generate $100 000$ random sequences of letters A, G, C, T. The learning task we are facing now is very similar to the MNIST classification, though in the present case, we only have two classes. Note, however, that we generated random sequences of bases and labeled them as random, even though we might have accidentally created sequences that do correspond to human DNA. This limits the quality of our data set and thus naturally also the final performance of the network.

The model we choose here has a standard architecture and can serve as a guiding example for supervised learning with neural networks that will be useful in many other scenarios. In particular, we implement the following architecture:

\begin{exbox}[DNA classification]{ex: DNA}
\begin{enumerate}
\item
\emph{Import the data} from \url{http://genome.uscs.edu}
\item
\emph{Define the model}:
\begin{itemize}
\item
\emph{Input layer}: The input layer has dimension $36\times 4$ ($36$ entries per DNA sequence, $4$ to encode each of 4 different bases A, G, C, T)\\
\emph{Example}:
[[1,0,0,0], [0,0,1,0], [0,0,1,0], [0,0,0,1]] = ACCT 
\item
 \emph{Convolutional layer}:  Kernel size $k= 4$, stride $s= 1$ and number of filters $N=64$.
\item
\emph{Pooling layer}: max pooling over $n=2$ neurons, which reduces the output of the previous layer by a factor of 2.
\item
\emph{Dense layer}: 256 neurons with a ReLU activation function.
\item
\emph{Output layer}: 2 neurons (DNA and non-DNA output) with softmax activation function.
\end{itemize}
\item
\emph{Loss function}: Cross-entropy between DNA and non-DNA.
\end{enumerate}
\end{exbox}

A schematic of the network structure as well as the evolution of the loss and the accuracy measured over the training and validation sets with the number of training steps are shown in Fig.~\ref{fig:DNA}. Comparing the accuracies of the training and validation sets is a standard way to avoid overfitting: On the examples from the training set we can simply check the accuracy during training. 
When training and validation accuracy reach the same number, this indicates that we are not overfitting on the training set since the validation set is never used to adjust the weights and biases. A decreasing validation accuracy despite an increasing training accuracy, on the other hand, is a clear sign of overfitting.

We see that this simple convolutional network is able to achieve around 80\% accuracy. By downloading a larger training set, ensuring that only truly random sequences are labeled as such, and by optimizing the hyper-parameters of the network, it is likely that an even higher accuracy can be achieved. We also encourage you to test other architectures: one can try to add more layers (both convolutional and dense), adjust the size of the convolution kernel or stride, add dropout layers, and finally, test whether it is possible to reach higher accuracies without over-fitting on the training set.

\subsubsection{Example: advanced MNIST}

We can now revisit the MNIST example and approach the classification with the more advanced neural network structures of the previous section. In particular, we use the following architecture
\begin{exbox}
[Advanced MNIST]{ex: adv MNIST}
\begin{enumerate}
\item
\emph{Input layer}: $28\times28$ ($=784$) neurons.
\item
\emph{Convolutional layer 1}: Kernel size $k= 3\times3$, stride $s= 1$ and number of filters $N=32$ with a ReLU activation function.
\item
\emph{Pooling layer}: max pooling over $n=2\times 2$ neurons.
\item
\emph{Convolutional layer 2}:  Kernel size $k= 3\times3$, stride $s= 1$ and number of filters $N=64$ with a ReLU activation function.
\item
\emph{Pooling layer}: max pooling over $n=2\times 2$ neurons.
\item
\emph{Dropout}: dropout layer for regularization with a 50\% dropout probability.
\item
\emph{Dense layer}: 100 neurons with a ReLU activation function.
\item
\emph{Output layer}: 10 neurons with softmax activation function.
\end{enumerate}
\end{exbox}
For the loss function, we again use cross-entropy between the output and the labels.
Notice here the repeated structure of convolutional layers and pooling layers. This is a very common structure for deep convolutional networks. With this model, we achieve an accuracy on the MNIST test set of 99.1\%, a massive improvement over the simple dense network.

	\subsection{Recurrent neural networks}
\label{sec:rnn}
\begin{figure}[bb!]
	\centering
	\includegraphics{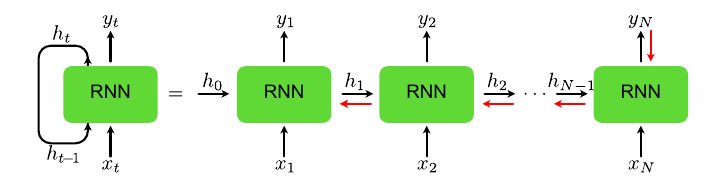}
	\caption{{\bf Recurrent neural network architecture.} The input $x_t$ is fed into the recurrent cell together with the (hidden) memory $h_{t-1}$ of the previous step to produce the new memory $h_t$ and the output $y_t$. One can understand the recurrent structure via the ``unwrapped'' depiction of the structure on the right hand side of the figure. The red arrows indicate how gradients are propagated back in time for updating the network parameters.}
\label{fig:RNN}
\end{figure}

We have seen in the previous section how the convolutional neural network allows to retain local information through the use of filters. While this context-sensitivity of the CNN is applicable in many situations, where geometric information is important, there are situations we have more than neighborhood relations, namely sequential order. An element is before or after an other, not simply next to it. A common situation, where the order of the input is important, is with time-series data. Examples are measurements of a distant star collected over years or the events recorded in a detector after a collision in a particle collider. The classification task in these examples could be the determination whether the star has an exoplanet, or whether a Higgs boson was observed, respectively. Another example without any temporal structure in the data is the prediction of a protein's functions from its primary amino-acid sequence.

A property that the above examples have in common is that the length of the input data is not necessarily always fixed for all samples. This emphasizes again another weakness of both the dense network and the CNN: The networks only work with fixed-size input and there is no good procedure to decrease or increase the input size. While we can in principle always cut our input to a desired size, of course, this finite window is not guaranteed to contain the relevant information.

In this final section on supervised learning, we introduce one more neural network architecture that solves both problems discussed above: \emph{recurrent neural networks}\index{recurrent neural networks} (RNNs). The key idea behind a recurrent neural network is that input is passed to the network one element after another---unlike for other neural networks, where an input 'vector' is given the network all at once---and to recognize context, the network keeps an internal state, or memory, that is fed back to the network together with the next input.
Recurrent neural networks were developed in the context of \emph{natural language processing}~(NLP), the field of processing, translating and transforming spoken or written language input, where clearly, both context and order are crucial pieces of information. However, over the last couple of years, RNNs have found applications in many fields in the sciences.

The special structure of a RNN is depicted in Fig.~\ref{fig:RNN}. At step $t$, the input $\bm{x}_t$ and the (hidden) internal state of the last step $\bm{h}_{t-1}$ are fed to the network to calculate $\bm{h}_t$. The new hidden memory of the RNN is finally connected to the output layer $\bm{y}_t$. As shown in Fig.~\ref{fig:RNN}, this is equivalent to having many copies of the input-output architecture, where the hidden layers of the copies are connected to each other.  The RNN cell itself can have a very simple structure with a single activation function. 

Concretely, in each step of a simple RNN we update the hidden state as 
\begin{equation}
  \bm{h}_{t} = \tanh(W_{hh} \bm{h}_{t-1} + W_{xh} \bm{x}_{t} + \bm{b}_h),
  \label{eq:rnn first step}
\end{equation}
where we used for the nonlinearity the hyperbolic tangent, a common choice, which is applied element-wise.
Further, if the input data $\bm{x}_t$ has dimension $n$ and the hidden state $\bm{h}_t$ dimension $m$, the weight matrices $W_{hh}$ and $W_{xh}$ have dimensions $m\times m$ and $m\times n$, respectively. 
Finally, the output at step $t$ can be calculated using the hidden state $\bm{h}_t$, 
\begin{equation}
  \bm{y}_{t} = W_{ho} \bm{h}_t.
  \label{eq:rnn output}
\end{equation}
A schematic of this implementation is depicted in Fig.~\ref{fig:lstm}(a).
Note that in this simplest implementation, the output is only a linear operation on the hidden state. A straight forward extension---necessary in the case of a classification problem---is to add a non-linear element to the output as well. Concretely, 
\begin{equation}
  \bm{y}_{t} = \bm{g}(W_{ho}\bm{h}_t + \bm{b}_y) 
\end{equation}
with $\bm{g}(\bm{q})$ some activation function, such as a softmax.
Note that while in principle an output can be calculated at every step, this is only done after the last input element in a classification task.

An interesting property of RNNs is that the weight matrices and biases, the parameters we learn during training, are the same for each input element. This property is called \emph{parameter sharing} and is in stark contrast to dense networks. In the latter architecture, each input element is connected through its own weight matrix. While it might seem that this property could be detrimental in terms of representability of the network, it can greatly help extracting sequential information: Similar to a filter in a CNN, the network does not need to learn the exact location of some specific sequence that carries the important information, it only learns to recognize this sequence somewhere in the data. Note that the way each input element is processed differently is instead implemented through the hidden memory.

Parameter sharing is, however, also the root of a major problem when training a simple RNN. To see this, remember that during training, we update the network parameters using gradient descent. As in the previous sections, we can use backpropagation to achieve this optimization. Even though the unwrapped representation of the RNN in Fig.~\ref{fig:RNN} suggests a single hidden layer, the gradients for the backpropagation have to also propagate back through time. This is depicted in Fig.~\ref{fig:RNN} with the red arrows~\footnote{In the context of RNNs, backpropagation is thus referred to as \emph{backpropagation through time} (BPTT).}.
Looking at the backpropagation algorithm in Sec.~\ref{sec:training}
, we see that to use data points from $N$ steps back, we need to multiply $N-1$ Jacobi matrices $D\bm{f}^{[t']}$ with $t-N < t' \leq t$. Using Eq.~\eqref{eq:rnn output}, we can write each Jacobi matrix as a product of the derivative of the activation function, $\partial_q \tanh(q)$, with the weight matrix. If either of these factors~\footnote{For the weight matrix this means the singular values.} is much smaller than $1$, the gradients  decrease exponentially. This is known as the problem of \emph{vanishing gradients}. Alternatively, if the factors are much larger than $1$, the gradients grow exponentially, known as the problem of \emph{exploding gradients}. Both problems make learning long-term dependencies with simple RNNs practically impossible. 

Note that the problem of exploding gradients can be mitigated by clipping the gradients, in other words scaling them to a fixed size. Furthermore, we can use the ReLU activation function instead of a hyperbolic tangent, as the derivative of the ReLU for $q>0$ is always 1. However, the problem of the shared weight matrices can not so easily be resolved. In order to learn long-time dependencies, we have to introduce a different architecture. In the following, we will discuss the long short-term memory (LSTM) network. This architecture and its variants are used in most applications of RNNs nowadays. 

\subsubsection{Long short-term memory}
The key idea behind the LSTM is to introduce another state to the RNN, the so-called \emph{cell state}, which is passed from cell to cell, similar to the hidden state. However, unlike the hidden state, no matrix multiplication takes place, but information is added or removed to the cell state through \emph{gates}. The LSTM then commonly comprises four gates which correspond to three steps: the forget step, the input and update step, and finally the output step. We will in the following go through all of these steps individually.
\vspace{5pt}
\begin{figure}[tb]
  \centering
  \includegraphics{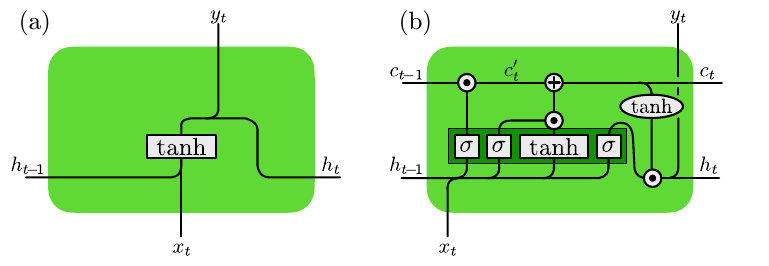}
  \caption{{\bf Comparison of (a) a simple RNN and (b) a LSTM.} The boxes denote neural networks with the respective activation function, while the circles denote element-wise operations. The dark green box indicates that the four individual neural networks can be implemented as one larger one.}
  \label{fig:lstm}
\end{figure}

\noindent
{\bf Forget step}\\
In this step, specific information of the cell state is forgotten. Specifically, we update the cell state as
\begin{equation}
  \bm{c}'_t = \sigma(W_{hf} \bm{h}_{t-1} + W_{xf} \bm{x}_t + \bm{b}_f)\odot \bm{c}_{t-1}.
  \label{eq:forget}
\end{equation}
where $\sigma$ is the sigmoid function (applied element-wise) and $\odot$ denotes element-wise multiplication.
Note that this step multiplies each element of the gate state with a number $\in(0,1)$, in other words elements multiplied with a number close to $0$ forget their previous memory.
\vspace{5pt}

\noindent
{\bf Input and update step}\\
In the next step, we decide what and how much to add to the cell state. 
For this purpose, we first decide what to add to the state. We first define what we would like to add to the cell,
\begin{equation}
  \bm{g}_t = \tanh(W_{hu}\bm{h}_{t-1} + W_{xu} \bm{x}_t + \bm{b}_u),
  \label{eq:gate gate}
\end{equation}
which due to the hyperbolic tangent, $-1 < g^\alpha_t < 1$ for each element. However, we do not necessarily update each element of the cell state, but rather we introduce another gate, which determines whether to actually write to the cell,
\begin{equation}
  \bm{i}_t = \sigma(W_{hi} \bm{h}_{t-1} + W_{xi} \bm{x}_t + \bm{b}_i),
  \label{eq:input}
\end{equation}
again with $0<i^\alpha_t < 1$. Finally, we update the cell state
\begin{equation}
  \bm{c}_t = \bm{c}'_t +  \bm{i}_t \odot \bm{g}_t.
  \label{eq:update}
\end{equation}
\vspace{5pt}

\noindent
{\bf Output step}\\
In the final step, we decide how much of the information stored in the cell state should be written to the new hidden state,
\begin{equation}
  \bm{h}_t = \sigma(W_{ho} \bm{h}_{t-1} + W_{xo} \bm{x}_t + \bm{b}_o) \odot \tanh (\bm{c}_t).
  \label{eq:output step}
\end{equation}
\vspace{5pt}

The full structure of the LSTM with the four gates and the element-wise operations is schematically shown in Fig.~\ref{fig:lstm}(b).
Note that we can concatenate the input $\bm{x}_t$ and hidden memory $\bm{h}_{t-1}$ into a vector of size $n+m$ and write one large weight matrix $W$ of size $4m \times (m+n)$.

So far, we have only used the RNN in a supervised setting for classification purposes, where the input is a sequence and the output a single class at the end of the full sequence. A network that performs such a task is thus called a many-to-one RNN. We will see in the next section, that unlike the other network architectures encountered in this section, RNNs can straight-forwardly be used for unsupervised learning, usually as one-to-many RNNs.

%-----Unsupervised Learning------------------------------%
	\section{Unsupervised Learning}
\label{sec:unsupervized}

In Sec.~\ref{sec:supervized}, we discussed supervised learning tasks, for which datasets consist of input-output pairs, or data-label pairs. More often than not, however, we have data without labels and would like to extract information from such a dataset. Clustering problems fall in this category, for instance: We suspect that the data can be divided into different types, but we do not know which features distinguish these types.

Mathematically, we can think of the data $\bm{x}$ as samples that were drawn from a probability distribution  $P_{\rm data}(\bm{x})$. The unsupervised learning task is to represent this distribution with a model $P_{\theta}(\bm{x})$, for example represented by a neural network and parametrized by $\theta$.  The model can then be used to study properties of the distribution or to generate new `artificial' data. The models we encounter in this chapter are thus also referred to as \emph{generative models}\index{generative models}. In general, unsupervised learning is conceptually more challenging than supervised learning. At the same time, unsupervised algorithms are highly desirable, since unlabelled data is much more abundant than labelled data. Moreover, we can in principle use a generative model for a classification task by learning the joint probability distribution of the data-label pair.

In this chapter, we will introduce three types of neural networks that are specific to unsupervised learning tasks: \emph{Restricted Boltzmann machines}\index{Restricted Boltzmann machines}, \emph{autoencoders}\index{autoencoders}, and \emph{generative adversarial networks}\index{generative adversarial network}. Furthermore, we will discuss how the RNN introduced in the previous chapter can also be used for an unsupervised task.

\subsection{Maximum Likelihood Estimation}
\label{sec:maxlikelihood}
Before we discuss the specific neural network architectures, we need to formalize the goal of the training. As we have discussed in the previous section, we need to define a loss function, whose minimization encodes this goal. In the case of generative models, we will focus in the following on the \emph{maximum likelihood principle}: The role of the model is to provide an estimate $P_{\theta}(\bm{x})$ of a probability distribution parametrized by parameters $\theta$. The \emph{likelihood} of a given data point $\bm{x}$ is then the probability that the model assigns to this data point, $\mathcal{L}(\theta;\bm{x}) =  P_{\theta}(\bm{x})$. The likelihood of the whole training data set is given by
\begin{equation}
	\mathcal{L}(\theta;\{\bm{x}_i\}) = \prod_{i=1}^m P_{\theta}(\bm{x}_{i}),
\end{equation}
where $m$ is the number of samples in the data $\{\bm{x}_i\}$. The goal is to find the parameters $\theta$ that maximize the likelihood. Thanks to the monotonicity of the logarithm function, we can work with the \emph{log-likelihood},
\begin{equation}
\begin{split}
	\hat{\theta}=&\,\underset{\theta}{\text{argmax}}\prod_{i=1}^mP_{\theta}(\bm{x}_{i})\\
=&\,\underset{\theta}{\text{argmax}}\sum_{i=1}^m\mathrm{log}\,P_{\theta}(\bm{x}_{i}),
\end{split}
\end{equation}
which is easier to handle. The maximization of the log-likelihood is equivalent to the minimization of \emph{negative log-likelihood}, which in turn can be interpreted as the cross-entropy between two probability distributions: the `true' distribution $P_{\mathrm{data}}(\bm{x})$, from which the data has been drawn, and $P_{\theta}(\bm{x})$. While we do not have access to $P_{\mathrm{data}}(\bm{x})$ in principle, we estimate it empirically as a distribution peaked at the $m$ data points we have in our data set. 

	\subsection{Restricted Boltzmann machine}
\label{sec:rbm}
\emph{Restricted Boltzmann Machines}\index{restricted Boltzmann machine} (RBM) are a class of generative stochastic neural networks. More specifically, given some (binary) input data $\bm{x}\in\{0,1\}^{n_v}$, an RBM can be trained to approximate the probability distribution of this input. Moreover, once the neural network is trained to approximate the distribution of the input, we can sample from the network, in other words we generate new instances from the learned probability distribution. 

\begin{figure}
	\centering
	\includegraphics{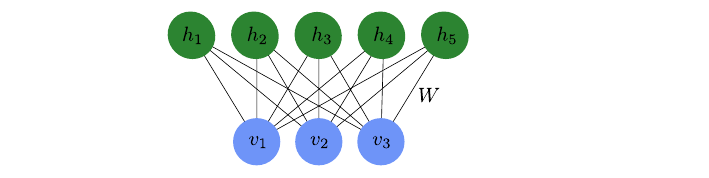}
	\caption{{\bf Restricted Boltzmann machine.} Each of the three visible units and five hidden units represents a variable that can take the values $\pm1$ and the connections between them represent the entries $W_{ij}$ of the weight matrix that enters the energy function~\eqref{eqn: RBM Energy}.}
	\label{fig:RBM}
\end{figure}

The RBM consists of two layers (see Fig.~\ref{fig:RBM}) of \emph{binary units}. Each binary unit is a variable which can take the values $0$ or $1$.
We call the first (input) layer visible and the second layer hidden. 
The visible layer with input variables $\lbrace v_{1}, v_{2}, \dots v_{n_{\mathrm{v}}}\rbrace $, which we collect in the vector $\bm{v}$, is connected to the hidden layer with variables $\{ h_{1}, h_{2}, \dots h_{n_{\mathrm{h}}}\}$, which we collect in the vector $\bm{h}$. 
The role of the hidden layer is to mediate correlations between the units of the visible layer. 
In contrast to the neural networks we have seen in the previous chapter, the hidden layer is not followed by an output layer. Instead, the RBM represents a probability distribution $P^{\text{rbm}}_{\theta}(\bm{v})$, which depends on variational parameters $\theta$ represented by the weights and biases of a neural network. 
The RBM, as illustrated by the graph in Fig.~\ref{fig:RBM}, is a special case of a network structure known as a Boltzmann machine with the restriction that a unit in the visible layer is only connected to hidden units and vice versa, hence the name \emph{restricted} Boltzmann machine.

The structure of the RBM is motivated from statistical physics:
To each choice of the binary vectors $\bm{v}$ and $\bm{h}$, we assign a value we call the energy
\begin{equation}\label{eqn: RBM Energy}
    E(\bm{v},\bm{h}) = -\sum_{i}a_{i}v_{i} - \sum_{j}b_{j}h_{j} - \sum_{ij} v_{i}W_{ij}h_{j},
\end{equation}
where the vectors $\bm{a}$, $\bm{b}$, and the matrix $W$ constitute the variational parameters $\theta$ of the model. Given the energy, the probability distribution over the configurations $(\bm{v}, \bm{h})$ is defined as
\begin{equation}\label{eqn: RBM Joint Probability}
    P_\theta^{\textrm{rbm}}(\bm{v},\bm{h}) = \frac{1}{Z}e^{-E(\bm{v},\bm{h})},
\end{equation}
where
\begin{equation}
    Z = \sum_{\bm{v},\bm{h}} e^{-E(\bm{v},\bm{h})}
    \label{eq: partition function}
\end{equation}
is a normalization factor called the partition function. 
The sum in Eq.~\eqref{eq: partition function} runs over all binary vectors $\bm{v}$ and $\bm{h}$, in other words vectors with entries $0$ or $1$.
The probability that the model assigns to a visible vector $\bm{v}$ is then the marginal over the joint probability distribution Eq.~\eqref{eqn: RBM Joint Probability},
\begin{equation}\label{eqn: RBM visible probability}
    P_\theta^{\textrm{rbm}}(\bm{v}) = \sum_{\bm{h}} P_\theta^{\textrm{rbm}}(\bm{v},\bm{h}) = \frac{1}{Z}\sum_{\bm{h}}e^{-E(\bm{v},\bm{h})},
\end{equation}
where the sum runs over all hidden-layer configurations $\bm{h} \in \{0,1\}^{n_h}$.

As a result of the restriction, the visible units, with the hidden units fixed, are mutually independent: given a choice of the hidden units $\bm{h}$, we have an \textbf{independent} probability distribution for \textbf{each} visible unit given by
\begin{equation}
    P_\theta^{\textrm{rbm}}(v_{i} = 1 | \bm{h}) = \sigma(a_{i} + \sum_{j}W_{ij}h_{j}),
    \qquad i=1,\ldots, n_{\mathrm{v}},
\end{equation}
where $\sigma(x) = 1/(1+e^{-x})$ is the sigmoid function. Similarly, with the visible units fixed, the individual hidden units are also mutually independent with the probability distribution
\begin{equation}\label{eqn: RBM P(h|v)}
     P_\theta^{\textrm{rbm}}(h_{j} = 1 | \bm{v}) = \sigma(b_{j} + \sum_{i}v_{i}W_{ij})
     \qquad j=1,\ldots, n_{\mathrm{h}}.
\end{equation}
The visible (hidden) units can thus be interpreted as artificial neurons connected to the hidden (visible) units with sigmoid activation function and bias $\bm{a}$ ($\bm{b}$).
A direct consequence of this mutual independence is that sampling a vector $\bm{v}$ or $\bm{h}$ reduces to sampling every component individually.
Notice that this simplification comes about due to the restriction that visible (hidden) units do not directly interact amongst themselves, meaning there are no terms proportional to $v_i v_j$ or $h_i h_j$ in Eq.~\eqref{eqn: RBM Energy}.
In the following, we explain how one can train an RBM and discuss possible applications of RBMs.

\subsubsection{Training an RBM}
Consider a set of binary input data $\bm{x}_k$, $k=1,\ldots,m$, drawn from a probability distribution $P_{\textrm{data}}(\bm{x})$. The aim of the training is to tune the parameters $\theta=\lbrace \bm{a}, \bm{b}, W \rbrace$ in an RBM such that after training $P_\theta^{\textrm{rbm}}(\bm{x}) \approx  P_{\textrm{data}}(\bm{x})$.
To solve this problem, we use the maximum likelihood principle introduced in Sec.~\ref{sec:maxlikelihood}.
As we discussed there, maximizing the likelihood is equivalent to training the RBM using the negative log-likelihood as a loss function,
\begin{equation}
    L(\theta) = - \sum_{k=1}^{m} \log P_\theta^{\textrm{rbm}}(\bm{x}_{k}).
\end{equation}
For the gradient descent, we need derivatives of the loss function of the form
\begin{equation}\label{eqn: log-likelihood derivative}
    \frac{\partial L(\theta)}{\partial W_{ij}} = -\sum_{k=1}^{m} \frac{\partial\log P_\theta^{\textrm{rbm}}(\bm{x}_{k})}{\partial W_{ij}}.
\end{equation}
This derivative consists of two terms,
\begin{equation} \label{eqn: RBM derivatives}
\begin{split}
        \frac{\partial\log P_\theta^{\textrm{rbm}}(\bm{x}_k)}{\partial W_{ij}}
        &= x_{k,i}P_\theta^{\textrm{rbm}}(h_{j}=1|\bm{x}_k) - \sum_{\bm{v}} v_{i} P_\theta^{\textrm{rbm}}(h_{j}=1|\bm{v}) P_\theta^{\textrm{rbm}}(\bm{v})
\end{split}
\end{equation}
with the second one containing a sum over all possible visible-layer configurations $\bm{v}\in\{0,1\}^{n_v}$.
Note that similarly simple forms are found for the derivatives with respect to the components of $\bm{a}$ and $\bm{b}$.
We can then iteratively update the parameters just as we have done in Sec.~\ref{sec:supervized},
\begin{equation}
    W_{ij} \rightarrow W_{ij} - \eta \frac{\partial L(\theta)}{\partial W_{ij}}
\end{equation}
with a sufficiently small learning rate $\eta$.
As we have seen in the previous chapter in the context of backpropagation, we can reduce the computational cost by replacing the summation over the whole data set in Eq.~\eqref{eqn: log-likelihood derivative} with a summation over a small randomly chosen batch of samples. This reduction in the computational cost comes at the expense of noise, but at the same time it can help to improve generalization.

However, there is one more problem: The second summation in Eq.~\eqref{eqn: RBM derivatives}, which contains $2^{n_v}$ terms, cannot be efficiently evaluated exactly. Instead, we have to approximate the sum by sampling the visible layer $\bm{v}$ from the marginal probability distribution $P_\theta^{\textrm{rbm}}(\bm{v})$. This sampling can be done using \index{Gibbs sampling}\emph{Gibbs sampling} as follows:
\begin{algbox}[Gibbs sampling]{alg: Gibbs Sampling}
\begin{algorithm}[H]
\SetAlgoLined
\KwIn{Any visible vector $\bm{v}(0)$}
\KwOut{Visible vector $\bm{v}(r)$}
 \For{$n=1$\dots $r$}{
  sample $\bm{h}(n)$ from $P_\theta^{\rm rbm}(\bm{h}|\bm{v}=\bm{v}(n-1))$\;
  sample  $\bm{v}(n)$ from $P_\theta^{\rm rbm}(\bm{v}|\bm{h}=\bm{h}(n))$\;
 }
\end{algorithm}
\end{algbox}
\noindent With sufficiently many steps $r$, the vector $\bm{v}(r)$ is an unbiased sample drawn from $P_\theta^{\textrm{rbm}}(\bm{v})$. By repeating the procedure, we can obtain multiple samples to estimate the summation. Note that this is still rather computationally expensive, requiring multiple evaluations on the model.

The key innovation which allows the training of an RBM to be computationally feasible was proposed by Geoffrey Hinton (2002). Instead of obtaining multiple samples, we simply perform the Gibbs sampling with $r$ steps and estimate the summation with a single sample, in other words we replace the second summation in Eq.~\eqref{eqn: RBM derivatives} with
\begin{equation}
    \sum_{\bm{v}} v_{i} P_\theta^{\textrm{rbm}}(h_{j}=1|\bm{v}) P_\theta^{\textrm{rbm}}(\bm{v}) \rightarrow v'_{i} P_\theta^{\textrm{rbm}}(h_{j}=1|\bm{v}'),
\end{equation}
where $\bm{v}' = \bm{v}(r)$ is simply the sample obtained from $r$-step Gibbs sampling. With this modification, the gradient, Eq.~\eqref{eqn: RBM derivatives}, can be approximated as
\begin{equation}
    \frac{\partial\log P_\theta^{\textrm{rbm}}(\bm{x})}{\partial W_{ij}} \approx x_{i}P_\theta^{\textrm{rbm}}(h_{j}=1|\bm{x}) -  v'_{i} P_\theta^{\textrm{rbm}}(h_{j}=1|\bm{v}').
\end{equation}

This method is known as \textit{contrastive divergence}\index{contrastive divergence}. Although the quantity computed is only a biased estimator of the gradient, this approach is found to work well in practice. 
The complete algorithm for training a RBM with $r$-step contrastive divergence can be summarized as follows:
\begin{algbox}[Contrastive divergence]{alg: contrastive divergence}
\begin{algorithm}[H]
\SetAlgoLined
\KwIn{Dataset $\mathcal{D} = \lbrace \bm{x}_{1}, \bm{x}_{2}, \dots \bm{x}_{m} \rbrace$ drawn from a distribution $P_{\rm data}(\bm{x})$}
initialize the RBM weights $\theta=\lbrace \bm{a},\bm{b},W \rbrace$\;
Initialize $\Delta W_{ij} = \Delta a_{i} = \Delta b_{j} =0$\;
 \While{not converged}{
    select a random batch $\mathcal{S}$ of samples from the dataset $\mathcal{D}$ \;
    \ForAll{$\bm{x} \in \mathcal{S}$}{
        Obtain $\bm{v}'$ by $r$-step Gibbs sampling starting from $\bm{x}$
        $\Delta W_{ij} \leftarrow \Delta W_{ij} - x_{i}P_\theta^{\textrm{rbm}}(h_{j}=1|\bm{x}) +  v'_{i} P_\theta^{\textrm{rbm}}(h_{j}=1|\bm{v}')$
    }
    $W_{ij} \leftarrow W_{ij} - \eta\Delta W_{ij}$\\
    (and similarly for $\bm{a}$ and $\bm{b}$)
 }
\end{algorithm}
\end{algbox}

Having trained the RBM to represent the underlying data distribution $P_{\rm data}(\bm{x})$, there are a few ways one can use the trained model:
\begin{enumerate}
    \item \textbf{Pretraining} We can use $W$ and $\bm{b}$ as the initial weights and biases for a deep network (c.f. Chapter 4), which is then fine-tuned with gradient descent and backpropagation.
    \item \textbf{Generative Modelling} As a generative model, a trained RBM can be used to generate new samples via Gibbs sampling (Alg.~\ref{alg: Gibbs Sampling}). Some potential uses of the generative aspect of the RBM include \textit{recommender systems} and \textit{image reconstruction}. In the following subsection, we provide an example, where an RBM is used to reconstruct a noisy signal.
\end{enumerate}

\subsubsection{Example: signal or image reconstruction/denoising}
A major drawback of the simple RBMs for their application is the fact that they only take binary data as input. As an example, we thus look at simple periodic waveforms with 60 sample points. In particular, we use sawtooth, sine, and square waveforms. In order to have quasi-continuous data, we use eight bits for each point, such that our signal can take values from 0 to 255. Finally, we generate samples to train with a small variation in the maximum value, the periodicity, as well as the center point of each waveform. 

\begin{figure}
	\centering
	\includegraphics{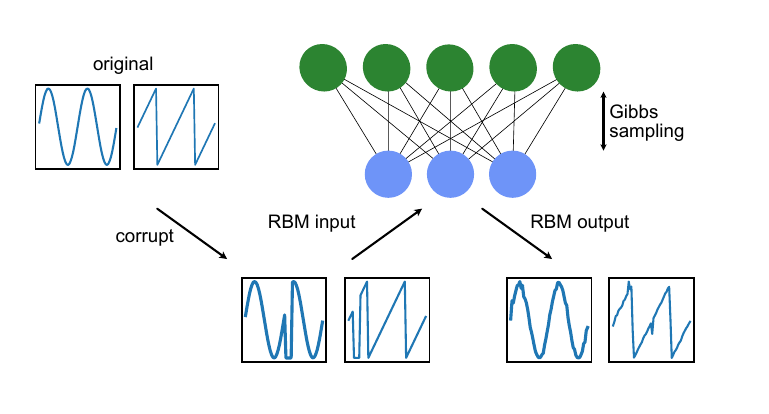}
	\caption{{\bf Signal reconstruction.} Using an RBM to repair a corrupted signal, here a sine and a sawtooth waveform.}
	\label{fig:RBM_reconstruction}
\end{figure}

After training the RBM using the contrastive divergence algorithm, we now have a model which represents the data distribution of the binarized waveforms. Consider now a signal which has been corrupted, meaning some parts of the waveform have not been received and have thus been set to 0. By feeding this corrupted data into the RBM and performing a few iterations of Gibbs sampling (Alg.~\ref{alg: Gibbs Sampling}), we can obtain a reconstruction of the signal, where the missing part has been repaired, as can been seen at the bottom of Fig.~\ref{fig:RBM_reconstruction}.

Note that the same procedure can be used to reconstruct or denoise images. Due to the limitation to binary data, however, the picture has to either be binarized, or the input size to the RBM becomes fairly large for high-resolution pictures. It is thus not surprising that while RBMs have been popular in the mid-2000s, they have largely been superseded by more modern architectures such as \textit{generative adversarial networks}\index{generative adversarial network} which we shall explore later in the chapter. However, they still serve a pedagogical purpose and could also provide inspiration for future innovations, in particular in science. A recent example is the idea of using an RBM to represent a quantum mechanical state~\footnote{\href{https://www.science.org/doi/full/10.1126/science.aag2302}{Carleo and Troyer, Science {\bf 355}, 602 (2017)}}.

	\subsection{Training an RNN without supervision}
\label{sec:rnn2}
\begin{figure}
	\centering
	\includegraphics{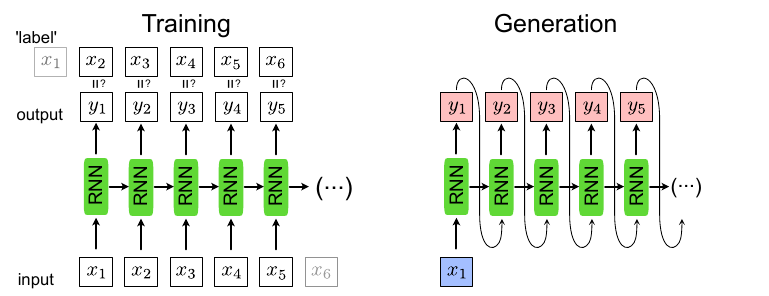}
	\caption{{\bf RNN used as a generator.} For training, left, the input data shifted by one, $\bm{x}_{t+1}$, are used as the label. For the generation of new sequences, right, we input a single data point $\bm{x}_1$ and the RNN uses the recurrent steps to generate a new sequence.}
	\label{fig:RNN_gen}
\end{figure}
In Sec.~\ref{sec:supervized}, the RNN was introduced as a classification model. Instead of classifying sequences of data, such as time series, the RNN can also be trained to generate valid sequences itself. Given the RNN introduced in Sec.~\ref{sec:rnn} with variational parameters $\theta$, the implementation of such a generator is straight-forward and does not require a new architecture. The main difference is that the output $\bm{y}_t$ of the network given the data point $\bm{x}_t$ is a guess of the subsequent data point $\bm{x}_{t+1}$ instead of the class to which the whole sequence belongs to. This means in particular that the input and output size are now the same. 
For training this network, we can once again use the cross-entropy as a loss function, which for an individual data point reads
\begin{equation}
    L_{\mathrm{ent}}(\theta)
    =-\sum_{t=1}^{N} \bm{x}_{t+1}\cdot
    \log \left(
    \bm{y}_{t}
    \right),
    \label{eq:unsup_RNN}
\end{equation}
where $\bm{x}_{t+1}$ is now the `label' for the input $\bm{x}_{t}$ and $\bm{y}_{t}$ is the output of the network and $t$ runs over the input sequence with length $N$, where one usually uses an `end' label for $\bm{x}_{N+1}$ to signal the end of the sequence. This training is schematically shown in Fig.~\ref{fig:RNN_gen}. 

For generating a new sequence, it is enough to have one single input point $\bm{x}_1$ to start the sequence. Note that since we now can start with a single data point $\bm{x}_1$ and generate a whole sequence of data points $\{\bm{y}_t\}$, this mode of using an RNN is referred to as \emph{one-to-many}. This sequence generation is shown on the right side of Fig.~\ref{fig:RNN_gen}.

\subsubsection{Example: generating molecules with an RNN}
\label{sec:rnn_gen}

\begin{figure}
	\centering
	\includegraphics{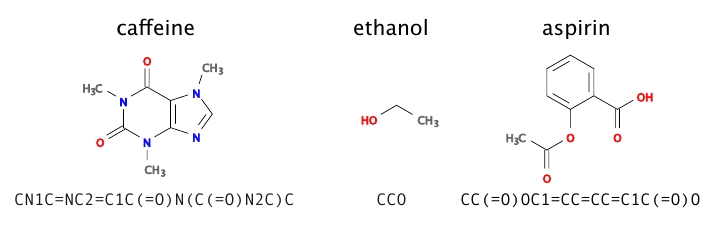}
	\caption{{\bf SMILES.} Examples of molecules and their representation in SMILES.}
	\label{fig:smiles}
\end{figure}

To illustrate the concept of sequence generation using recurrent neural networks, we use an RNN to generate new molecules. 
The first question we need to address is how to encode a chemical structure into input data---of sequential form no less---that a machine learning model can read. A common representation of molecular graphs used in chemistry is the \emph{simplified molecular-input line-entry system}, or SMILES. 
Figure~\ref{fig:smiles} shows examples of such SMILES strings for the caffeine, ethanol, and aspirin molecules. We can use the dataset \textit{Molecular Sets} \footnote{\url{https://github.com/molecularsets/moses}}, which contains $\sim 1.9$M molecules written in the SMILES format.

Using the SMILES dataset, we create a dictionary to translate each character that appears in the dataset into an integer. We further use one-hot-encoding to feed each character separately to the RNN. This creates a map from characters in SMILES strings onto an array of numbers.
Finally, in order to account for the variable size of the molecules and hence, the variable length of the strings, we can introduce a `stop' character such that the network learns and later generates sequences of arbitrary length. 

We are now ready to use the SMILES strings for training our network as described above, where the input is a one-hot-encoded vector and the output is again a vector of the same size. Note, however, that similar to a classification task, the output vector is a probability distribution over the characters the network believes could come next. Unlike a classification task, where we consider the largest output the best guess of the network, here we sample in each step from the probability distribution $\bm{y}_t$ to again have a one-hot-encoded vector for the input of the next step.

	\subsection{Autoencoders}\label{sec:vae}
Autoencoders\index{autoencoder} are neuron-based generative models, initially introduced for dimensionality reduction. The original purpose, thus, is similar to that of PCA  or t-SNE that we already encountered in Sec.~\ref{sec:structuring_data}, namely the reduction of the number of features that describe our input data. Unlike for PCA, where we have a clear recipe how to reduce the number of features, an autoencoder learns the best way of achieving the dimensionality reduction. An obvious question, however, is how to measure the quality of the compression, which is essential for the definition of a loss function and thus, training. 
In the case of t-SNE, we introduced two probability distributions based on the distance of samples in the original and feature space, respectively, and minimized their difference using the Kullback-Leibler divergence. 

\begin{figure}
\centering
\includegraphics{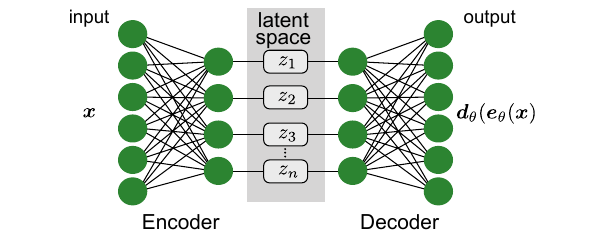}
\caption{{\bf General autoencoder architecture.} A neural network is used to contract a compressed representation of the input in the latent space. A second neural network is used to reconstruct the original input.}
\label{fig:AE_scheme}
\end{figure}

The solution the autoencoder uses is to have a neural network do first, the dimensionality reduction, or encoding to the \emph{latent space}\index{latent space}, $\bm{x}\mapsto \bm{e}_{\theta}(\bm{x})=\bm{z}$, and then, the decoding back to the original dimension, $\bm{z} \mapsto \bm{d}_{\theta}(\bm{z})$, see Fig.~\ref{fig:AE_scheme}. For the purpose of our discussion here, we assume that both the encoder and decoder are deterministic. This architecture allows us to directly compare the original input $\bm{x}$ with the reconstructed output $\bm{d}_\theta(\bm{e}_\theta(\bm{x}))$, such that the autoencoder trains itself unsupervised by minimizing the difference. A good example of a loss function that achieves successful training and that we have encountered already is the L2 loss,
\begin{equation}
	L_{\rm ae}(\theta) = \frac{1}{2m}\sum_{i=1}^m \| \bm{x}_i - \bm{d}_\theta(\bm{e}_\theta(\bm{x_i}))\|_2^2.
\end{equation}
In other words, we compare point-wise the difference between the input to the encoder with the decoder's output. 

Intuitively, the latent space with its lower dimension presents a bottleneck for the information propagation from input to output. The goal of training is to find and keep the most relevant information for the reconstruction to be optimal. The latent space then corresponds to the reduced space in PCA and t-SNE. Note that much like in t-SNE but unlike in PCA, the new features are in general not independent.

\subsubsection{Variational autoencoders}
A major problem of the approach introduced in the previous section is its tendency to overfitting\index{overfitting}. As an extreme example, a sufficiently complicated encoder-decoder pair could learn to map all data in the training set onto a single variable and back to the data. Such a network would indeed accomplish completely lossless compression and decompression. However, the network would not have extracted any useful information from the dataset and thus, would completely fail to compress and decompress previously unseen data. Moreover, as in the case of the dimensionality-reduction schemes discussed in Sec.~\ref{sec:structuring_data}, we would like to analyze the latent space images and extract new information about the data. Finally, we might also want to use the decoder part of the autoencoder as a generator for new data. For these reasons, it is essential that we combat overfitting as we have done in the previous chapters by regularization.

The question then becomes how one can effectively regularize the autoencoder. First, we need to analyze what properties we would like the latent space to fulfil. We can identify two main properties: 
\begin{enumerate}
	\item If two input data points are close (according to some measure), their images in the latent space should also be close. We call this property \emph{continuity}. 
	\item Any point in the latent space should be mapped through the decoder onto a meaningful data point, a property we call \emph{completeness}. 
\end{enumerate}
While there are principle ways to achieve regularization along similar paths as discussed in the previous section on supervised learning, we will discuss here a solution that is particularly useful as a generative model: the \emph{variational autoencoder}\index{variational autoencoder} (VAE).

\begin{figure}
\centering
\includegraphics{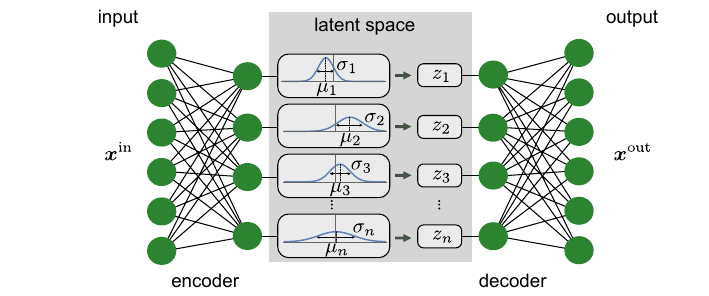}
\caption{{\bf Architecture of variational autoencoder.} Instead of outputting a point $z$ in the latent space, the encoder provides a distribution $N(\bm{\mu}, \bm{\sigma})$, parametrized by the means $\bm{\mu}$ and the standard deviations $\bm{\sigma}$. The input $\bm{z}$ for the decoder is then drawn from $N(\bm{\mu}, \bm{\sigma})$.}
\label{fig:VAE}
\end{figure}

The idea behind VAEs is for the encoder to output not just an exact point $\bm{z}$ in the latent space, but a (factorized) Normal distribution of points, $\mathcal{N}(\bm{\mu}, \bm{\sigma})$. In particular, the output of the encoder comprises two vectors, the first representing the means $\bm{\mu}$, and the second the standard deviations $\bm{\sigma}$. The input for the decoder is then sampled from this distribution, $\bm{z} \sim \mathcal{N}(\bm{\mu}, \bm{\sigma})$,  and the original input is reconstructed and compared to the original input for training. In addition to the standard loss function comparing input and output of the VAE, we further add a regularization term to the loss function such that the distributions from the encoder are close to a standard normal distribution $\mathcal{N}(\bm{0}, \bm{1})$. Using the Kullback-Leibler divergence, Eq.~\eqref{eq:KL},
to measure the deviation from the standard normal distribution, the full loss function then reads
\begin{align}
	L_{\rm vae}(\theta) &= \frac{1}{2m}\sum_{i=1}^m \| \bm{x}^{\rm in}_i - \bm{x}^{\rm out}_i\|_2^2 + {\rm KL} (\mathcal{N}(\bm{\mu}_i, \bm{\sigma}_i)|| \mathcal{N}(\bm{0}, \bm{1}))\nonumber\\
	&=\frac{1}{2m}\sum_{i=1}^m \| \bm{x}^{\rm in}_i - \bm{x}^{\rm out}_i\|_2^2 + \frac12 \sum_k [\sigma_{i,k}^2 + \mu_{i,k}^2 -1 -2 \log\sigma_{i,k}].
	\label{eq:loss_vae}
\end{align}
In this expression, the first term quantifies the reconstruction loss with $\bm{x}_i^{\rm in}$ the input to and $\bm{x}_i^{\rm out}$ the reconstructed data from the VAE. The second term is the regularization on the latent space for each input data point, which for two (diagonal) Normal distributions can be expressed in a closed form, see second line of Eq.~\eqref{eq:loss_vae}.
This procedure regularizes the training through the introduction of noise, similar to the dropout layer in Sec.~\ref{sec:supervized}. However, the regularization here not only generically increases generalization, but also enforces the desired structure in the latent space.

The structure of a VAE is shown in Fig.~\ref{fig:VAE}. By enforcing the mean and variance structure of the encoder output, the latent space fulfills the requirements outlined above. This type of structure can then serve as a generative model for many different data types: anything from human faces to complicated molecular structures. Hence, the variational autoencoder goes beyond extracting information from a dataset, but can be used for scientific discovery. Note, finally, that the general structure of the variational autoencoder can be applied beyond the simple case introduced above. As an example, a different distribution function can be enforced in the latent space other than the standard Normal distribution, or a different neural network can be used as encoder and decoder, such as a RNN.

	\subsection{Generative adversarial networks}

In this last part on unsupervised learning, we will introduce a final type of generative neural network, the generative adversarial network (GAN), which gained high popularity in recent years. 
Before getting into the details of this method, we summarize the chapter with a brief systematic overview over the types of generative methods we covered so far to place GANs in proper relation to them.

\subsubsection{Types of generative models}
In this section, we have restricted ourselves to methods that are based on the maximum likelihood principle, see Sec.~\ref{sec:maxlikelihood}. The role of a model is to provide an estimate $P_{\theta}(\bm{x})$ parametrized by parameters $\theta$ of the probability distribution the data set $\{\bm{x}_i\}$ is drawn from. The likelihood is the probability that the model assigns to the data set 
\begin{equation}
	\mathcal{L}(\theta; \{\bm{x}_i\}) = \prod_{i=1}^m P_\theta(\bm{x}_{i}),
\end{equation}
where $m$ is the number of samples in the data set $\{\bm{x}_i\}$. The goal is to choose the parameters $\theta$ such as to maximize the likelihood. While we have used the maximum likelihood principle explicitly only in the case of the RBM in Sec.~\ref{sec:rbm}, both the RNN and the VAE of Secs~\ref{sec:rnn2} and \ref{sec:vae} can be understood in this framework .

\begin{figure}
\centering
\includegraphics{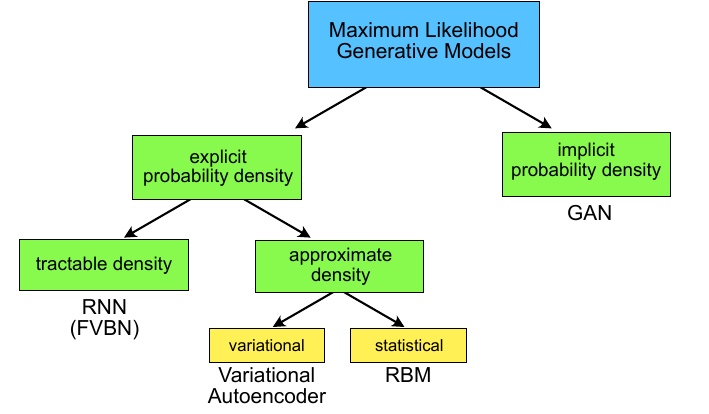}
\caption{{\bf Maximum likelihood approaches to generative modeling.} Unlike the author generative models we discussed in this chapter, GANs model the probability density only implicitly, in other words we can sample from $P_\theta(\bm{x})$ without an actual model for it.s}
\label{fig: generative taxonomy}
\end{figure}

The methods and their respective models can be distinguished by the way $P_\theta^{\mathrm{model}}(\bm{x})$ is defined, trained, and evaluated (see Fig.~\ref{fig: generative taxonomy}). We differentiate between models that define $P_{\theta}(\bm{x})$ \emph{explicitly}  through some functional form and those, where the probability distribution is defined implicitly. The explicit models have the general advantage that maximization of the likelihood is rather straight-forward, since we have direct access to this function. The disadvantage is that the functional forms are generically limiting the ability of the model to fit the data distribution or become computationally intractable.  

Among the explicit density models, we thus further distinguish between those that represent a computationally tractable density and those that do not. An example for tractable explicit density models are so-called \emph{fully visible belief networks} (FVBNs) that decompose the probability distribution over an $n$-dimensional vector $\bm{x}$ into a product of conditional probabilities
\begin{equation}
P_\theta^{\mathrm{FVBN}}(\bm{x})=\prod_{j=1}^n\, p_\theta^{\mathrm{FVBN}}(x_j|x_1,\cdots,x_{j-1}).
\end{equation}
We can already see that once we use the model to draw new samples, this sampling is done one entry of the vector $\bm{x}$ at a time (first $x_1$ is drawn, then, knowing $x_1$, $x_2$ is drawn etc.). This is computationally costly and not parallelizable but is useful for data with a sequential structure. An example of a fully visible belief network is the RNN we discussed in Section~\ref{sec:rnn2}.

Models that encode an explicit density, but require approximations to maximize the likelihood can either be variational in nature or use stochastic methods. We have seen examples of both. Variational methods define a lower bound to the log-likelihood which can be maximized,
\begin{equation}
	l(\theta; \bm{x})\leq \mathrm{log}\,P_\theta^{\text{var}}(\bm{x}).
\end{equation}
The algorithm produces a maximum value of the log-likelihood that is at least as high as the value for $l(\theta; \bm{x})$ obtained (when summed over all data points). Variational autoencoders belong to this category when the output layer is interpreted as a probability density $p_\theta(\bm{x}|\bm{z})$. Their most obvious shortcoming is that  $l(\theta; \bm{x})$ may represent a very bad lower bound to the log-likelihood (and is in general not guaranteed to converge to it for infinite model size), so that the distribution represented by the model can be very different from $P_{\mathrm{data}}(\bm{x})$.
Stochastic methods, in contrast, often rely on a Markov-chain process for their evaluation and training: The model is defined by a  probability $q(\bm{x}'|\bm{x})$ from which the next sample $\bm{x}'$ is drawn, which depends on the current sample $\bm{x}$ but not any others. RBMs are an example for this class of models, where Gibbs sampling realizes the Markov chain. RBMs are guaranteed to converge to a random sample drawn from $P^{\rm rbm}_\theta(\bm{x})$, but this convergence may be slow and as such drawing samples inefficient. 
    
All the above classes of models allow for an explicit representation of a model probability density function. 
In contrast, GANs and related models, only allow indirect access to the probability density: The model only allows us to sample from it. While this makes optimization potentially harder, the architecture circumvents many of the other previously mentioned problems. In particular, GANs can generate samples in parallel with no need for a Markov chain; There are few restrictions on the form of the generator function (as compared to Boltzmann machines, for instance, which have a restricted form to make Markov chain sampling work); no variational bound is needed; and some GAN model families are known to be asymptotically consistent (meaning that for a large enough model they are approximations to any probability distribution).

GANs have been immensely successful in several application scenarios, most of which in the field of image generation, and largely on the ImageNet database. Some of the standard tasks in this context are generating an image from a sentence or phrase that describes its content (``a blue flower''), generating realistic images from sketches, generating abstract maps from satellite photos generating a high-resolution (``super-resolution'') image from a lower resolution one, or predicting a next frame in a video. Note that for the actual assessment of the output, the likelihood might not be relevant, as especially for images, the quality of an output is subjective.
As far as more science-related applications are concerned, GANs have been used to predict the impact of climate change on individual houses or to generate new molecules that have been later synthesized.

In the light of these examples, it is of fundamental importance to understand that GANs enable (and excel at) problems with multi-modal outputs. That means the problems are such that a single input corresponds to many different `correct' or `likely' outputs. (In contrast to a mathematical function, which would always produce the same output.) This is important to keep in mind in particular in scientific applications, where we often search for \emph{the one answer}. Only if that is not the case, GANs can actually play out their strengths. 

Let us consider image super-resolution as an illustrative example: Conventional (deterministic) methods of increasing image resolution would necessarily lead to some blurring or artifacts, because the information that can be encoded in the finer pixel grid simply is not existent in the input data. A GAN, in contrast, will provide a possibility how a realistic image could have looked if it had been taken with higher resolution. This way they add information that may differ from the true scene of the image that was taken -- a process that is obviously not yielding a unique answer since many versions of the information added may correspond to a realistic image. 

\begin{figure}
\centering
\includegraphics{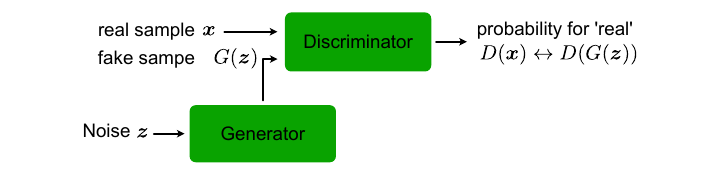}
\caption{{\bf Architecture of a GAN.} A generator $G(\bm{z})$ creates new samples (usually with $\bm{z}$ drawn randomly) and a discriminator $D$ tries to distinguish generated from real data.}
\label{fig:GAN_scheme}
\end{figure}

\subsubsection{The working principle of GANs}
So far, we have formulated the learning of neural network models by defining a single cost function---in this section the negative log-likelihood--- which we then minimize. While such a formulation is indeed possible for GANs, a more illuminating perspective is viewing the GAN as a \emph{game} between two players, the \emph{generator} ($G$) and the \emph{discriminator} ($D$), see Fig.~\ref{fig:GAN_scheme}. The role of $G$ is to generate from some random input $\bm{z}$  drawn from a simple distribution samples that could be mistaken from being drawn from $P_{\mathrm{data}}(\bm{x})$. The task of $D$ is to classify its input as generated by $G$ or coming from the data. Training should improve the performance of both $D$ and $G$ at their respective tasks simultaneously. After training is completed, $G$ can be used to draw samples that closely resembles those drawn from $P_{\mathrm{data}}(\bm{x})$.  
The two functions then are
\begin{equation}
\begin{split}
D_{\theta_D}(\bm{x})&:\ \bm{x}\mapsto \text{binary: true/false},\\
G_{\theta_G}(\bm{z})&:\ \bm{z}\mapsto \bm{x},
\end{split}
\end{equation}
where we have also indicated the two sets of parameters on which the two functions depend: $\theta_D$ and $\theta_G$, respectively.
The game is then defined by \emph{two} cost functions: The discriminator wants to minimize $L_D(\theta_D;\theta_G)$ by only changing $\theta_D$, while the generator $L_G(\theta_G;\theta_D)$ by only changing $\theta_G$. So, each player's cost function depends on both their and the other player's parameters, the latter of which cannot be controlled by the player. The solution to this game optimization problem is a (local) minimum, in other words a point in $(\theta_D,\theta_G)$-space where $L_D(\theta_D;\theta_G)$ has a local minimum with respect to $\theta_D$ and $L_G(\theta_G;\theta_D)$ has a local minimum with respect to $\theta_G$. In game theory such a solution is called a Nash equilibrium. 
Let us now specify possible choices for the cost functions as well as for $D$ and $G$.

The most important requirement of $G$ is that it is differentiable. In contrast to VAEs, it can thus not have discrete variables on the output layer. A typical representation is a deep (possibly convolutional) neural network~\footnote{A popular Deep Conventional architecture is called DCGAN}.  Then $\theta_G$ are the networks weights and biases. The input $\bm{z}$ is drawn from some simple prior distribution, such as the uniform distribution or a normal distribution. The space, where $\bm{z}$ is living in, is called the latent space as in the case of the autoencoder. It is important that the latent space ($\bm{z}$) has at least as high a dimension as the feature space ($\bm{x}$) if the model $P_{\theta}(\bm{x})$ should have full support on the feature space.

The training proceeds in steps. At each step, a minibatch of $\bm{x}$ is drawn from the data set and a minibatch of $\bm{z}$ is sampled from the prior distribution. Using this, gradient descent-type updates are performed: One update of $\theta_D$ using the gradient of $L_D(\theta_D;\theta_G)$ and one of $\theta_G$ using the gradient of $L_G(\theta_G;\theta_D)$.
 
 \subsubsection{The cost functions}

For the discriminator $D$, the cost function of choice is the cross-entropy as with standard binary classifiers that have a sigmoid output. Given that the labels are `1' for data and `0' for samples from $G$, the const function reads
\begin{equation}
L_D(\theta_D;\theta_G)
=-\frac{1}{2 N_1}\sum_i\,\log\,D_{\theta_D}(\bm{x}_i)-\frac{1}{2 N_2}\sum_j\log (1-D_{\theta_D}(G_{\theta_G}(\bm{z}_j))),
\end{equation}
where the sums over $i$ and $j$ run over the respective minibatches, which in general contain $N_1$ and $N_2$ points, respectively. 

For the generator $G$, more variations of the cost functions have been explored. Maybe the most intuitive one is 
\begin{equation}
L_G(\theta_G;\theta_D)=-L_D(\theta_D;\theta_G),
\end{equation}
which corresponds to the so-called \emph{zero-sum} or \emph{minimax} game. Its solution is formally given by
\begin{equation}
\theta_G^\star=\underset{\theta_G}{\text{arg min}}\ \ \underset{\theta_D}{\text{max}}
\left[-L_D(\theta_D;\theta_G)\right].
\label{eq: GAN Minmax}
\end{equation}
This form of the cost is convenient for theoretical analysis, because there is only a single target function, which helps drawing parallels to conventional optimization. However, other cost functions have been proven superior in practice. The reason is that minimization can get trapped very far from an equilibrium: When the discriminator manages to learn rejecting generator samples with high confidence, the gradient of the generator will be very small, making its optimization very hard. 

Instead, we can use the cross-entropy also for the generator cost function (but this time from the generator's perspective)
\begin{equation}
L_G(\theta_G;\theta_D)=-\frac{1}{2 N_2}\sum_j\log\, D_{\theta_D}(G_{\theta_G}(\bm{z}_j)).
\end{equation}
 Now, the generator maximizes the probability of the discriminator being mistaken. This way, each player still has a strong gradient when the player is loosing the game. We observe that this version of $L_G(\theta_D;\theta_G)$ has no direct dependence of the training data. Of course, such a dependence is implicit via $D$, which has learned from the training data. This indirect dependence also acts like a regularizer, preventing overfitting: $G$ has no possibility to directly `fit' its output to training data.

 \subsubsection{Final Remarks on GANs}

In closing, we comment on a few properties and issues of GANs, which also mark frontiers for improvements. A general problem is that GANs are typically difficult to train: they require large training sets and are highly susceptible to hyperparameter fluctuations. It is a current topic of research to compensate for this with the structural modification and novel loss function formulations.
\vspace{5pt}

\noindent\textbf{Mode collapse}\\ Mode collapse refers to one of the most obvious problems of GANs: If the generator $G$ does not explore the full space covered in the data set, but rather maps several inputs $\bm{z}$ to the same output $\bm{x}$. Full mode collapse is rather rare, but a generator $G$ trained on generating images may always resort to certain fragments, colors, or patterns of images. The formal reason for mode collapse is that the simultaneous gradient descent gravitates towards a solution, which can schematically be written as
\begin{equation}
	G^\star=\underset{D}{\text{max}}\ \ \underset{G}{\text{min}}
	\left[V(D,G)\right],
\end{equation}
instead of the order in Eq.~\eqref{eq: GAN Minmax}, where V(D,G) is the overall `loss function'~\footnote{The function is rather a \emph{value function}, as we have not really defined a single loss function. In Eq.~\eqref{eq: GAN Minmax}, we would have $V(D,G) = L_D(\theta_D;\theta_G)$.}. While the interchange of min and max might seem innocent, the reversed order corresponds to a different solution: It is now sufficient for $G$ to always produce one (and the same) output that the discriminator $D$ classified as data with very high probability. Due to the mode collapse problem, GANs are not good at exploring ergodically the full space of possible outputs. They rather produce few very good possible outputs. 

One strategy to fight mode collapse is called \emph{minibatch features}. Instead of letting $D$ rate one sample at a time, a minibatch of real and generated samples  is considered at once. It then detects whether the generated samples are unusually close to each other.
\vspace{5pt}

\noindent\textbf{One-sided label smoothing}\\ 
Often, the discriminator $D$ gives proper results, but shows a too confident probability~\footnote{Note that this is not only a problem of the discriminator in GANs, but more general for neural-network classifiers. The technique is thus useful also other binary classification problems with neural networks.}. This overshooting confidence can be counteracted by one-sided label smoothing and is a kind of regularization. The idea  is to simply replace the target value for the real examples in the loss function with a value $1-\alpha < 1$ . This results in a smoother distribution of the discriminator. Why do we only perform this off-set one-sided and not also give a small nonzero value $\beta$ to the fake samples target values? In this case, the optimal function for $D$ would be
\begin{equation}
D^\star(\bm{x})=\frac{(1-\alpha)P_{\mathrm{data}}(\bm{x})+\beta P_{\theta}(\bm{x})}{P_{\mathrm{data}}(\bm{x})+  P_{\theta}(\bm{x})}.
\end{equation}
Consider now a range of $\bm{x}$ for which $P_{\mathrm{data}}(\bm{x})$ is small but  $P_{\theta}(\bm{x})$ is large (a ``spurious mode''). $D^\star(\bm{x})$ will have a peak near this spurious mode. This means $D$ reinforces incorrect behavior of $G$ and encourages $G$ to reproduce samples that it already makes (irrespective of whether they are anything like real data). 
\vspace{5pt}

\noindent\textbf{Using GANs with labelled data}\\
If (at least partially) labeled data is available, using the labels when training the discriminator $D$ may improve the performance of $G$. In this constellation, $G$ still has the same task as before and does not interact with the labels. However, for data with $n$ classes $D$ is constructed as a classifier for $(n+1)$ classes, where the extra class corresponds to `fake' data that $D$ attributes to coming from $G$. If a data point has a label, then this  label is used as a reference in the cost function.  If a datapoint has no label, then the first $n$ outputs of $D$ are summed up. 
\vspace{5pt}

\noindent\textbf{Arithmetics with GANs}\\
GANs can do linear arithmetics with inputs to add or remove abstract features from the output. In other words, the latent space, in which $\bm{z}$ is a point, has structure that can be exploited. This has been demonstrated using a DCGAN trained on images of faces: The gender and the feature `wearing glasses' can be added or subtracted and thus changed at will. Of course, such a result is only empirical, there is no formal mathematical theory why it works.

%-----Vulnerabilities-------------------------------------------%
	\section{Interpretability of Neural Networks}
\sectionmark{Interpretability}
\label{sec:interpretability}
In particular for applications in science, we not only want to obtain a neural network that excels at performing a given task, but we also seek an understanding of how the problem was solved. Ideally, we want to know the underlying principles the network has learned, deduce causal relations, or identify abstracted notions. This is the topic of \emph{interpretability}. We will in the following first discuss how information can be extracted in generative models from the latent space if its dimension is sufficiently small. Then, we discuss supervised learning and how \emph{dreaming} helps us understand the learning of neural networks better. 

\subsection{Interpreting latent-space representations}
\label{sec:latent-space}
In the previous chapter, we have learned about the broad scope of applications of generative models. We have seen how generative models can be used in the scientific context to extract a compressed representation of the input data and use this latent-space representation to generate new instances of the data. In the case of GANs, we have already seen that there can be enough structure in the latent space to perform simple arithmetics. It turns out that in simple enough problems with a latent space with sufficiently small dimension, we can find a meaningful interpretation of the representation that may help us get new insight into a given problem.

A neat example of such an interpretation of latent-space variables concerns one of the most famous models in the history of physics: Copernicus' heliocentric system of the solar orbits. Via a series of precise measurements of the position of objects in the night sky seen from Earth, Copernicus conjectured that the Sun is at the center of the solar system and all the planets are orbiting around it. We can now ask whether is it possible for a neural network that receives the same observed angles to reach the same conclusion.

\begin{figure}
\centering
\includegraphics{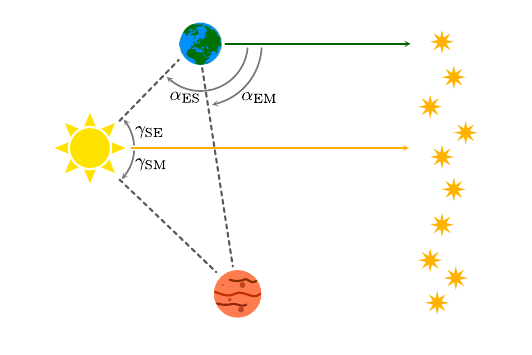}
\caption{{\bf The Copernicus problem.} Relation between angles to describe the position of Mars in the heliocentric and geocentric coordinate system within the reference frame of the fixed stars.}
\label{fig:copernicus}
\end{figure}

In a recent work, a neural network resembling an autoencoder was trained to predict the position of Mars at a future time~\footnote{Iten, Metger, \emph{et al.}, \href{https://journals.aps.org/prl/abstract/10.1103/PhysRevLett.124.010508}{PRL {\bf 124}, 010508 (2020)}}. For the purpose of training, the network received observations of the angles of Mars and the Sun as seen from Earth ($\alpha_{\rm ES}$ and $\alpha_{\rm EM}$) and the time of observation, see Fig.~\ref{fig:copernicus}. In analogy to an autoencoder, the network could choose a `parametrization' of the relevant information in the form of a (two-dimensional) latent space. When analyzing this latent-space representation of the trained model, it turned out that the two latent neurons were storing information in heliocentric coordinates. Specifically, the information was stored in the latent space as a linear combination of angles between Sun and Mars, $\gamma_{\rm SM}$ and Sun and Earth $\gamma_{\rm SE}$. Just like Copernicus, the network had learned that the most efficient way to store the data was to transform them into the heliocentric coordinate system.

While this intriguing example shows how generative models can be interpreted in some simple cases, the general question of interpretability of the latent space is still very much open and subject to ongoing research. In the instances discussed earlier in this lecture, such as generation of molecules, where the input data is compressed through several layers of a neural network to a high-dimensional latent space, interpreting the latent space becomes intractably challenging.

\subsection{Dreaming and the problem of extrapolation}
Exploring the weights and intermediate activations of a neural network in order to understand what the network has learnt very quickly becomes unfeasible or uninformative for large network architectures. Here, we adopt a different approach, where we focus on the inputs to the neural network, rather than the intermediate activations, to better understand what the network has learned.

More precisely, let us consider a neural network classifier $\bm{F}_\theta(\bm{x})$, depending on the parameters $\theta = \{W, b\}$ with weights $W$ and biases $b$, which maps an input $\bm{x}$ to a probability distribution over $n$ classes $\bm{F}_\theta(\bm{x}) \in \mathbb{R}^{n}$, in other words
\begin{equation}
    F_{\theta, i}(\bm{x}) \geq 0 \ \ \  \mathrm{and} \ \ \  \sum_{i} F_{\theta, i}(\bm{x}) = 1.
\end{equation}
In the previous sections, We minimized the distance between the output of the network $\bm{F}_\theta(\bm{x})$ and a chosen target output $\bm{y}$, which we did by minimizing a loss function, such as
\begin{equation} \label{eqn: dreaming loss}
    L = |\bm{F}_\theta(\bm{x}) - \bm{y}|^{2}.
\end{equation}
However, unlike in supervised learning, where we minimized the loss minimization with respect to the parameters $\theta$ of the network, we are here interested in minimizing with respect to the input $\bm{x}$ while keeping the parameters $\theta$ fixed. For this purpose, we again use gradient descent, 
\begin{equation} \label{eqn: dreaming update}
    \bm{x} \mapsto \bm{x} - \eta \frac{\partial L}{\partial \bm{x}},
\end{equation}
where $\eta$ is the learning rate. With a sufficient number of iterations, an initial input $\bm{x}^{0}$ will be transformed into a final input $\bm{x}^{*}$, such that
\begin{equation}
    \bm{F}_\theta(\bm{x}^*) \approx \bm{y}.
\end{equation}
By choosing the target output to correspond to a particular class, $\bm{y}^* = (1, 0, 0, \dots)$ as an example, we are essentially finding examples of inputs which the network would classify as belonging to the chosen class. This procedure is called \emph{dreaming}\index{dreaming}.

\begin{figure}[t]
\centering
\includegraphics[width=0.75\textwidth]{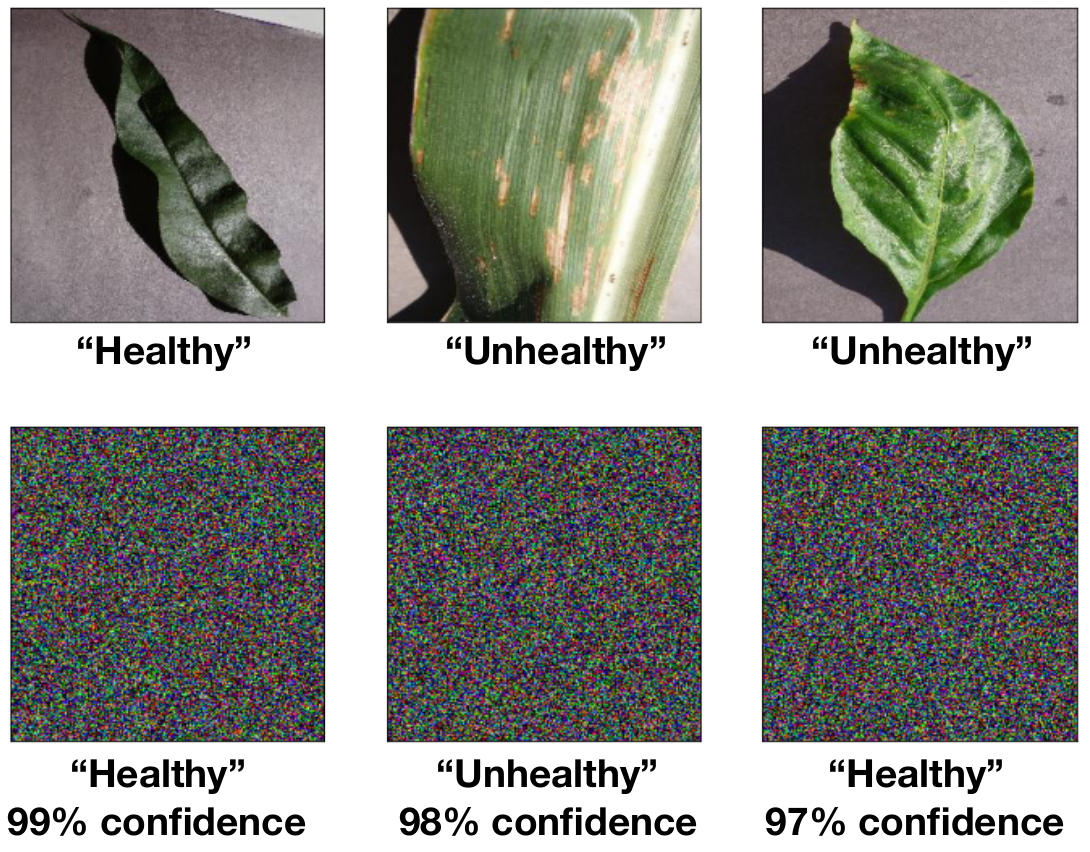}
\caption{\textbf{Plant leaves.} Top: Some samples from the plants dataset. Bottom: Samples generated by using the "dreaming" procedure starting from random noise.}
\label{fig: dreaming}
\end{figure}

We will apply this technique to a binary classification example. We consider a dataset consisting of images of healthy and unhealthy plant leaves~\footnote{Source: https://data.mendeley.com/datasets/tywbtsjrjv/1}. Some samples from the dataset are shown in the top row of Fig.~\ref{fig: dreaming}. After training a deep convolutional network to classify the leaves (reaching a test accuracy of around $95\%$), we start with a random image as our initial input 
$\bm{x}^{0}$ and perform gradient descent on the input, as described above, to arrive at the final image $\bm{x}^{*}$ which our network confidently classifies.

In the bottom row of Fig.~\ref{fig: dreaming}, we show three examples produced using the `dreaming' technique. On first sight, it might be astonishing that the final image actually does not even remotely resemble a leaf. How could it be that the network has such a high accuracy of around $95\%$, yet we have here a confidently classified image which is essentially just noise. While this seems surprising at first, a closer inspection reveals the problem: The noisy image $\bm{x}^{*}$ looks nothing like the samples in the dataset with which we trained our network. By feeding this image into our network, we are asking the network to make an extrapolation, which, as can be seen, leads to uncontrolled behavior. This is a key issue which plagues most data-driven machine-learning approaches. With few exceptions, it is very difficult to train a model capable of performing reliable extrapolations. Since scientific research is often in the business of making extrapolations, this is an extremely important point of caution to keep in mind.

While it might seem obvious that any model should only be predictive for data that `resembles' those in the training set, the precise meaning of `resembles' is actually more subtle than one imagines. For example, if one trains a ML model using a dataset of images captured using a Canon camera but subsequently decide to use the model to make predictions on images taken with a Nikon camera, we could actually be in for a surprise. Even though the images may `resemble' each other to our naked eye, the different cameras can have a different noise profile which might not be perceptible to the human eye. We will see in the next section that even such minute image distortions can already be sufficient to completely confuse a model.

\subsection{Adversarial attacks}
As we have seen, it is possible to modify the input $\bm{x}$ so that the corresponding model approximates a chosen target output. This concept can also be applied to generate \textit{adverserial examples}\index{adverserial examples}, in other words images which have been intentionally modified to cause a model to misclassify them. As a further requirement, we usually want the modification to be minimal, almost imperceptible to the human eye. 

One common method for generating adversarial examples is known as the \textit{fast gradient sign method} (FSGM). Starting from an input $\bm{x}^{0}$ which our model correctly classifies, we choose a target output $\bm{y}^{*}$ which corresponds to a wrong classification, and follow the procedure described in the previous section with a slight modification. Instead of updating the input according to Eq.~\eqref{eqn: dreaming update} we use the following update rule:
\begin{equation}
    \bm{x} \mapsto \bm{x} - \eta \  \textrm{sign}\left(\frac{\partial L}{\partial \bm{x}}\right),
\end{equation}
where $L$ is given be Eq.~\eqref{eqn: dreaming loss}. The $\textrm{sign}(\dots) \in \lbrace -1, 1 \rbrace$ both serves to enhance the signal and also acts as constraint to limit the size of the modification. By choosing $\eta = \frac{\epsilon}{T}$ and performing only $T$ iterations, we can then guarantee that each component of the final input $\bm{x}^{*}$ satisfies
\begin{equation}
    |x^{*}_{i} - x^{0}_{i}| \leq \epsilon.
\end{equation}
Consequently, our final image $\bm{x}^{*}$ is only minimally modified. We summarize this algorithm as follows:
\begin{algbox}[Fast Gradient Sign Method]{alg: FGSM}
	\begin{algorithm}[H]
	\SetAlgoLined
	\KwIn{
	A classification model $\bm{F}_\theta$, a loss function $L$, an initial image $\bm{x}^{0}$, a target label $\bm{y}^*$, perturbation size $\epsilon$ and number of iterations $T$}
	\KwOut{
	Adversarial example $\bm{x}^{*}$ with $|x^{*}_{i} - x^{0}_{i}| \leq \epsilon$}
 	$\eta = \epsilon/T$ \;
 	\For{i=1\dots T}{
  	$\bm{x} = \bm{x} - \eta \ \textrm{sign}\left(\frac{\partial L}{\partial \bm{x}}\right)$ \;
 	}
 	\end{algorithm}
\end{algbox}

This process of generating adversarial examples is called an \textit{adversarial attack}\index{adversarial attack}, which we can classify under two broad categories: \textit{white box}\index{white box} and \textit{black box}\index{black box} attacks. In a white box attack, the attacker has full access to the network $\bm{F}_\theta$ and is thus able to compute or estimate the gradients with respect to the input. On the other hand, in a black box attack, the adversarial examples are generated without using the target network $\bm{F}_\theta$. In this case, a possible strategy for the attacker is to train his own model $\bm{G}_\theta$, find an adversarial example for his model and use it against his target $\bm{F}_\theta$ without actually having access to it. Although it might seem surprising, this strategy has been found to work albeit with a lower success rate as compared to white box methods. We illustrate these concepts in the example below.

\begin{figure}[t]
\centering
\includegraphics[width=1.0\textwidth]{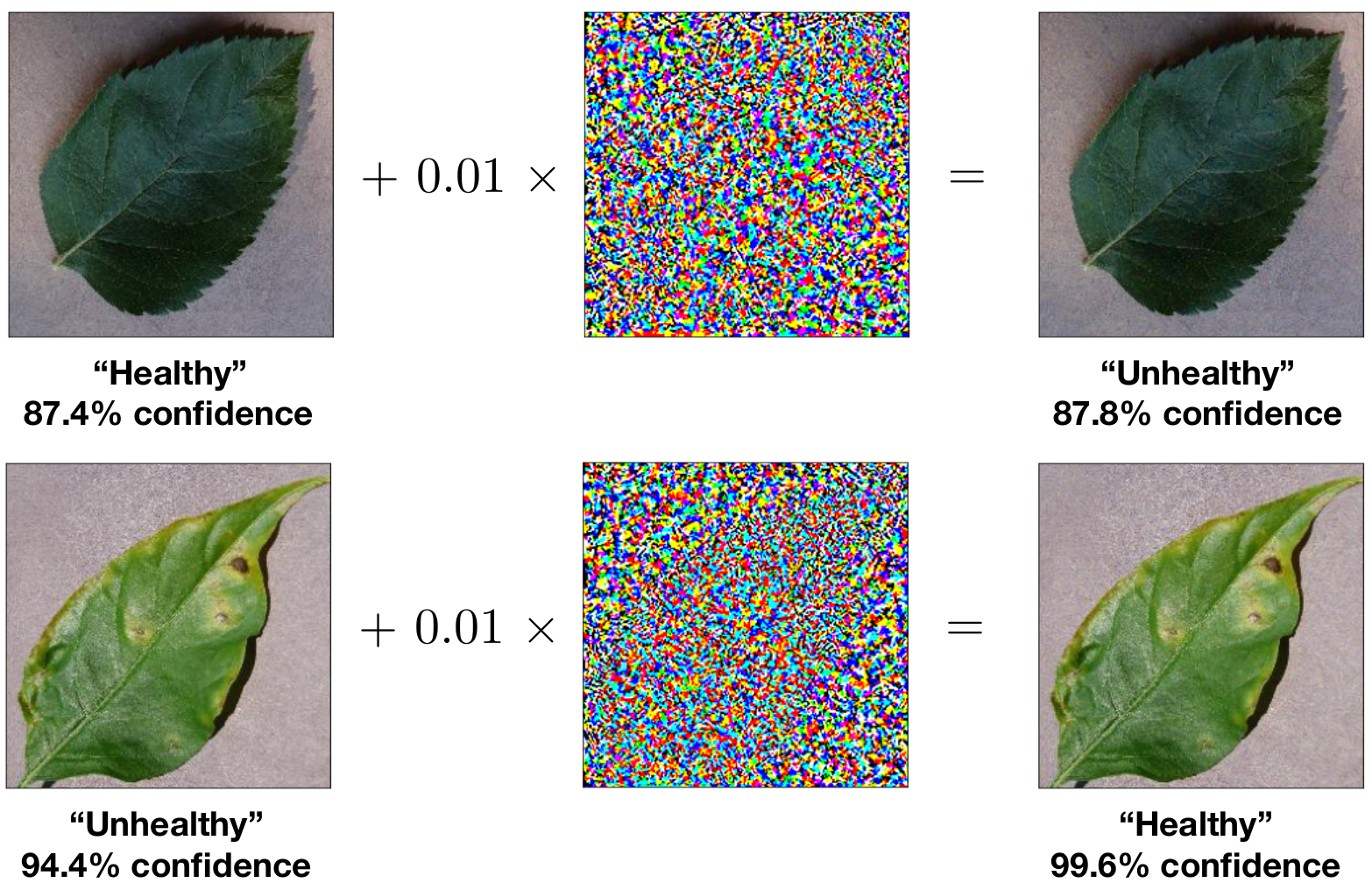}
\caption{\textbf{Adversarial examples.} Generated using the fast gradient sign method with $T=1$ iteration and $\epsilon = 0.01$. The target model is Google's \textit{InceptionV3} deep convolutional network with a test accuracy of $\sim 95\%$ on the binary ("Healthy" vs "Unhealthy") plants dataset. }
\label{fig: white box attack}
\end{figure}

\subsubsection{Example}
We use the same plant-leaves-classification example as above. The target model $\bm{F}_\theta$ which we want to 'attack' is a \textit{pretrained}\index{pretrained model} model using Google's well known \textit{InceptionV3}\index{InceptionV3} deep convolutional neural network containing over $20$ million parameters\footnote{This is an example of \textit{transfer learning}\index{transfer learning}. The base model, InceptionV3, has been trained on a different classification dataset, \textit{ImageNet}\index{ImageNet}, with over $1000$ classes. To apply this network to our binary classification problem, we simply replace the top layer with a simple duo-output dense softmax layer. We keep the weights of the base model fixed and only train the top layer.}. The model achieved a test accuracy of $\sim 95\%$. Assuming we have access to the gradients of the model $\bm{F}_\theta$, we can consider a white box attack. Starting from an image in the dataset which the target model correctly classifies and applying the fast gradient sign method (Alg.~\ref{alg: FGSM}) with $\epsilon=0.01$ and $T=1$, we obtain an adversarial image which differs from the original image by almost imperceptible amount of noise as depicted on the left of Fig.~\ref{fig: white box attack}. Any human would still correctly identify the image but yet the network, which has around $95\%$ accuracy has completely failed.

\begin{figure}
\centering
\includegraphics[width=1.0\textwidth]{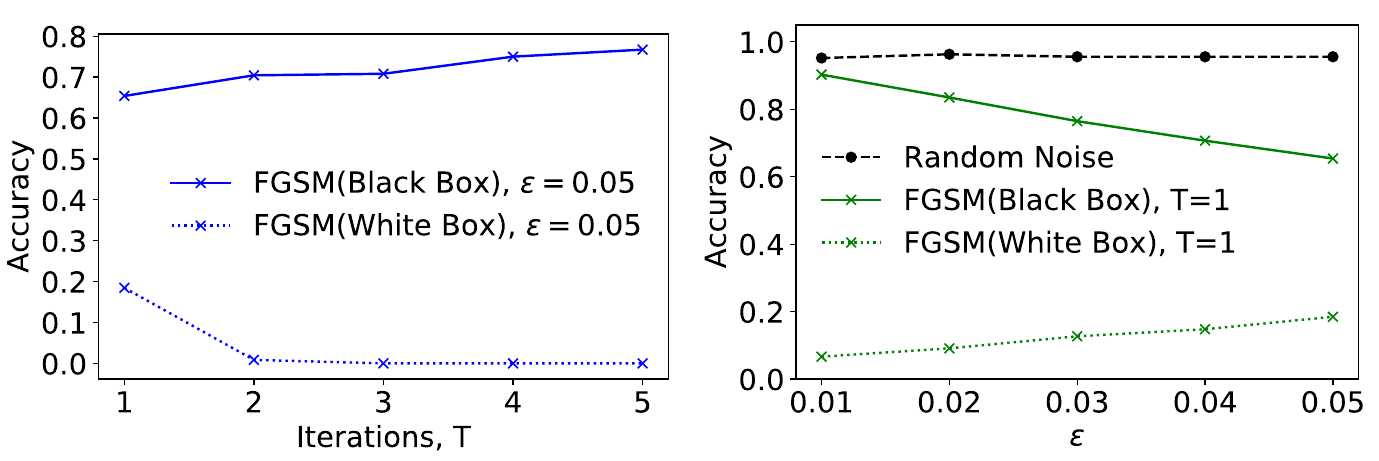}
\caption{\textbf{Black Box Adversarial Attack.}  }
\label{fig: black box attack}
\end{figure}

If, however, the gradients and outputs of the target model $\bm{F}_\theta$ are hidden, the above white box attack strategy becomes unfeasible.  In this case, we can adopt the following `black box attack' strategy. We train a secondary model $\bm{G}_{\theta'}$, and then applying the FGSM algorithm to $\bm{G}_{\theta'}$ to generate adversarial examples for $\bm{G}_{\theta'}$. Note that it is not necessary for $\bm{G}_{\theta'}$ to have the same network architecture as the target model $\bm{F}_\theta$. In fact, it is possible that we do not even know the architecture of our target model.

Let us consider another pretrained network based on \textit{MobileNet}\index{MobileNet} containing about $2$ million parameters. After retraining the top classification layer of this model to a test accuracy of $\sim 95\%$, we apply the FGSM algorithm to generate some adversarial examples. If we now test these examples on our target model $\bm{F}_\theta$, we notice a significant drop in the accuracy as shown on the graph on the right of Fig.~\ref{fig: black box attack}. The fact that the drop in accuracy is greater for the black box generated adversarial images as compared to images with random noise (of the same scale) added to it, shows that adversarial images have some degree of transferability between models. As a side note, on the left of Fig.~\ref{fig: black box attack} we observe that black box attacks are more effective when only $T=1$ iteration of the FGSM algorithm is used, contrary to the situation for the white box attack. This is because, with more iterations, the method has a tendency towards overfitting the secondary model, resulting in adversarial images which are less transferable.

These forms of attacks highlight not only a problem for the scientific application of neural networks, but a serious vulnerability of data-driven machine-learning techniques in general. Defending against such attack is thus an active area of research, buts is largely a cat and mouse game between the attacker and defender.

%-----Reinforcement Learning------------------------------%
	\section{Reinforcement Learning}
\label{sec:RL}
In the previous sections, we have introduced data-based learning, where we are given a dataset $\{\bm{x}_i\}$ for training. Depending on whether we are given labels $y_i$ with each data point, we have further divided our learning task as either being supervised or unsupervised, respectively. The aim of machine learning is then to classify unseen data (supervised), or extract useful information from the data and generate new data resembling the data in the given dataset (unsupervised). However, the concept of learning as commonly understood certainly encompasses other forms of learning that are not falling into these data-driven categories. 

An example for a form of learning not obviously covered by supervised or unsupervised learning is learning how to walk: in particular, a child that learns how to walk does not first collect data on all possible ways of successfully walking to extract rules on how to walk best. Rather, the child performs an action, sees what happens, and then adjusts their actions accordingly. This kind of learning thus happens best `on-the-fly', in other words while performing the attempted task. \emph{Reinforcement learning} formalizes this different kind of learning and introduces suitable (computational) methods.

As we will explain in the following, the framework of reinforcement learning considers an \emph{agent} that interacts with an \emph{environment} through actions, which, on the one hand, changes the \emph{state} of the agent and on the other hand, leads to a \emph{reward}. Whereas we tried to minimize a loss function in the previous sections, the main goal of reinforcement learning is to maximize this reward by learning an appropriate \emph{policy}. One way of reformulating this task is to find a \emph{value function}, which associates to each state (or state-action pair) a value, or expected total reward.
Note that, importantly, to perform our learning task we do not require knowledge, a \emph{model}, of the environment. All that is needed is feedback to our actions in the form of a reward signal and a new state. We stress again that we study in the following methods that learn at each time step~\footnote{Note that alternatively, one could also devise methods, where an agent tries a policy many times and judges only the final outcome.}.

The framework of reinforcement learning is very powerful and versatile. Examples include:
\begin{itemize}
	\item We can train a robot to perform a task, such as using an arm to collect samples. The state of the agent is the position of the robot arm, the actions move the arm, and the agent receives a reward for each sample collected.
	\item We can use reinforcement learning to optimize experiments, such as chemical reactions. In this case, the state contains the experimental conditions, such as temperature, solvent composition, or pH and the actions are all possible ways of changing these state variables. The reward is a function of the yield, the purity, or the cost. Note that reinforcement learning can be used at several levels of this process: While one agent might be trained to target the experimental conditions directly, another agent could be trained to reach the target temperature by adjusting the electrical current running through a heating element.
	\item We can train an agent to play a game, with the state being the current state of the game and a reward is received once for winning. The most famous example for such an agent is Google's AlphaGo, which outperforms humans in the game of Go. A possible way of applying reinforcement learning in the sciences is to phrase a problem as a game. An example, where such rephrasing was successfully applied, is error correction for (topological) quantum computers.
	\item In the following, we will use a toy example to illustrate the concepts introduced:  We want to train an agent to help us with the plants in our lab: in particular, the state of the agent is the water level. The agent can turn on and off a growth lamp and it can send us a message if we need to show up to water the plants. Obviously, we would like to optimize the growth of the plants and not have them die. 
\end{itemize}

As a full discussion of reinforcement learning goes well beyond the scope of this lecture, we will focus in the following on the main ideas and terminology with no claim of completeness.

\subsection{Exploration versus exploitation}\label{sec:expl_v_expl}
We begin our discussion with a simple example that demonstrates some important aspects of reinforcement learning. In particular, we discuss a situation, where the reward does not depend on a state, but only on the action taken. The agent is a doctor, who has to choose from $n$ actions, the treatments, for a given disease, with the reward depending on the recovery of the patient. The doctor `learns on the job' and tries to find the best treatment. The \emph{value} of a treatment $a\in\mathcal{A}$ is denoted by $q_* (a) = E( r ) $, the expectation value of our reward. 

Unfortunately, there is an uncertainty in the outcome of each treatment, such that it is not enough to perform each treatment just once to know the best one. Rather, only by performing a treatment many times we find a good estimate $Q_t(a) \approx q_*(a)$. Here, $Q_t(a)$ is our estimate of the value of $a$ after $t$ (time-) steps. Obviously, we should not perform a bad treatment many times, only to have a better estimate for its failure. We could instead try each action once and then continue for the rest of the time with the action that performed best. This strategy is called a \emph{greedy} method and \emph{exploits} our knowledge of the system. However, also this strategy bears risks, as the uncertainty in the outcome of the treatment means that we might use a suboptimal treatment. It is thus crucial to \emph{explore} other actions. This dilemma is called the `conflict between exploration and exploitation'. A common strategy is to use the best known action $a_* = {\rm argmax}_a Q_t(a)$ most of the time, but with probability $\epsilon$ chose randomly one of the other actions. This strategy of choosing the next action is called \emph{$\epsilon$-greedy}.

\subsection{Finite Markov decision process}
\begin{figure}[t]
	\centering
	\includegraphics{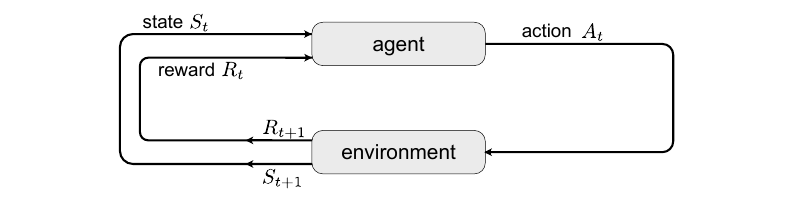}
	\caption{{\bf Markov decision process.} Schematic of the agent-environment interaction.}
	\label{fig:mdp}
\end{figure}

After this introductory example, we introduce the idealized form of reinforcement learning with a Markov decision process (MDP). 
At each time step $t$, the agent starts from a state $S_t\in \mathcal{S}$, performs an action $A_t\in\mathcal{A}$, which, through interaction with the environment, leads to a reward $R_{t+1}\in \mathcal{R}$ and moves the agent to a new state $S_{t+1}$. This agent-environment interaction is schematically shown in Fig.~\ref{fig:mdp}. Note that we assume the space of all actions, states, and rewards to be finite, such that we talk about a \emph{finite} MDP.

For our toy example, the sensor we have only shows whether the water level is high (h) or low (l), so that the state space of our agent is $\mathcal{S} = \{ {\rm h}, {\rm l} \}$. In both cases, our agent can choose to turn the growth lamps on or off, or in the case of low water, he can choose to send us a message so we can go and water the plants. The available actions are thus $\mathcal{A} = \{{\rm on}, {\rm off}, {\rm text}\}$. When the growth lamps are on, the plants grow faster, which leads to a bigger reward, $r_{\rm on} > r_{\rm off}>0$. Furthermore, there is a penalty for texting us, but an even bigger penalty for letting the plants die, $0 > r_{\rm text} > r_{\rm fail}$.

A model of the environment provides the probability of ending in state $s'$ with reward $r$, starting from a state $s$ and choosing the action $a$, $p(s', r | s, a)$. In this case, the dynamics of the Markov decision process is completely characterized. Note that the process is a Markov process, since the next state and reward only depend on the current state and chosen action. 

\begin{figure}[t]
	\centering
	\includegraphics{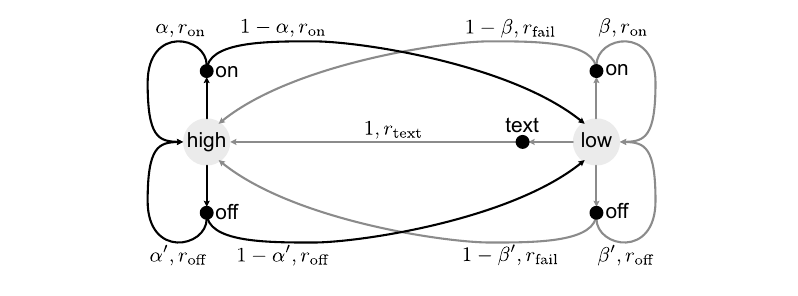}
	\caption{{\bf Transition graph of the MDP for the plant-watering agent.} The states `high' and `low' are denoted with large circles, the actions with small black circles, and the arrows correspond to the probabilities and rewards.}
	\label{fig:mdp_example}
\end{figure}

In our toy example, being in state `high' and having the growth lamp on will provide a reward of $r_{\rm on}$ and keep the agent in `high' with probability $p({\rm h}, r_{\rm on} | \rm {h}, {\rm on}) = \alpha$, while with $1-\alpha$ the agent will end up with a low water level. However, if the agent turns the lamps off, the reward is $r_{\rm off}$ and the probability of staying in state `high' is $\alpha' > \alpha$. For the case of a low water level, the probability of staying in low despite the lamps on is $p({\rm l}, r_{\rm on} | \rm {l}, {\rm on}) = \beta$, which means that with probability $1 - \beta$, our plants run out of water. In this case, we will need to get new plants and we will water them, of course, such that $p({\rm h}, r_{\rm fail} | \rm {l}, {\rm on}) = 1-\beta$. As with high water levels, turning the lamps off reduces our rewards, but increases our chance of not losing the plants, $\beta' > \beta$. Finally, if the agent should choose to send us a text, we will refill the water, such that $p({\rm h}, r_{\rm text} | {\rm l}, {\rm text}) = 1$. The whole Markov process is summarized in the \emph{transition graph} in Fig.~\ref{fig:mdp_example}.

From the probability for the next reward and state, we can also calculate the expected reward starting from state $s$ and choosing action $a$, namely
\begin{equation}
	r(s, a) = \sum_{r\in\mathcal{R}} r \sum_{s'\in\mathcal{S}} p(s', r | s, a).
\end{equation}
Obviously, the value of an action now depends on the state the agent is in, such that we write for the optimal total reward $q_* (s, a)$. Alternatively, we can also assign to each state a value $v_*(s)$, which quantizes the optimal reward from this state. 

Finally, we can define what we want to accomplish by learning: knowing our current state $s$, we want to know what action to choose such that our future total reward is maximized. Importantly, we want to accomplish this without any prior knowledge of how to optimize rewards directly. This poses yet another question: what is the total reward? We usually distinguish tasks with a well-defined end point $t=T$, so-called \emph{episodic tasks}, from \emph{continuous tasks} that go on for ever. The total reward for the former is simply the total \emph{return}
\begin{equation}
	G_t = R_{t+1} + R_{t+2} + R_{t+3} + \cdots + R_T.
\end{equation}
As such a sum is not guaranteed to converge for a continuous task, the total reward is the \emph{discounted return}
\begin{equation}
	G_t = R_{t+1} + \gamma R_{t+2} + \gamma^2 R_{t+3} + \cdots = \sum_{k=0}^\infty \gamma^k R_{t+k+1},
	\label{eq:disc_return}
\end{equation}
with $0 \leq \gamma <  1$ the discount rate. Equation~\eqref{eq:disc_return} is more general and can be used for an episodic task by setting $\gamma = 1$ and $R_t = 0$ for $t>T$.
Note that for rewards which are bound, this sum is guaranteed to converge to a finite value. 

\subsection{Policies and value functions}
A policy $\pi(a | s)$ is the probability of choosing the action $a$ when in state $s$. We can thus formulate our learning task as finding the policy that maximizes our reward and reinforcement learning as adapting an agent's policy as a result of its experience. For a given policy, we can define the value function of a state $s$ as the expected return from starting in that state and using the policy function $\pi$ for choosing all our future actions. We can write this as
\begin{equation}
	v_\pi(s) \equiv E_\pi (G_t | S_t = s).
	\label{eq:value_function}
\end{equation}
Alternatively, we can define the action-value function of $\pi$ as
\begin{equation}
	q_\pi (s, a) \equiv E_\pi(G_t | S_t = s, A_t = a).
\end{equation}
This is the expectation value for the return starting in state $s$ and choosing action $a$, but using the policy $\pi$ for all future actions. Note that one of the key ideas of reinforcement learning is to use such value functions, instead of the policy, to organize our learning process.

The value function of Eq.~\eqref{eq:value_function} satisfies a self-consistency equation,
\begin{align}
	v_\pi(s) &= E_\pi ( R_{t+1} + \gamma G_{t+1} | S_t = s)\label{eq:BE_expect}\\
	&= \sum_a \pi(a | s) \sum_{s', r} p(s', r | s, a) [ r + \gamma v_\pi (s')].
\end{align}
This equation, known as the \emph{Bellman equation}, relates the value of state $s$ to the expected reward and the (discounted) value of the next state after having chosen an action under the policy $\pi(a|s)$. 

As an example, we can write the Bellman equation for the strategy of always leaving the lamps on in our toy model. Then, we find the system of linear equations
\begin{align}
	v_{\rm on}({\rm h}) &= p({\rm h}, r_{\rm on} | {\rm h}, {\rm on}) [r_{\rm on} + \gamma v_{\rm on} ({\rm h})] + p({\rm l}, r_{\rm on} | {\rm h}, {\rm on}) [r_{\rm on} + \gamma v_{\rm on} ({\rm l})] \nonumber\\
	& = r_{\rm on} + \gamma [\alpha v_{\rm on}({\rm h}) + (1-\alpha) v_{\rm on}({\rm l})],\vspace{3pt}\\
	v_{\rm on}({\rm l}) &=  \beta[r_{\rm on} + \gamma v_{\rm on}({\rm l})] + (1-\beta) [r_{\rm fail} + \gamma v_{\rm on}({\rm h})],
\end{align}
from which we can solve easily for $v_{\rm on}({\rm h})$ and $v_{\rm on}({\rm l})$.

Instead of calculating the value function for all possible policies, we can directly try and find the optimal policy $\pi_*$, for which $v_{\pi_*}(s) > v_{\pi'}(s)$ for all policies $\pi'$ and $s\in\mathcal{S}$. For this policy, we find the \emph{Bellman optimality equations}
\begin{align}
	v_*(s) &= \max_a q_{*}(s, a)\nonumber\\
	&= \max_a E(R_{t+1} + \gamma v_*(S_{t+1}) | S_t = s, A_t = a)\label{eq:bellman-optimality-1a}\\
	&=\max_a \sum_{s', r} p(s', r | s, a) [ r + \gamma v_* (s')].
	\label{eq:bellman-optimality-1b}
\end{align}
Importantly, the Bellman optimality equations do not depend on the actual policy anymore. As such, Eq.~\eqref{eq:bellman-optimality-1b} defines a non-linear system of equations, which for a sufficiently simple MDP can be solved explicitly. 
For our toy example, the two equations for the value functions are
\begin{equation}
	v_*({\rm h}) = \max \left\{\begin{array}{l} r_{\rm on} + \gamma [\alpha v_*({\rm h}) + (1-\alpha) v_*({\rm l})] \\ 
	r_{\rm off} + \gamma [\alpha' v_*({\rm h}) + (1-\alpha') v_*({\rm l})] \end{array}\right. 
\end{equation}
and 
\begin{equation}
	v_*({\rm l}) = \max \left\{\begin{array}{l} \beta[r_{\rm on} + \gamma v_*({\rm l})] + (1-\beta) [r_{\rm fail} + \gamma v_*({\rm h})] \\ 
	\beta'[r_{\rm off} + \gamma v_*({\rm l})] + (1-\beta') [r_{\rm fail} + \gamma v_*({\rm h})]\\
	r_{\rm text} + \gamma v_*({\rm h})\end{array}\right. .
\end{equation}
Note that equivalent equations to Eqs.~\eqref{eq:bellman-optimality-1a} and \eqref{eq:bellman-optimality-1b} hold for the state-action value function
\begin{align}
	q_*(s, a) &= E(R_{t+1} + \gamma \max_{a'} q_*(S_{t+1},a'))\\
	&= \sum_{s', r} p(s', r | s, a) [ r + \gamma \max_{a'} q_* (s', a')].
	\label{eq:bellman-optimality-2}
\end{align}

Once we know $v_*$, the optimal policy $\pi_* (a| s)$ is the greedy policy that chooses the action $a$ that maximizes the right-hand side of Eq.~\eqref{eq:bellman-optimality-1b}. If, instead, we know $q_*(s,a)$, then we can directly choose the action which maximizes $q_*(s,a)$, namely $\pi_*(a | s) = {\rm argmax}_{a'} q_*(s, a')$, without looking one step ahead.

While Eqs~\eqref{eq:bellman-optimality-1b} or \eqref{eq:bellman-optimality-2} can be solved explicitly for a sufficiently simple system, such an approach, which corresponds to an exhaustive search, is often not feasible. In the following, we distinguish two levels of complexity: First, if the explicit solution is too hard, but we can still keep track of all possible value functions---we can choose either the state or the state-action value function---we can use a \emph{tabular} approach. A main difficulty in this case is the evaluation of a policy, or prediction, which is needed to improve on the policy. While various methods for \emph{policy evaluation} and \emph{policy improvement} exist, we will discuss in the following an approach called \emph{temporal-difference learning}. Second, in many cases the space of possible states is much too large to allow for a complete knowledge of all value functions. In this case, we additionally need to approximate the value functions. For this approximation, we can use the methods encountered in the previous chapters, such as (deep) neural networks.

	\subsection{Temporal-difference learning}
If we cannot explicitly solve the Bellman optimality equations---the case most often encountered---we need to find the optimal policy by some other means. If the state space is still small enough to keep track of all value functions, we can tabulate the value function for all the states and a given policy and thus, speak of \emph{tabular methods}. The most straight-forward approach, referred to as \emph{policy iteration}, proceeds in two steps: First, given a policy $\pi(a|s)$, the value function $v_{\pi}(s)$ is evaluated. Second, after this \emph{policy evaluation}, we can improve on the given policy $\pi(a|s)$ using the greedy policy 
\begin{equation}
	\pi'(a|s) = {\rm argmax}_a  \sum_{s', r} p(s', r| s, a) [r + \gamma v_\pi(s')].
	\label{eq:greedy_improvement}
\end{equation}
This second step is called \emph{policy improvement}. The full policy iteration then proceeds iteratively
\begin{equation}
	\pi_0 \rightarrow v_{\pi_0} \rightarrow \pi_1 \rightarrow v_{\pi_1} \rightarrow \pi_2 \rightarrow \cdots
\end{equation}
until convergence to $v_*$ and hence $\pi_*$. Note that, indeed, the Bellman optimality equation~\eqref{eq:bellman-optimality-1b} is the fixed-point equation for this procedure.

Policy iteration requires a full evaluation of the value function of $\pi_k$ for every iteration $k$, which is usually a costly calculation. Instead of fully evaluating the value function under a fixed policy, we can also directly try and calculate the optimal value function by iteratively solving the Bellman optimality equation, 
\begin{equation}
	v^{[k+1]} (s) = \max_a \sum_{s', r} p(s', r| s, a) [r + \gamma v^{[k]}(s')].
\end{equation}
Note that once we have converged to the optimal value function, the optimal policy is given by the greedy policy corresponding to the right-hand side of Eq.~\eqref{eq:greedy_improvement} An alternative way of interpreting this iterative procedure is to perform policy improvement every time we update the value function, instead of finishing the policy evaluation each time before policy improvement.
This procedure is called \emph{value iteration} and is an example of a \emph{generalized policy iteration}, the idea of allowing policy evaluation and policy improvement to interact while learning.

In the following, we want to use such a generalized policy iteration scheme for the (common) case, where we do not have a model for our environment. In this model-free case, we have to perform the (generalized) policy improvement using only our interactions with the environment. It is instructive to first think about how to evaluate a policy. We have seen in Eqs.~\eqref{eq:value_function} and \eqref{eq:BE_expect}
 that the value function can also be written as an expectation value,
\begin{align}
	v_{\pi} (s) &= E_\pi (G_t | S_t = s)\\
	&= E_\pi (R_{t+1} + \gamma v_{\pi}(S_{t+1})| S_t = s).\label{eq:policy_evaluation}
\end{align}
We can thus either try to directly sample the expectation value of the first line---this can be done using \emph{Monte Carlo sampling} over possible state-action sequences---or we try to use the second line to iteratively solve for the value function. In both cases, we start from state $S_t$ and choose an action $A_t$ according to the policy we want to evaluate. The agent's interaction with the environment results in the reward $R_{t+1}$ and the new state $S_{t+1}$. Using the second line, Eq.~\eqref{eq:policy_evaluation}, goes under the name \emph{temporal-difference learning} and is in many cases the most efficient method. In particular, we make the following updates
\begin{equation}
	v_\pi^{[k+1]} (S_t) = v_\pi^{[k]}(S_t) + \alpha [R_{t+1} + \gamma v_\pi^{[k]} (S_{t+1}) - v_\pi^{[k]} (S_{t}) ].
	\label{eq:policy_evaluation_modelfree}
\end{equation}
 The expression in the brackets is the difference between our new estimate and the old estimate of the value function and $\alpha<1$ is a learning rate. As we look one step ahead for our new estimate, the method is called one-step temporal difference method.

We now want to use generalized policy iteration to find the optimal value. We already encountered a major difficulty when improving a policy using a value function based on experience in Sec.~\ref{sec:expl_v_expl}: it is difficult to maintain enough exploration over possible action-state pairs and not end up exploiting the current knowledge. However, this sampling is crucial for both Monte Carlo methods and the temporal-difference learning we discuss here. In the following, we will discuss two different methods of performing the updates, both working on the state-action value function, instead of the value function. Both have in common that we look one step ahead to update the state-action value function. A general update should then be of the form
\begin{equation}
	q^{[k+1]} (S_t, a) = q^{[k]}(S_t, a) + \alpha [R_{t+1} + \gamma q^{[k]} (S_{t+1}, a') - q^{[k]} (S_{t}, a) ]
\end{equation}
and the question is then what action $a$ we should take for the state-action pair and what action $a'$ should be taken in the new state $S_{t+1}$. 

Starting from a state $S_0$, we first choose an action $A_0$ according to a policy derived from the current estimate of the state-action value function~\footnote{We assume here an episodic task. At the very beginning of training, we may initialize the state-action value function randomly.}, such as an $\epsilon$-greedy policy. For the first approach, we perform updates as
\begin{equation}
	q^{[k+1]} (S_t, A_t) = q^{[k]}(S_t, A_t) + \alpha [R_{t+1} + \gamma q^{[k]} (S_{t+1}, A_{t+1}) - q^{[k]} (S_{t}, A_t) ].
\end{equation}
As above, we are provided a reward $R_{t+1}$ and a new state $S_{t+1}$ through our interaction with the environment. To choose the action $A_{t+1}$, we again use a policy derived from $q^{[k]}(s=S_{t+1}, a)$. Since we are using the policy for choosing the action in the next state $S_{t+1}$, this approach is called \emph{on-policy}. Further, since in this particular case we use the quintuple $S_t, A_t, R_{t+1}, S_{t+1}, A_{t+1}$, this algorithm is referred to as \emph{Sarsa}. Finally, note that for the next step, we use $S_{t+1}, A_{t+1}$ as the state-action pair for which $q^{[k]}(s,a)$ is updated. 

Alternatively, we only keep the state $S_t$ from the last step and first choose the action $A_t$ for the update using the current policy. Then, we choose our action from state $S_{t+1}$ in greedy fashion, which effectively uses $q^{[k]}(s=S_t, a)$ as an approximation for $q_*(s=S_t, a)$. This leads to 
\begin{equation}
	q^{[k+1]} (S_t, A_t) = q^{[k]}(S_t, A_t) + \alpha [R_{t+1} + \gamma \max_a q^{[k]} (S_{t+1}, a) - q^{[k]} (S_{t}, A_t) ].
\end{equation}
and is a so-called \emph{off-policy} approach. The algorithm, a variant of which is used in AlphaGo, is called \emph{Q-learning}.

\subsection{Function approximation}
When the state-action space becomes very large, we face two problems: First, we can not use tabular methods anymore, since we can not store all values. Second and more important, even if we could store all the values, the probability of visiting all state-action pairs with the above algorithms becomes increasingly unlikely, in other words most states will never be visited during training.
Ideally, we should thus identify states that are `similar', assign them `similar' value, and choose `similar' actions when in these states. This grouping of similar states is exactly the kind of \emph{generalization} we tried to achieve in the previous sections.
Not surprisingly, reinforcement learning is most successful when combined with neural networks. 

In particular, we can parametrize a value function $\hat{v}_\pi(s; \theta)$ and try to find parameters $\theta$ such that $\hat{v}_\pi(s; \theta) \approx v_\pi(s)$. This approximation can be done using the supervised-learning methods encountered in the previous sections, where the target, or label, is given by the new estimate. In particular, we can use the \emph{mean squared error} to formulate a gradient descent method for an update procedure analogous to Eq.~\eqref{eq:policy_evaluation_modelfree}.
Starting from a state $S$ and choosing an action $A$ according to the policy $\pi(a|S)$, we update the parameters
\begin{equation}
	\theta^{[k+1]} = \theta^{[k]} + \alpha [R +\gamma \hat{v}_\pi(S'; \theta^{[k]}) - \hat{v}_\pi(S; \theta^{[k]}) ] \nabla \hat{v}_\pi (S;\theta^{[k]})
\end{equation}
with $0< \alpha < 1$ again the learning rate.
Note that, even though the new estimate also depends on $\theta^{[k]}$, we only take the derivative with respect to the old estimate. This method is thus referred to as \emph{semi-gradient method}. In a similar fashion, we can reformulate the Sarsa algorithm introduced for generalized gradient iteration.

%-----Concuding Remarks----------------------------------%
	\section{Concluding remarks}
In this lecture, `Machine Learning for the Sciences', we have discussed common structures and algorithms of machine learning to analyze data or learn policies to achieve a given goal.
Even though machine learning is often associated with neural networks, we have first introduced methods commonly known from statistical analysis, such as linear regression. Neural networks, which we used for most of this lecture, are much less controlled than these conventional methods. As an example, we do not try to find an absolute minimum in the optimization procedure, but one of many almost degenerate minima. This uncertainty might feel like a loss of control to a scientist, but it is crucial for the successful generalization of the trained network

The goal of our discussions was not to provide the details needed for an actual implementation---as all standard algorithms are provided by standard libraries such as TensorFlow or PyTorch, this is indeed not necessary---but to give an overview over the most important terminology and the common algorithms. We hope that such an overview is helpful for reading the literature and deciding, whether a given method is suitable for your own problems. 

To help with the use of machine learning in your own research, here a few lessons for a successful machine learner:
\begin{enumerate}
    \item 
    Your learning result can only be as good as your data set.
    \item
    Understand your data, its structure and biases.
    \item
    Try simpler algorithms first.
    \item
    Don't be afraid of lingo. Not everything that sounds fancy actually is.
    \item
    Neural networks are better at interpolating  than  extrapolating.
    \item
    Neural networks represent smooth functions well, not discontinuous or spiky ones.
\end{enumerate}

Regarding the use of machine learning in a scientific setting, several points should be kept in mind.
First, unlike in many other applications, scientific data often exhibits specific structure, correlations, or biases, which are known beforehand. It is thus important to use our prior knowledge in the construction of the neural network and the loss function. There are also many situations, where the output of the network has to satisfy conditions, such as symmetries, to be meaningful in a given context. This should ideally be included in the definition of the network.
Finally, scientific analysis needs to be well defined and reproducible. Machine learning, with its intrinsic stochastic nature, does not easily satisfy these conditions. It is thus crucial to document carefully the architecture and all hyperparameters of a network and training. The results should be compared to conventional statistical methods, their robustness to variations in the structure of the network and the hyperparameters should be checked.

\newpage
\begin{figure}
	\label{fig:final_overview}
	\includegraphics{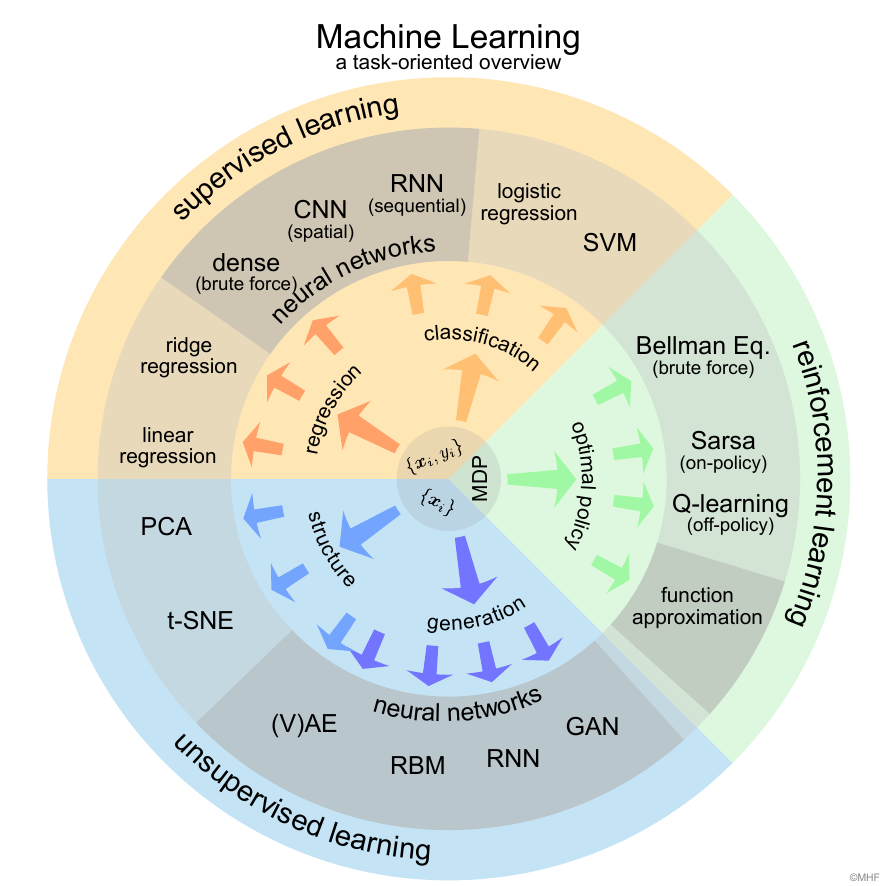}
	\caption{{\bf Machine Learning overview.} Methods covered in this lecture from the point of view of tasks.}
\end{figure}

\end{document}